\documentclass{ectj}

\usepackage{amsfonts,amssymb,graphics,epsfig,verbatim,bm,latexsym,amsmath,url,amsbsy}
\usepackage{mathrsfs}

\newtheorem{theorem}{Theorem}

\newtheorem{proposition}{Proposition}
\newtheorem{corollary}{Corollary}
\newtheorem{lemma}{Lemma}

\newtheorem{remark}{Remark}

\renewcommand{\thesection}{\arabic{section}}
\renewcommand{\theequation}{\arabic{section}.\arabic{equation}}

\renewcommand{\thelemma}{\arabic{section}.\arabic{lemma}}

\def\L{\scriptscriptstyle L}
\def\R{\scriptscriptstyle R}

\def\N{\scriptscriptstyle N}

\def\B{\scriptscriptstyle B}
\def\H{\scriptscriptstyle H}

\def\EL{\scriptscriptstyle EL}
\def\GMM{\scriptscriptstyle GMM}
\def\CUE{\scriptscriptstyle CUE}

\def\GEL{\scriptscriptstyle GEL}

\def\mS{\mathcal{S}}

\def\mF{\mathcal{F}}

\def\U{\mbox{\tiny U}}
\def\N{\mbox{\tiny N}}

\received{July 2022}
\accepted{November 2017}
\volume{20}
  
\title[ ]{Augmented two-step estimating equations with  nuisance functionals and complex survey data}

\author[ ]{Puying Zhao$^{\dagger}$ and Changbao Wu$^{\ddagger}$}

\address{$^{\dagger}$Yunnan University, Kunming, 650091, China.}
\email{pyzhao@live.cn}

\address{$^{\ddagger}$University of Waterloo, Waterloo, ON, N2L3G1, Canada.}
\email{cbwu@uwaterloo.ca}

\def\AmSTeX{$\cal A$\kern-.1667em\lower.5ex\hbox{$\cal M$}\kern-.125em
            $\cal S$-\TeX}
\def\BibTeX{{\rm B\kern-.05em{\sc i\kern-.025em b}\kern-.08em
            T\kern-.1667em\lower.7ex\hbox{E}\kern-.125emX}}

\begin{document}

  \begin{abstract}
    Statistical inference in the presence of nuisance functionals with complex survey data is an important topic in social and economic studies. The Gini index, Lorenz curves and quantile shares are among the commonly encountered examples. The nuisance functionals are usually handled by a plug-in nonparametric estimator and the main inferential procedure can be carried out through a two-step generalized empirical likelihood method. Unfortunately, the resulting inference is not efficient and the nonparametric version of the Wilks' theorem breaks down even under simple random sampling. We propose an augmented estimating equations method with nuisance functionals and complex surveys. The second-step augmented estimating functions obey the Neyman orthogonality condition  and automatically handle the impact of the first-step plug-in estimator, and the resulting estimator of the main parameters of interest is invariant to the first step method. More importantly, the generalized empirical likelihood based Wilks' theorem holds for the main parameters of interest under the design-based framework for commonly used survey designs, and the maximum generalized empirical likelihood estimators achieve the semiparametric efficiency bound. Performances of the proposed methods are demonstrated through simulation studies and an application using the dataset from the New York City Social Indicators Survey. 


  \end{abstract}

 \section{Introduction}
 
 \setcounter{equation}{0}
 
\noindent
In the era of big data, survey sampling remains one of the most important data collection vehicles for many fields of scientific investigations. Population health research, social and economic studies such as inequality measures and other policy related issues focus on a particular finite population, and design-based framework  with complex survey data is well suited for the inferential problems. Regression analysis and estimating equations with survey data have become a standard tool for statistical inference (Wu and Thompson, 2020). Empirical likelihood (EL), first proposed by Owen (1988) for independent samples, has been adapted successfully for survey data analysis through the pseudo EL approach (Chen and Sitter, 1999; Wu and Rao, 2006).  
Zhong and Rao  (2000)
studied EL inferences on population mean under stratified sampling.
Finite population parameters defined through the so-called census estimating equations and the related inferential procedures have been discussed by Chen and Kim (2014) and  Zhao et al. (2022) through the sample EL approach as well as the pseudo EL approach (Zhao and Wu, 2019). For parameters defined through U-statistics, jackknife EL can be used to reduce the computational complexities (Chen and Tabri, 2021).

Nuisance parameters of a finite dimension are typically handled through profiling; see, for instance, Berger and Torres (2016), Oguz-Alper and Berger (2016), Zhao {\em et al.} (2022), among others. Statistical inferences in the presence of nuisance functionals, i.e., nuisance parameters with infinite dimension, are an important problem, especially in social and economic studies. The most commonly used strategy is to use a two-step procedure where a consistent nonparametric estimator for the nuisance functional is constructed first and then used in the second step as a plug-in estimator for inferences on the main parameters of interest. Zhao et al. (2020) is among the first to discuss the design-based two-step EL method and the generalized method of moments (GMM)  method for complex survey data in the presence of nuisance functionals. The two-step survey weighted estimating equations (SWEE) approach discussed by Zhao et al. (2020),  however, has two major limitations. First, the maximum EL or GMM estimators are sensitive to the plug-in estimator for the nuisance functional and do not achieve the semiparametric efficiency bound. Second, the two-step EL ratio statistic does not lead to the nonparametric version of the Wilks' theorem even for simple random sampling. Applications of the results require tedious evaluation of the limiting distributions and design-based variance estimation and therefore are very difficult.
There has been a well developed statistical and econometric literature with non-survey data on
semiparametric efficiency bounds and Wilks' theorem for semiparametric models; see, for instance, Newey (1990),   Chen et al. (2008), Cattaneo (2010), Ackerberg et al. (2014),  Frazier and Renault (2017), Chernozhukov et al. (2018), Bravo et al. (2020),  Matsushita and Otsu (2020), 
Chernozhukov et al. (2022), among others.
Unfortunately, these  model-based efficient analytical procedures do not apply directly to complex survey data for design-based inference on finite population parameters.

This article presents an augmented two-step survey weighted estimating equations approach with nuisance functionals and complex survey data. The proposed methods are formulated through the generalized empirical likelihood (GEL, Newey and Smith, 2004; Parente and Smith, 2011) and represent a major advance over the usual two-step SWEE approach as discussed in Zhao et al. (2020). The GEL methods cover a large class of estimators as special cases, including the EL estimators (Owen  1988; Qin and Lawless 1994;  Chen and Sitter 1999; and Zhao et al., 2022), the continuous updating estimators (CU, Hansen et al., 1996), and the exponential tilting estimators (ET, Kitamura and Stutzer, 1997; and Imbens et al., 1998).
Under our proposed methods, the second-step augmented estimating functions  obey the Neyman orthogonality condition (Chernozhukov et al., 2018)  and  automatically handle the impact of the first-step plug-in estimator, and the resulting estimators of the main parameters of interest are invariant to the first step method for the plug-in estimator for the nuisance functional. Our methods are bias-corrected for the main parameters of interest in the sense that the nonparametric Wilks' theorem with standard chi-square limiting distributions holds under commonly used survey designs, and the maximum GEL estimators achieve the semiparametric efficiency bound. Our results are established  under the design-based framework for complex survey data, and our setting is very general, allowing the estimation equations system to be over-identified, the estimating functions to be nonsmooth and the plug-in estimator of the nuisance functional to be slower than root-$n$-consistent. In other words, our results allow the nuisance functional to be estimated through
any consistent nonparametric procedures in the first step, including the non-parametric series-based method (Newey, 1994b; Chen, 2007). These features have theoretical and practical importance since the estimating equations under study can be semiparametric and encompass a large class of econometric and statistical models.

Our proposed methods have immediate applications to inequality measures widely used in social and economic studies. Popular income inequality measures, such as the Lorenz curve, income shares and the Gini index, all involve nuisance functionals.
The measurement and analysis of income inequality have been well documented  in econometric  literature; see, for instance, Atkinson (1970), Beach and
Davidson (1983), Davidson and Duclos (2000), among others.
Income data are usually collected through complex surveys. Design-based approach to estimation and inference for income inequality measures with the focus on a particular finite population has been addressed by several authors; see, for instance,
 Nyg\aa rd and  Sandstr\"{o}m (1989),  Zheng (2002),  Bhattacharya (2007),
Goga and Ruiz-Gazen (2014), Zhao et al. (2020), among others.
Our proposed augmented two-step SWEE approach provides a powerful inference tool for this important topic in statistics and econometrics.

The rest of the paper is organized as follows.
In Section \ref{sec.method}, we first describe basic setup and the conventional two-step method of Zhao et al. (2020), and then present our proposed augmented two-step method with the GEL approach.
In Section \ref{sec.thm}, we examine the theoretical properties of the proposed point estimators and general hypothesis test problems. In Section \ref{sec.exam}, we discuss general procedures with illustrating examples on the construction of the augmentation terms. Complex survey designs and asymptotic variance estimation are discussed in
Section \ref{sec.design}.
Results from simulation studies are reported in Section \ref{sec.sim},
and an application to income share using the New York City Social Indicators Survey data is presented  in Section \ref{sec.data}.
Some concluding remarks are given in Section \ref{sec.dis}.
Technical details and proofs of the main theoretical results are presented  in Appendices A and B.

\section{Proposed Methods}\label{sec.method}
\setcounter{equation}{0}

\subsection{Preliminaries}
\label{sec.pre}

\noindent
Consider a survey population $\mathcal {U}_{\N}=\{1,\cdots,N\}$ with $N$ labelled units.
Let $Z\in\mathsf {R}^{d_z}$ be a $d_z$-dimensional vector of variables, and let   $Z_i$ be the value of $Z$ associated with the $i$th unit.  Denote by $\mathcal {F}_{\N}=(Z_1,\cdots, Z_{\N})$  the full set of vectors for the finite population. Let $\mS$ be the set of $n$ sampled units selected from $\mathcal {U}_{\N}$ by a probability survey design.
For asymptotic development, we assume there is a sequence of finite populations and a sequence of survey samples which allow $n$ and $N$ go to infinity; see Fuller (2009) for further details.
The sample size $n$ could be a random number under certain sampling designs.
Let $\pi_i = {\rm Pr}(i\in \mS)$ and $\pi_{ij} = {\rm Pr}(i,j \in \mS)$ be the first and second order inclusion probabilities.
A detailed discussion on the probability space induced by the survey design is given in Section \ref{subsec.ident}.

Let $\Theta \subseteq \mathsf {R}^{p}$ be the parameter space and assume it is a compact set. Let $\Psi$ be the space for the nuisance functional and assume it is a linear subspace of the space of square integrable functions with respect to $Z$.
Consider a vector of  $r$ real-valued functions $g(Z,\theta,\varphi)$ with a known form up to the unknown parameters of interest $\theta\in\Theta$ and the nuisance functional $\varphi\in\Psi$.
The main assumption on  $\mathcal {F}_{\N}$ is that    for some $\theta_{\N}\in\Theta$,
 \begin{equation}  \label{pee}
U_{\N}(\theta_{\N},\varphi_{\N})=\dfrac{1}{N}\sum\limits_{i=1}^Ng(Z_i,\theta_{\N},\varphi_{\N})= 0,
\end{equation}
where $\varphi_{\N}=\varphi_{\N}(\cdot,\theta_{\N})\in\Psi$ is the true value of the nuisance functional with the given finite population.
We assume that the census estimating equations (\ref{pee}) may be an over-identified
system in the sense that $r \geq p$, and that the estimating functions  $g(Z,\theta,\varphi)$ can be non-smooth in $\theta$  and/or $\varphi$. As in Chen et al. (2003) and Zhao et al. (2020), the nuisance functional $\varphi\in\Psi$ is allowed to depend on the parameters $\theta$ and the population data on $Z$.
For ease of presentation, we use the notation $(\theta,\varphi)\equiv(\theta,\varphi(\cdot,\theta))$, $(\theta,\varphi_{\N})\equiv(\theta,\varphi_{\N}(\cdot,\theta))$,
and $(\theta_{\N},\varphi_{\N})\equiv(\theta_{\N},\varphi_{\N}(\cdot,\theta_{\N}))$.

The empirical likelihood (EL) of Owen (1988) is a popular tool for effectively combining available auxiliary information and parameters of interest through a system of estimating equations (Qin and Lawless, 1994).  Assume that we have at hand a suitable ``plug-in" estimator
$\hat{\varphi}$ for $\varphi_{\N}$.
Let $(p_1,\cdots,p_n)$ be the discrete probability measure assigned to the  sampled units in $\mS$.
For any $\theta\in\Theta$ and the given $\hat{\varphi}$, the  two-step  survey-weighted EL ratio statistic is defined as  (Zhao et al., 2020)
\[
L_{\N}(\theta,\hat\varphi)=\sup\left\{\prod_{i\in \mS} (np_i) \; \Big| \; p_i\geq0,\sum_{i\in \mS}p_i=1, \sum_{i\in \mS}p_i \big[\pi_i^{-1}g(Z_i,\theta,\hat\varphi)\big] = 0\right\} \,.
\]
Note that the survey weights $\pi_i^{-1}$ are part of the parameter constraints. Using the standard Lagrange multiplier method, we can rewrite the EL ratio statistic as
$L_{\N}(\theta,\lambda,\hat\varphi) = \prod_{i\in \mS} \{1+\lambda^{\top}\pi_i^{-1}g(Z_i,\theta,\hat\varphi)\}^{-1}$,
where the Lagrange multiplier $\lambda=\lambda(\theta, \hat\varphi)$ is the solution to
$
\sum_{i\in \mS}\pi_i^{-1}g(Z_i,\theta,\hat\varphi)\{1+\lambda^{\top}\pi_i^{-1}g(Z_i,\theta,\hat\varphi)\}^{-1} = 0
$
with the given $\theta$ and $\hat\varphi$. The  two-step maximum EL estimator $\hat{\theta}_{\EL}$ for $\theta_{\N}$ is given by
\[
\hat\theta_{\EL}=\arg\min_{\theta\in\Theta}\sup_{\lambda\in\hat{\Lambda}_{\N,g}(\theta,\hat\varphi)}l_{\N}(\theta,\lambda,\hat\varphi) \,,
\]
where  $l_{\N}(\theta,\lambda,\varphi)=-\log L_{\N}(\theta,\lambda,\varphi)$ and $\hat{\Lambda}_{\N,g}(\theta,\hat\varphi) = \{\lambda: \lambda^{\top}\pi_i^{-1}g(Z_i,\theta,\hat\varphi)> -1, i\in \mS\}$ for the given $\theta$ and $\hat\varphi$.
The estimator $\hat{\theta}_{\EL}$  is also called the maximum sample EL estimator by Zhao et al. (2020).

Suppose that $\Psi$ is a vector space of   functions  endowed with the sup-norm metric
$\|\varphi\|_{\Psi}=\sup_{\theta}\|\varphi(\cdot,\theta)\|_{\infty}=\sup_{\theta}\sup_{{z}}|\varphi({z},\theta)| \,.
$
Define $\Theta(\delta)=\{\theta: \, \theta\in\Theta, \, \|\theta-\theta_{\N}\|\leq\delta\}$ and
$\Psi(\delta)=\{\varphi: \, \varphi\in\Psi, \; \|\varphi-\varphi_{\N}\|_{\Psi}\leq\delta\}$. Throughout the paper, we denote $E(\cdot \mid \mF_{\N})$ and ${\rm Var}(\cdot \mid \mF_{\N})$ to be the expectation and variance with respect to the design probability space, which will be  discussed in detail in Section \ref{subsec.ident}.
Let $n_{\B} = E(n \mid \mF_{\N})$ be the expected sample size under the sampling design.
Let $\|A\|=\{{\rm trace}(A^{\top}A)\}^{1/2}$ and $A^{\otimes 2}=AA^{\top}$ for any matrix or vector $A$.
We use $\stackrel{{\cal L}}{\rightarrow}$ to denote convergence in distribution.

The asymptotic properties of $\hat{\theta}_{\EL}$ under the design-based framework were investigated by
Zhao et al.  (2020) under the regularity conditions presented in Appendix A. In particular,  Condition A2 states that there exists vector-valued functions $U(\theta,\varphi)$ such that $\sup_{\theta \in \Theta,\varphi\in\Psi(\delta_{\N})}\|U_{\N}(\theta,\varphi) - U(\theta,\varphi)\|=o(1)$ with $\delta_{\N}=o(1)$; Condition A4 indicates that
for any  $(\theta,\varphi)\in\Theta(\delta)\times\Psi(\delta)$,
the  ordinary derivative $\Gamma_1(\theta,\varphi)$ in $\theta$ of  the limiting  functions $U(\theta,\varphi)$ exists and
  satisfies
$\Gamma_1(\theta,\varphi)(\bar{\theta}-\theta)=\lim_{t\rightarrow 0}[U(\theta+t(\bar{\theta}-\theta),\varphi(\cdot,\theta+t(\bar{\theta}-\theta)))
-U(\theta,\varphi(\cdot,\theta))]/t$
for $\bar{\theta}\in\Theta$;  and  Condition A5 requires that  for any $\theta\in \Theta(\delta)$,  the limiting function $U(\theta,\varphi)$ is pathwise differentiable at $\varphi\in \Psi(\delta)$ in the direction $[\bar{\varphi}-\varphi]$   in the sense that  the limit $D(\theta,\varphi)[\bar{\varphi}-\varphi]=\lim_{t\rightarrow 0}[U(\theta,\varphi(\cdot,\theta)+t(\bar{\varphi}(\cdot,\theta)-\varphi(\cdot,\theta)))
-U(\theta,\varphi(\cdot,\theta))]/t$ exists for $\{\varphi+t(\bar{\varphi}-\varphi):t\in[0,1]\}\subset \Psi$. Moreover, Condition A6 specifies that the pathwise derivative $D(\theta_{\N},\varphi_{\N})[\hat{\varphi}-\varphi_{\N}]$      is of the  following form:
\begin{eqnarray}  \label{first-step}
D(\theta_{\N},\varphi_{\N})[\hat{\varphi}-\varphi_{\N}]=\dfrac{1}{N}\sum_{i\in \mS}\pi_i^{-1}\Xi(Z_i,\theta_{\N},\varphi_{\N})+o_p(n_{\B}^{-1/2}),
\end{eqnarray}
where $\Xi(Z,\theta_{\N},\varphi_{\N})$ \ has finite fourth population moments and  $\sum_{i\in \mS}\pi_i^{-1}\Xi(Z_i,\theta_{\N},\varphi_{\N})$ is asymptotically normally distributed with mean zero and variance-covariance matrix at the order $O(n_{\B}^{-1}N^2)$. The following results were established by Zhao et al. (2020).

\begin{proposition}
\label{prop1}
Under the regularity conditions A1--A8 specified in Appendix A and as $N \rightarrow \infty$,
\begin{itemize}
\item [(a)]
$
n_{\B}^{1/2}(\hat{\theta}_{\EL}-\theta_{\N}) \stackrel{{\cal L}}{\rightarrow}  N(0,V_1),
$
where 
$
V_1=\Sigma_1\Gamma_1^{\top}W_1^{-1}\Omega W_1^{-1}\Gamma_1 \Sigma_1,
\label{V1}
$ 
$\Sigma_1=(\Gamma_1^{\top}W_1^{-1}\Gamma_1)^{-1}$,
$\Gamma_1=\Gamma_1(\theta_{\N},\varphi_{\N})$,
 $
\Omega= (n_{\B}/ N^2){\rm Var}\{\sum_{i\in \mS}\pi_i^{-1}[g(Z_i,\theta_{\N},\varphi_{\N})+\Xi(Z_i,\theta_{\N},\varphi_{\N})]\mid \mF_{\N}\}
$, and $W_1= (n_{\B}/ N^2)\sum_{i=1}^{N}\pi_i^{-1}g(Z_i,\theta_{\N},\varphi_{\N})^{\otimes 2}$.
\item [(b)]
$-2\log L_{\N}(\theta_{\N},\hat\varphi)\stackrel{{\cal L}}{\rightarrow}\delta_1\chi_{1}^{2}+\cdots+\delta_r\chi_{r}^{2}$,
where the $\chi_{j}^{2}$'s  are independent $\chi^2$ random variables with one degree of freedom and the weights $\delta_j$
are the eigenvalues of   $W_1^{-1}\Omega$.
\end{itemize}
\end{proposition}

Proposition \ref{prop1} shows that for complex survey data the Wilks' theorem breaks down with the two-step EL approach even under simple random sampling.
When using the two-step EL ratio statistic $-2\log L_n(\theta,\hat\varphi)$  to construct confidence regions or conduct hypothesis tests on $\theta_{\N}$, one needs to
approximate the distribution of a weighted $\chi^2$ random variable, and finding the weights $\delta_j$ requires estimation of the matrix $W_1$ and the design-based variance-covariance matrix $\Omega$. The last component is especially cumbersome for complex surveys.
A bootstrap calibration procedure could be an option but the method is computationally intensive and theoretical justifications are not available for general survey designs. Moreover, inferences based on the two-step EL approach do not use information on the main parameters and on the nuisance functionals simultaneously and therefore are not efficient, which motivates the research presented in the current paper. For an in-depth discussion on weighted 
chi-squared statistic, see Rao  and Scott (1981).

\subsection{An augmented survey weighted estimating equations approach}
\label{sec.gel}

\subsubsection{Neyman orthogonal score}
\label{sec.orthogonal-score}
We first investigate the key condition for restoring Wilks’ phenomenon in  two-step survey weighted  EL inferences. 
We refer to $\pi_i^{-1}\Xi(Z_i,\theta,\varphi)$ defined in (\ref{first-step}) as the first step survey weighted influence function (FSSWIF). 
 It follows from  the arguments of Zhao et al. (2020)   that
\begin{equation}  \label{swee-taylor}
\frac{1}{N}\sum\limits_{i\in \mS}\pi_i^{-1}g(Z_i,\theta_{\N},\hat{\varphi})=\frac{1}{N}\sum\limits_{i\in \mS}\pi_i^{-1}\{g(Z_i,\theta_{\N},\varphi_{\N})+\Xi(Z_i,\theta_{\N},\varphi_{\N})\}+o_p(n_{\B}^{-1/2}).
\end{equation}
Therefore, we conclude from (\ref{swee-taylor})  and the arguments of Zhao et al. (2020)   that the     two-step survey weighted  EL ratio statistic  satisfies a nonparametric version of Wilks' theorem if the 
FSSWIF at $(\theta_{\N},\varphi_{\N})$  is zero, or equivalently $\Xi(Z,\theta_{\N},\varphi_{\N})=0$. This motivates us to propose an augmented survey weighted estimating equations approach  to mitigate the impact of the plug-in estimator $\hat{\varphi}$ of the nuisance functional in the usual two-step survey weighted estimating equations through an ingenious augmentation term for the main estimating functions.
Motivated by Chernozhukov et al. (2018)  and Chernozhukov et al. (2022), we define the  augmented estimating functions as, 
 \begin{eqnarray}\label{nee}
\psi(Z,\theta,\varphi)=g(Z,\theta,\varphi)+\Xi(Z,\theta,\varphi).
\end{eqnarray}

With  the given finite population $\mathcal {F}_{\N}=(Z_1,\cdots, Z_{\N})$, we define the following  augmented population (census) estimating functions
 \begin{equation}  \label{pee-new}
\mathbb{U}_{\N}(\theta,\varphi)=\dfrac{1}{N}\sum\limits_{i=1}^N\psi(Z_i,\theta,\varphi).
\end{equation}
It follows from the original population estimating equations given in (\ref{pee}) and  Condition A3(ii) above that   $\mathbb{U}_{\N}(\theta,\varphi) = 0$ has a unique root at
$(\theta, \varphi)  = (\theta_{\N},\varphi_{\N})$.  

We analogously impose some conditions on the augmented population  estimating functions:
(i)
 there exists real-valued functions $\mathbb{U}(\theta,\varphi)$  such that 
 $\sup_{(\theta,\varphi)\in\Theta\times\Psi(\delta_{\N})}\|\mathbb{U}_{\N}(\theta,\varphi)-\mathbb{U}(\theta,\varphi)\|=o(1)$  for all sequences of positive numbers $\{\delta_{\N}\}$ with $\delta_{\N} = o(1)$;  (ii) for any $\theta\in \Theta(\delta)$,  the limiting function $\mathbb{U}(\theta,\varphi)$ is pathwise differentiable at $\varphi\in \Psi(\delta)$ in the direction $[\bar{\varphi}-\varphi]$   in the sense that   $\mathbb{D}(\theta,\varphi)[\bar{\varphi}-\varphi]=\lim_{t\rightarrow 0}[\mathbb{U}(\theta,\varphi(\cdot,\theta)+t(\bar{\varphi}(\cdot,\theta)-\varphi(\cdot,\theta)))
-\mathbb{U}(\theta,\varphi(\cdot,\theta))]/t$ exists for $\{\varphi+t(\bar{\varphi}-\varphi):t\in[0,1]\}\subset \Psi$.  The   augmented population estimating functions    has the orthogonality property in the sense that
\begin{equation}  \label{orthg-eq}
\mathbb{D}(\theta_{\N},\varphi_{\N})[\varphi-\varphi_{\N}]=0,\,\,\,\mbox{for all}\,\,\,\varphi\in\Psi.
\end{equation}
Given the  set of sampled units $\mS$ and the set of survey weights $\{\pi_i^{-1},i\in\mS\}$,  the  augmented survey weighted estimating functions are defined as
\begin{equation}  \label{swee-ad}
\hat{\mathbb{U}}_{\N}(\theta,\varphi)=\frac{1}{N}\sum\limits_{i\in \mS}\pi_i^{-1}\psi(Z_i,\theta,\varphi).
\end{equation}
It is clear  that  $E\{\hat{\mathbb{U}}_{\N}(\theta,\varphi)\mid\mF_{\N}\}=\mathbb{U}_{\N}(\theta,\varphi)$ for any $(\theta,\varphi)\in\Theta\times\Psi$.  
 The orthogonality property  in (\ref{orthg-eq})   implies that, modulo some regularity conditions, the following invariance property holds:
\begin{equation}  \label{swee-invariance}
\hat{\mathbb{U}}_{\N}(\theta_{\N},\hat{\varphi})=\hat{\mathbb{U}}_{\N}(\theta_{\N},\varphi_{\N})
+o_p(n_{\B}^{-1/2}).
\end{equation}
In this sense, the  augmented estimating functions defined in (\ref{nee})  are also referred to as  Neyman orthogonal score (Chernozhukov et al., 2018; Chernozhukov et al., 2022).

\subsubsection{Generalized empirical likelihood}
  For scenarios where $r=p$,  a design-based estimator of $\theta_{\N}$ may be obtained by solving $\hat{\mathbb{U}}_{\N}(\theta,\hat{\varphi})=0$.
The resulting estimator for $\theta_{\N}$ is bias-corrected due to  the invariance property (\ref{swee-invariance}).
In other words, the estimation of the nuisance functional has no impact asymptotically on the estimation of the main parameters of interest. Note that the discussions of Binder (1983) and Godambe and Thompson (1986) on survey weighted estimating equations based inferences do not apply to the augmented estimating equations proposed here.

For general cases where $r\geq p$, we consider the generalized empirical likelihood (GEL) approach. 
GEL has a well-know dual representation that facilitates computations and analysis of higher-order properties (Newey and Smith, 2004). 
Let $\rho(v)$ be a concave function of the scalar $v\in \mathcal{V}$ (an open interval containing zero);
let $\rho_j(v)=\partial^j\rho(v)/\partial v^j$ and $\rho_j=\rho_j(0)$ for $j=0,1,2,\ldots$.
Following Newey and Smith (2004), we impose a normalization constraint on $\rho(v)$ such that $\rho_1=\rho_2=-1.$
Define  the re-centred  GEL objective function as
\begin{equation*}
\hat{P}_{\N}(\theta,\eta,\varphi)
=\sum\limits_{i\in \mS}\big\{\rho\big(\eta^{\top}\pi_i^{-1}\psi(Z_i,\theta,\varphi)\big)-\rho_0\big\},
\end{equation*}
where $\eta$ is an $r$-vector of ``pseudo parameters'' related to the Lagrange multipliers.

Given the first step  plug-in estimator
$\hat{\varphi}$ for the nuisance functional $\varphi_{\N}$,   a class of  augmented design-based  two-step  GEL estimators for $\theta_{\N}$ can be defined as the solution to the following saddle-point problem
\begin{equation}  \label{gele}
\hat{\theta}_{\GEL}=\arg\inf_{\theta\in\Theta}
\sup_{\eta\in\hat{\Lambda}_{\N,\psi}(\theta,\hat{\varphi})}\hat{P}_{\N}(\theta,
\eta,\hat{\varphi}),
\end{equation}
where
$\hat{\Lambda}_{\N,\psi}(\theta,\varphi)=\{\eta:\eta^{\top}\pi_i^{-1}
\psi(Z_i,\theta,\varphi)\in\mathcal {V}, i\in\mS\}$.
For  nonsmooth estimating functions, the augmented design-based
two-step  GEL  estimators $\hat{\theta}_{\GEL}$ are  no longer required to be defined by (\ref{gele}) but satisfy
\[
\hat{P}_{\N}(\hat{\theta}_{\GEL},\hat{\eta}_{\GEL},\hat{\varphi})
\leq\arg\inf_{\theta\in\Theta}\sup_{\eta\in\hat{\Lambda}_{\N,\psi}(\theta,\hat{\varphi})}
\hat{P}_{\N}(\theta,\eta,\hat{\varphi})+o_p(1),
\]
where    $\hat{\eta}_{\GEL}=\eta(\hat\theta_{\GEL},\hat{\varphi})$ and
$\eta(\theta,\varphi)=\arg\max_{\eta\in\hat{\Lambda}_{\N,\psi}(\theta,\varphi)}
\hat{P}_{\N}(\theta,\eta,\varphi)$.
Specific choices of the function $\rho(\cdot)$ for the GEL estimators lead to specific types of estimators.
The EL estimator is obtained by taking $\rho(v)=\log(1-v)$ and $\mathcal {V}=(-\infty,1)$;
the  ET estimator is constructed by setting $\rho(v)=-\exp(v)$.
The CU estimator is defined as
$$
\hat{\theta}_{\CUE}=\mathop{\arg\min}_{\theta} n_{\B}\hat{\mathbb{U}}_{\N}(\theta,\hat{\varphi})^{\top}
\big\{\hat{W}_{\N}(\theta,\hat{\varphi})\big\}^{-1} \hat{\mathbb{U}}_{\N}(\theta,\hat{\varphi}),
$$
where $\hat{\mathbb{U}}_{\N}(\theta,\varphi)$ is defined in (\ref{swee-ad}) and  $\hat{W}_{\N}(\theta,\varphi)=
n_{\B}N^{-2}\sum_{i\in \mS}\pi_i^{-2}\psi(Z_i,\theta,\varphi)^{\otimes 2}$.
Using the arguments of Newey and Smith (2004), we can show that $\hat{\theta}_{\CUE}=\hat{\theta}_{\GEL}$
if $\rho(\cdot)$ is quadratic.  
A dual representation to the augmented design-based two-step GEL estimators is described in detail in supplementary material.

Let
$\tilde{\theta}$ be an initial design-consistent estimator for $\theta_{\N}$. Then the augmented  design-based  two-step  GMM estimator  of $\theta_{\N}$ is obtained as
$$
\hat{\theta}_{\GMM}=\mathop{\arg\min}_{\theta} \hat{\mathbb{U}}_{\N}(\theta,\hat{\varphi})^{\top}
\{\hat{W}_{\N}(\tilde{\theta},\hat{\varphi})\}^{-1} \hat{\mathbb{U}}_{\N}(\theta,\hat{\varphi}).
$$
Detailed discussion on the regular design-based two-step  GMM estimator can be found in Zhao et al.  (2020).

The maximum GEL-based estimators for the empirical probabilities $(p_1,\cdots,p_n)$ are given by
\begin{eqnarray}\label{gel-ep}
\hat{p}_i=\dfrac{
\rho_1\big(\hat{\eta}_{\GEL}^{\top}\pi_i^{-1}\psi(Z_i,\hat{\theta}_{\GEL},
\hat{\varphi})\big)}{\sum\limits_{j\in\mS}\rho_1\big(\hat{\eta}_{\GEL}^{\top}\pi_j^{-1}
\psi(Z_j,\hat{\theta}_{\GEL},\hat{\varphi})\big)},~~ i \in\mS \,,
\end{eqnarray}
which satisfy the sample moment condition $\sum_{i\in\mS}\hat{p}_i\psi(Z_i,\hat{\theta}_{\GEL},\hat{\varphi})=0$.

The invariance property (\ref{swee-invariance}),  together with some regularity conditions, imply   that
$
n_{\B}^{1/2}\hat{\mathbb{U}}_{\N}(\theta_{\N},\hat{\varphi})\stackrel{{\cal L}}{\rightarrow} N(0,\Omega),
$
where  $
\Omega= (n_{\B}/ N^2){\rm Var}\{\sum_{i\in \mS}\pi_i^{-1}[g(Z_i,\theta_{\N},\varphi_{\N})+\Xi(Z_i,\theta_{\N},\varphi_{\N})]\mid \mF_{\N}\}
$. Under single-stage PPS sampling with replacement or single-stage PPS sampling without replacement with negligible sampling fractions, we have that 
$\Omega=W_2$,  where $
 W_2= (n_{\B}/ N^2)\sum_{i=1}^{N}\pi_i^{-1}\psi(Z_i,\theta_{\N},\varphi_{\N})^{\otimes 2}.$
This, coupled with  the fact   $\|\hat{W}_{\N}(\theta_{\N},\hat{\varphi})$ $ - W_2\|=o_p(1)$, intuitively implies that  the Wilks’ phenomenon is restored in the augmented design-based two-step GEL
inferences.  More details can be found in Sections    \ref{sec.thm} and  \ref{cee-approach}.

\section{Main Results}
\label{sec.thm}

\setcounter{equation}{0}

\noindent
We now present the main   results on the proposed methods.
We first present theorems regarding
the consistency and efficiency  of the augmented design-based two-step GEL estimators $\hat{\theta}_{\GEL}$. We then discuss the
construction of   confidence regions and general hypothesis testing problems
on $\theta_{\N}$  based on the GEL ratio statistic. The following regularity conditions are used for the establishment of the main results.

 \begin{itemize}
\item [B1.]
 There exists real-valued functions $\mathbb{U}(\theta,\varphi)$  such that $\sup_{(\theta,\varphi)\in\Theta\times\Psi(\delta_{\N})}\|\mathbb{U}_{\N}(\theta,\varphi)-\mathbb{U}(\theta,\varphi)\|=o(1)$  for all sequences of positive numbers $\{\delta_{\N}\}$ with $\delta_{\N} = o(1)$, and    $\mathbb{U}(\theta,\varphi)$  satisfies the following conditions:
 \begin{itemize}
\item [(i)]
The  ordinary derivative $\Gamma_2(\theta,\varphi)$ of    $\mathbb{U}(\theta,\varphi)$ with respect to $\theta$ 
  exists for $\theta\in\Theta(\delta)$,  and is continuous  at $\theta=\theta_{\N}$; the matrix $\Gamma_2(\theta,\varphi)$ has
full column rank $p$;
\item [(ii)]  There exists a unique $\theta_0\in\Theta$  such that   \ $\mathbb{U}(\theta_0,\varphi_0)=0$, where  $\varphi_0=\varphi_0(\cdot,\theta_0)\in\Psi$;
\item [(iii)]
For any  $\theta\in\Theta$, $\mathbb{U}(\theta,\varphi)$ is continuous (with respect to the metric $\|\cdot\|_{\Psi}$) in $\varphi$ at $\varphi = \varphi_0$.
 \end{itemize}

 \item [B2.] The  augmented estimating functions $\psi(Z,\theta,\varphi)$ defined in (\ref{nee}) satisfy the following conditions:
\begin{itemize}
\item[(i)] $\max_{i\in \mS}\sup_{\theta\in\Theta,\varphi\in\Psi}\|\psi(Z_i,\theta,\varphi)\|=o_p(n_{\B}^{1/\alpha})$ for some $\alpha>2$;
\item[(ii)]
  For any sequence $c_{\N}=O(N^{-\kappa})$ with $\kappa\in(1/4,1/2]$,
$$\sup_{(\theta,\varphi)\in\Theta\times\Psi}  \dfrac{1}{N}\sum_{i=1}^{N}\|\psi(Z_i,\theta,\varphi)-\psi(Z_i,\theta+c_{\N},\varphi+c_{\N})\|= O(|c_{\N}|) \, .$$
\end{itemize}

\item[B3.]
(i) For any sequence of positive numbers $\{\delta_{\N}\}$ with $\delta_{\N}=o(1)$,
$$\sup_{(\theta,\varphi),(\theta',\varphi') \in \Theta(\delta_{\N})\times\Psi(\delta_{\N})} \| \mathbb{U}_{\N}(\theta,\varphi)  - \mathbb{U}(\theta,\varphi)-[\mathbb{U}_{\N}(\theta',\varphi')  - \mathbb{U}(\theta',\varphi')] \| = o(N^{-1/2});$$
(ii) For all $\delta>0$ and some positive constant $c$,
$$
\sup_{(\theta,\varphi),(\theta',\varphi') \in \Theta(\delta)\times\Psi(\delta)}
{\rm Var}\Big\{[\hat{\mathbb{U}}_{\N}(\theta,\varphi)-\hat{\mathbb{U}}_{\N}(\theta',\varphi')]\mid \mathcal {F}_{\N}\Big\}\leq cn_{\B}^{-1}|\delta| \,.
$$

\item [B4.]
For all $(\theta, \varphi), (\theta', \varphi')\in\Theta(\delta_{\N})\times\Psi(\delta_{\N})$ with   $\delta_{\N}=o(1)$,
$\|\mathbb{U}(\theta,\varphi)-\mathbb{U}(\theta,\varphi')\|\leq c\|\varphi-\varphi'\|_{\Psi}^2$ for some constant $c\ge0$.

\end{itemize}

Condition B1  states that the limiting function of the augmented  population estimating equations $\mathbb{U}_{\N}(\theta,\varphi)$ defined in (\ref{pee-new}) exists with certain smoothness properties.
Condition B2(i) is commonly adopted in the literature on EL inference with estimating equations,
while condition B2(ii) gives  a bound
on the variation of the estimating functions. Condition B3(i) restricts
the   class of moment functions under study  by requiring that  the empirical process  $\{\mathbb{U}_{\N}(\theta,\varphi)  - \mathbb{U}(\theta,\varphi):\theta\in\Theta,\varphi\in\Psi\}$ is asymptotically stochastically equicontinuous, which can easily be verified under the model-based framework. Condition
B3(ii) is on the correlation between two Horvitz-Thompson estimators at two close points of $(\theta,\varphi)$. As discussed in the proof of  Theorem \ref{thm2}  below,  Condition B4 is key to  guaranteeing the invariance property  (\ref{swee-invariance}).
Moreover,  if  the orthogonality equation (\ref{orthg-eq}) holds, then
$\|\mathbb{U}(\theta_{\N},\varphi)-\mathbb{U}(\theta,\varphi_{\N})-\mathbb{D}(\theta_{\N},\varphi_{\N})[\varphi-\varphi_{\N}]\|=\|\mathbb{U}(\theta_{\N},\varphi)-\mathbb{U}(\theta_{\N},\varphi_{\N})\|\leq c\|\varphi-\varphi'\|_{\Psi}^2$, which is  a commonly used   condition in the  literatures of two-step semiparametric inferences,  see, e.g.,  Chen et al. (2003) and Chen (2007).   Condition B4, together with condition A6 presented in Appendix A, imply that  the first step plug-in estimator $\hat{\varphi}$
can attain rate of convergence that are faster than $n_{\B}^{-1/4}$.

\subsection{Consistency and efficiency}

\noindent
We first  study the design consistency and asymptotic normality of the proposed augmented design-based two-step  GEL estimators. The main results are presented in the following two theorems. Regularity conditions A1--A8 were used by Zhao et al. (2020) and are listed in Appendix A.

\begin{theorem}
\label{thm1}
Suppose that $\hat{\varphi}=\varphi_{\N}+o_p(1)$, and that conditions  A1, A7--A8, B1--B2  hold. Then the proposed  augmented design-based two-step GEL  estimator  is design-consistent for $\theta_{\N}$ in the sense that
$
\lim_{\N \rightarrow \infty}{\rm Pr}\{\|\hat{\theta}_{\GEL} - \theta_{\N} \| > \epsilon \mid \mathcal {F}_{\N}\} = 0$ for any $\epsilon >0$.
\end{theorem}

\begin{theorem}
\label{thm2}
Suppose that conditions  A1, A6--A8, B1 and B3--B4  hold, and that
$\hat{\theta}_{\GEL}=\theta_{\N}+o_p(1)$.
 Then, as $N \rightarrow \infty$,
$$
n_{\B}^{1/2}(\hat{\theta}_{\GEL}-\theta_{\N})\stackrel{{\cal L}}{\rightarrow}N(0,V_2),
$$
where
 $
V_2=\Sigma_2\Gamma_2^{\top}W_2^{-1}\Omega W_2^{-1}\Gamma_2 \Sigma_2 \,,
\label{V1}
$
$\Sigma_2=(\Gamma_2^{\top}W_2^{-1}\Gamma_2)^{-1}$,    $\Gamma_2=\Gamma_2(\theta_{\N},\varphi_{\N})$,
 $
 W_2= (n_{\B}/ N^2)\sum_{i=1}^{N}\pi_i^{-1}\psi(Z_i,\theta_{\N},\varphi_{\N})^{\otimes 2},$ with
 $\Omega$ defined in Proposition \ref{prop1}.
\end{theorem}

\begin{corollary}
\label{cor1}
Suppose that the assumptions for Theorem \ref{thm2} hold. Under single-stage PPS sampling with replacement or single-stage PPS sampling without replacement with negligible sampling fractions, the asymptotic variance-covariance matrix $V_2=\Sigma_2$.
\end{corollary}

\begin{remark}
One important observation from the results presented in Theorem \ref{thm2} is that the limiting distribution of our proposed estimator of $\theta_{\N}$ based on the augmented estimating equations is invariant to the first-step estimator of the nuisance functional. This leads to the earlier statement that the proposed augmented two-step GEL estimators are less sensitive to the estimation of nuisance functionals.
By combining the results from Corollary  \ref{cor1} with the arguments of Ackerberg et al. (2014, Lemma 1), we conclude that the proposed estimators also achieve the semiparametric efficiency bound under the survey designs specified in Corollary  \ref{cor1}.
As discussed further in Section  5, the estimator $\hat{\theta}_{\GEL}$ together with its standard errors can be used as bases for statistical inferences.
Note that if  the nuisance functional $\varphi_{\N}$ does not depend on the parameter of interest $\theta_{\N}$ and the estimating equations (\ref{pee}) is 
just-identified (i.e., $r=p$),  then the    proposed augmented two-step GEL estimators have the same limit distribution as the  two step
EL estimator proposed in Zhao et al. (2020).   However, Zhao et al.'s (2020) estimator  does not  satisfy   invariance
property similar to that stated in (\ref{swee-invariance}).
\end{remark}

\subsection{Hypothesis testing}

\noindent
We next consider the GEL ratio based confidence regions and hypothesis tests on $\theta_{\N}$.
The GEL ratio statistic for $\theta_{\N}$ is defined as
\begin{equation*}
\label{sel-test1}
{\rm T}_{\N}(\theta) = -2\{[\hat{P}_{\N}(\hat{\theta}_{\GEL},\hat{\eta}_{\GEL},\hat{\varphi})- \hat{P}_{\N}(\theta,\eta_{\theta},\hat{\varphi})\}
\end{equation*}
for the given $\theta$, where $\hat{\eta}_{\GEL}=\eta(\hat{\theta}_{\GEL},\hat{\varphi})$ and $\eta_{\theta}=\eta(\theta,\hat{\varphi})$.
The asymptotic distribution of ${\rm T}_{\N}(\theta)$ at $\theta = \theta_{\N}$ is given in the following theorem.

\begin{theorem}\label{thm3}
Suppose that the assumptions for Theorem \ref{thm2}  hold. Then, as $N \rightarrow \infty$,
\begin{eqnarray*}
{\rm T}_{\N}(\theta_{\N})\;\ \stackrel{{\cal L}}{\rightarrow} \;\; Q^{\top} \Delta Q,
\end{eqnarray*}
where
$Q\sim N(0, I_{r})$,
$
\Delta =\Omega^{1/2} W_2^{-1} \Gamma_2 (\Gamma_2^{\top}W_2^{-1}\Gamma_2)^{-1} \Gamma_2^{\top}W_2^{-1}\Omega^{1/2}
$, and $I_{r}$ is the $r\times r$ identity matrix.
\end{theorem}

\begin{corollary}
\label{cor2}
Suppose that the assumptions for Theorem \ref{thm3} hold. Under single-stage PPS sampling with replacement or single-stage PPS sampling without replacement with negligible sampling fractions, we have
$
{\rm T}_{\N}(\theta_{\N}) \; \stackrel{{\cal L}}{\rightarrow} \; \chi_p^2
$
as $N \rightarrow \infty$.
\end{corollary}

\begin{remark}
Theorem \ref{thm3}  indicates that, under general unequal probability sampling designs, the proposed augmented design-based two-step GEL  ratio statistic  converges in distribution to a weighted
chi-square random variable, with the weights independent of the first-step estimation of the nuisance functions. More importantly,  Corollary \ref{cor2} shows that the proposed GEL ratio statistics satisfy a nonparametric version of the Wilks' Theorem under commonly used single-stage unequal probability sampling designs.  Note that   the Wilks’ theorem breaks down in the   two-step survey weighted  EL approach proposed in Zhao et al. (2020) even under simple random sampling.
\end{remark}

The standard Wilks  phenomenon with the proposed two-step survey weighted GEL  provides a convenient way to construct  confidence regions for $\theta_{\N}$ defined via the population estimating equations (\ref{pee}) or test the hypothesis $H_0$: $\theta_{\N} = \theta_0$ with a pre-specified $\theta_0$.
The $(1-\alpha)100\%$ confidence region for $\theta_{\N}$ can be constructed as
\[
\mathcal {C}_\alpha = \Big\{\theta \mid -2\Big[\hat{P}_{\N}(\hat{\theta}_{\GEL},\hat{\eta}_{\GEL},\hat{\varphi})- \hat{P}_{\N}\Big(\theta,\eta(\theta,\hat{\varphi}),\hat{\varphi}\Big)\Big]\leq \chi^2_{p,\alpha}\Big\} \,,
\]
where $\chi^2_{p,\alpha}$ satisfies ${\rm Pr}(\chi_p^2\geq \chi^2_{p,\alpha})=\alpha$.
The empirical results from simulation studies presented in section \ref{sec.sim} provide strong evidence that the standard Wilks' Theorem is also a good approximation for stratified sampling and cluster sampling.

Auxiliary population information is often available for different sources.  Including the information for survey data analysis often leads to efficiency gains in estimation and hypothesis testing problems. Auxiliary information can often be formed through a set of population estimating equations as
 \begin{equation}  \label{pee-side}
\mathfrak{U}_{\N}(\theta_{\N})=\dfrac{1}{N}\sum\limits_{i=1}^N q(Z_i,\theta_{\N})= 0,
\end{equation}
where $q(Z,\theta)$ is a known $s$-vector of estimating  functions.  Under the proposed GEL framework, the side information in the form of (\ref{pee-side})   can easily be incorporated into the inferential problems.

A general hypothesis test problem on the unknown parameters $\theta_{\N}$ can often be imposed as $H_0$: $R(\theta_{\N}) = 0$, where $R(\theta)$ is a  $k\times 1$ vector of functions, with  $k\leq p$. We are interested in developing tests for the
general parametric hypotheses in the form of $R(\theta_{\N}) = 0$ under the proposed GEL inferential framework.

Let
$\Theta^{\R} = \big\{ \theta \mid \theta \in \Theta \; {\rm and} \; R(\theta) = 0\big\}$
be the restricted parameter space under $H_0$. Write the combined estimating functions as  $\phi(Z,\theta,\varphi)=(\psi(Z,\theta,\varphi)^{\top},q(Z,\theta)^{\top})^{\top}$.
We define the ``restricted'' maximum GEL  estimator as
 \begin{equation}  \label{rgele}
\hat{\theta}_{\GEL}^{\R}=\arg\inf_{\theta\in\Theta^{\R}}\sup_{\nu\in\hat{\Lambda}_{\N,\phi}(\theta,\hat{\varphi})}
\hat{P}_{\N}^{\R}(\theta,\nu,\hat{\varphi}),
\end{equation}
where
$\hat{P}_{\N}^{\R}(\theta,\nu,\varphi)=\sum_{i\in\mS}(\rho(\eta^{\top}\pi_i^{-1}\phi(Z_i,\theta,\varphi))-\rho_0)$,
$\nu$ is an $(r+s)$-vector of auxiliary parameters and
$\hat{\Lambda}_{\N,\phi}(\theta,\varphi) = \{\nu:\nu^{\top}\pi_i^{-1}\phi(Z_i,\theta,\varphi)\in\mathcal {V}, i\in\mS\}$. The GEL  ratio statistic for testing $H_0$: $R(\theta_{\N}) = 0$ is given by
\begin{equation}
\label{sel-test2}
{\rm T}_{\N}^{\R}(\theta) = -2\{\hat{P}_{\N}(\hat{\theta}_{\GEL},\hat{\nu}_{\GEL},\hat{\varphi})- \hat{P}_{\N}^{\R}(\hat{\theta}_{\GEL}^{\R},\hat{\nu}_{\GEL}^{\R},\hat{\varphi})\},
\end{equation}
where $\hat{\nu}_{\GEL}^{\R}=\nu^{\R}(\hat{\theta}_{\GEL}^{\R},\hat{\varphi})$
 and $\nu^{\R}(\theta,\varphi)
=\arg\max_{\nu\in\hat{\Lambda}_{\N,\phi}(\theta,\varphi)}
\hat{P}_{\N}^{\R}(\theta,\nu,\varphi)$.

Let \ \ 
 $\mathscr{U}_{\N}(\theta,\varphi) =\sum_{i=1}^{N}\phi(Z_i,\theta,\varphi)/N$, \ \ $\hat{\mathfrak{U}}_{\N}(\theta)=\sum_{i\in\mS}\pi_i^{-1} q(Z_i,\theta)/N$, \ \ and \ \ 
$\Phi(\theta)=\partial R(\theta)/\partial\theta^{\top}$. The following additional regularity conditions are used  to investigate the asymptotic properties of  the estimator $\hat{\theta}_{\GEL}^{\R}$ defined in (\ref{rgele}) and the test statistic ${\rm T}_{\N}^{\R}(\theta)$ defined in  (\ref{sel-test2}).

\begin{itemize}

\item [B5.]  The finite population parameter vector  $\theta_{\N}  \in \Theta$  is the unique solution to  $\mathscr{U}_{\N}(\theta,\varphi_{\N})$ $=0$.

\item [B6.]
(i)
 There exists a  function $\mathfrak{U}(\theta)$  such that $\sup_{\theta\in\Theta}\|\mathfrak{U}_{\N}(\theta)-\mathfrak{U}(\theta)\|=o(1)$; (ii)
for all $\theta\in\Theta$, the ordinary derivative of $\mathfrak{U}(\theta)$ with respect to $\theta$, denoted as $H(\theta)$,  exists and has
full column rank $p$.

\item[B7.]
(i)
$\max_{i\in \mS}\sup_{\theta\in\Theta}\|q(Z_i,\theta)\|=o_p(n_{\B}^{1/\alpha})$, where $\alpha$ is as defined in condition B2(i);
(ii)
 For any sequence $c_{\N}=O(N^{-\kappa})$ with $\kappa\in(1/4,1/2]$,
$$\sup_{\theta\in\Theta}  \dfrac{1}{N}\sum_{i=1}^{N}\|q(Z_i,\theta)-q(Z_i,\theta+c_{\N})\|= O(|c_{\N}|) \, ;$$
(iii) For any sequence of positive numbers $\{\delta_{\N}\}$ with $\delta_{\N}=o(1)$,
$$\sup_{\theta, \theta'  \in \Theta(\delta_{\N})} \|\mathfrak{U}_{\N}(\theta)  - \mathfrak{U}(\theta)-[\mathfrak{U}_{\N}(\theta')  - \mathfrak{U}(\theta')] \| = o(N^{-1/2});$$
(iv) For all $\delta>0$ and some positive constant $c$,
$$
\sup_{\theta, \theta'  \in \Theta(\delta)}
{\rm Var}\Big\{[\hat{\mathfrak{U}}_{\N}(\theta)-\hat{\mathfrak{U}}_{\N}(\theta')]\mid \mathcal {F}_{\N}\Big\}\leq cn_{\B}^{-1}|\delta|.
$$
\end{itemize}

\begin{theorem}
\label{thm4}
Suppose that  the assumptions for Theorem \ref{thm3} and the conditions B5-B7 hold. Then, as $N \rightarrow \infty$,
$$
n_{\B}^{1/2}\big(\hat{\theta}_{\GEL}^{\R}-\theta_{\N}\big) \;\; \stackrel{{\cal L}}{\rightarrow} \;\; N(0,V^{\R}),
$$
where
 $
V^{\R}=\mathscr{C}^{\R}\Pi^{\top} \mathscr{W}^{-1} \Omega^{\R} \mathscr{W}^{-1} \Pi \mathscr{C}^{\R},
$
with $\mathscr{C}^{\R}=\Sigma^{\R}-\Sigma^{\R} \Phi^{\top}(\Phi\Sigma^{\R} \Phi^{\top})^{-1}\Phi\Sigma^{\R}$, $\Sigma^{\R}=[\Pi^{\top}\mathscr{W}^{-1}\Pi]^{-1}$,
   $\Pi=(\Gamma_2^{\top},H^{\top})^{\top}$, $\Gamma_2$ is given in Theorem \ref{thm2},
   $H=H(\theta_{\N})$, $\Phi=\Phi(\theta_{\N})$,
$\mathscr{W}= (n_{\B}/ N^2)\sum_{i=1}^{N}\pi_i^{-1}\phi(Z_i,\theta_{\N},\varphi_{\N})^{\otimes 2},$ and
$
\Omega^{\R}= n_{\B}N^{-2}{\rm Var}\{\sum_{i\in \mS}\pi_i^{-1}\phi(Z_i,\theta_{\N},\varphi_{\N})\mid \mF_{\N}\}.
$
\end{theorem}

\begin{theorem}
\label{thm5}
Suppose that the assumptions for Theorem \ref{thm4} hold.
Then, as $N \rightarrow \infty$,
\begin{eqnarray*}
{\rm T}_{\N}^{\R}(\theta_{\N}) \;\; \stackrel{{\cal L}}{\rightarrow} \;\; \mathcal{Q}^{\top} \Delta^{\R} \mathcal{Q},
\end{eqnarray*}
where
$\mathcal{Q}\sim N(0, I_{r+s})$,
$
\Delta^{\R} = (\Omega^{\R})^{1/2}[\mathscr{P}^{\R}-\mathscr{S}_{\psi}
\mathscr{P}\mathscr{S}_{\psi}^{\top}](\Omega^{\R})^{1/2}$, $\mathscr{S}_{\psi}=(I_r,0)^{\top}$ is an $(r+s)\times r$ matrix, $\mathscr{P}^{\R}=\mathscr{W}^{-1}-\mathscr{W}^{-1}\Pi \mathscr{C}^{\R}\Pi^{\top}\mathscr{W}^{-1}$, and $\mathscr{P}=W_2^{-1}-W_2^{-1}\Gamma_2\Sigma_2\Gamma_2^{\top}W_2^{-1}$.
\end{theorem}

\begin{corollary}
\label{cor3}
Suppose that  the assumptions for Theorem \ref{thm5} hold.  Under single-stage PPS sampling with replacement or single-stage PPS sampling without replacement with negligible sampling fractions, we have
$V^{\R}=\mathscr{C}^{\R}$
and
$
{\rm T}_{\N}^{\R}(\theta_{\N}) \;\; \stackrel{{\cal L}}{\rightarrow} \;\; \chi_{s+k}^2
$
as $N \rightarrow \infty$.
\end{corollary}

The standard chi-square limiting distribution presented in Corollary \ref{cor3} under the commonly used survey designs provides a convenient tool for conducting general hypothesis tests and the construction of confidence regions for a subvector, say $\theta_{1\N}$, of $\theta_{\N}$ consisting of $k$ elements. Let $\theta = (\theta_1^{\top}, \theta_2^{\top} )^{\top}$ be the partition of $\theta_{\N}$ with the first $k$ components corresponding to $\theta_{1\N}$. A $(1-\alpha)$-level confidence region for $\theta_{1\N}$ using the proposed GEL ratio statistic is given by
\[
\mathcal {C}_\alpha^{\R} = \bigg\{\theta_1 \mid -2\Big[\hat{P}_{\N}(\hat{\theta}_{\GEL},\hat{\eta}_{\GEL},\hat{\varphi}) - \hat{P}_{\N}\big(\tilde{\theta}(\theta_1),\eta(\tilde{\theta}(\theta_1),\hat{\varphi}),\hat{\varphi}\big)\Big]
\leq \chi^2_{k,\alpha}\bigg\} \,,
\]
where $\tilde{\theta}(\theta_1) = (\theta_1^{\top},\hat{\theta}_2(\theta_1)^{\top})^{\top}$ and $\hat{\theta}_2(\theta_1)=\arg\inf_{\theta_2}
\sup_{\eta\in\hat{\Lambda}_{\N,\psi}((\theta_1,\theta_2),\hat{\varphi})}\hat{P}_{\N}((\theta_1,\theta_2),
\eta,\hat{\varphi})$ for the given $\theta_1$.

 \section{Derivations of Augmentation Terms}
 \label{sec.exam}
 
 \setcounter{equation}{0}

 \noindent
The augmentation term $\Xi$  specified in (\ref{nee})  plays the most crucial role in the proposed methods and needs to be derived for the particular nuisance functionals involved. In this section, we first discuss general procedures for identifying $\Xi$, and then illustrate the methods using three examples related to income inequality measures widely used in economic studies.

\subsection{The general identification condition}
\label{subsec.ident}

\noindent
We first discuss the general identification condition of $\Xi$ using a superpopulation-based approach. To facilitate the technical arguments, we introduce the notion of probability spaces associated with the sampling design and the superpopulation model.
We assume
that the vectors $\mathcal {F}_{\N}=(Z_1,\cdots, Z_{\N})\in\mathsf {R}^{d_z\times N}$  are an independent and identically distributed sample from a superpopulation
model over a probability space $(\Omega, \mathscr{A}, \mathbb{P}_m)$, as well as  from a distribution function $F_0$.   The values $Z_i$ can be viewed as a mapping $\Omega\mapsto\mathsf {R}^{d_z}$, and   we can write $Z_i$ as $Z_i(\omega)$ with $\omega\in\Omega$.
Denote a $d_x$-dimensional component of $Z_i$ as $X_i $ with $X_i\in \mathsf {R}_{+}^{d_x}$ and $1\leq d_x\leq d_z$.
Suppose that $\mathbf{X}^{N}=(X_1,\cdots, X_{\N})\in\mathsf {R}_{+}^{d_x\times N}$ contains all the variables used for the sampling design.
With the given sampling design, denote by $\mathbf{S}_{\N}=\{\mS:\mS\subset\mathcal {U}_{\N}\}$ the set of all possible samples.
The smallest $\sigma$-algebra containing all the sets of $\mathbf{S}_{\N}$ is denoted as
$\mathsf{C}_{\N}$  and is called the sigma-algebra generated by $\mathbf{S}_{\N}$.
Following Rubin-Bleuer and Schiopu-Kratina (2005), the sampling design is characterized by a function $P$: $\mathsf{C}_{\N}\times\mathsf {R}_{+}^{d_x\times N}\rightarrow[0,1]$ such that
(i)
for all $\mS$ in $\mathbf{S}_{\N}$, $P(\mS,\cdot)$ is Borel-measurable in $\mathsf {R}_{+}^{d_x}$;
(ii)
for $\mathbf{X}^{N}\in\mathsf {R}_{+}^{d_x\times N}$, $P(\cdot,\mathbf{X}^{N})$ is a probability measure on $\mathsf{C}_{\N}$.
For each $\omega\in\Omega$ and $B\subset\mathbf{S}_{\N}$, define $\mathbb{P}_d(B,\omega)=\sum_{s\in B}P(s, \mathbf{X}^{N}(\omega))$. We call the triple $(\mathbf{S}_{\N},\mathsf{C}_{\N}, \mathbb{P}_d)$  a design probability space.   The product probability space that includes the super-population and the design space is defined as  $(\mathbf{S}_{\N}\times\Omega,\mathsf{C}_{\N}\times\mathscr{A}, \mathbb{P}_{d,m})$, in which the probability measure
$\mathbb{P}_{m,d}$ defined on rectangles $\{s\}\times A\in\mathsf{C}_{\N}\times \mathscr{A}$ has the value
$$\mathbb{P}_{m,d}(\{s\}\times A)=\int_{A}P(s,\mathbf{X}^{N}(\omega)){\rm d}\mathbb{P}_m(\omega)=\int_{A}\mathbb{P}_d(\{s\},\mathbf{X}^{N}(\omega)){\rm d}\mathbb{P}_m(\omega).$$

In what follows, we use
$\mathbb{E}_m\{\cdot\}$ to denote  the expectation with respect to  the probability space $(\Omega, \mathscr{A}, \mathbb{P}_m)$ and  $\mathbb{E}_{d,m}\{\cdot\}$  to represent the expectation with respect to  the product probability space
$(\mathbf{S}_{\N}\times\Omega,\mathsf{C}_{\N}\times\mathscr{A}, \mathbb{P}_{d,m})$.
For any $(\theta, \varphi)\in\Theta\times\Psi$, we have
\begin{eqnarray*}
\mathbb{E}_{d,m}\Big\{\dfrac{1}{N}\sum_{i\in\mS}\pi_{i}^{-1}g(Z_i,\theta,\varphi)\Big\}=\mathbb{E}_m\{g(Z,\theta,\varphi)\},\\
\mathbb{E}_{d,m}\Big\{\dfrac{1}{N}\sum_{i\in\mS}\pi_{i}^{-1}\Xi(Z_i,\theta,\varphi)\Big\}=\mathbb{E}_m\{\Xi(Z,\theta,\varphi)\}.
\end{eqnarray*}
Let $\theta_0\in\Theta$ and $\varphi_0\in\Psi$ be the superpopulation version  of the parameter of interest $\theta_{\N}$ and the nuisance functions $\varphi_{\N}$, respectively. Then, in terms of probability space $(\Omega, \mathscr{A}, \mathbb{P}_m)$, we have  $\theta_{\N}\stackrel{\mathbb{P}_m}{\longrightarrow}\theta_0$ and $\varphi_{\N}\stackrel{\mathbb{P}_m}{\longrightarrow}\varphi_0$.  Here $``\stackrel{\mathbb{P}_m}{\longrightarrow}"$ denotes convergence in probability with respect to probability space $(\Omega, \mathscr{A}, \mathbb{P}_m)$. By the identification  of the super-population model, we  further have that  $\mathbb{E}_m\{g(Z,\theta_0,\varphi_0)\}=0$ and
$\mathbb{E}_m\{\Xi(Z,\theta_0,\varphi_0)\}=0$.

Let $\mathscr{F}=\{F\}$ be a general family  of distribution of $Z$ and $\varphi(\cdot)$ be a mapping $\mathscr{F}\mapsto\mathsf {R}^{{\rm dim}(\varphi)}$. \ Suppose that \ $\hat{\varphi}\stackrel{\mathbb{P}_{d,m}}{\longrightarrow}\varphi(F)$ \ if the distribution of $Z$ is $F \in \mathscr {F}$, where  $``\stackrel{\mathbb{P}_{d,m}}{\longrightarrow}"$ denotes convergence in probability with respect to the product probability space $(\mathbf{S}_{\N}\times\Omega,\mathsf{C}_{\N}\times\mathscr{A}, \mathbb{P}_{d,m})$.
Let $\{F_{\alpha}: F_{\alpha}\in\mathscr{F}\}$ be a  one-dimensional subfamily of  $\mathscr{F}$.
Following Newey (1994), the path $\{F_{\alpha}: \alpha\in(-\varepsilon,\varepsilon)\subset \mathsf {R}, \varepsilon >0, F_{\alpha}\in\mathscr{F}\}$   is assumed to be regular and  satisfies the following mean-squared differentiability condition
\[\lim_{\alpha\rightarrow 0} \int\left[{\alpha}^{-1}(dF_{\alpha}^{1/2} - dF_0^{1/2}) - \dfrac{1}{2}\mathfrak{F}(z)dF_0^{1/2}\right]^2dz = 0,\]
 where  $dF_{\alpha}$ is the density of  $F_{\alpha}$, and $\mathfrak{F}(z)=\partial\ln (F_{\alpha})/\partial \alpha$ is the corresponding score function satisfying $\mathbb{E}_m\{\mathfrak{F}(Z)\}=0$ and $\mathbb{E}_m\{\mathfrak{F}(Z)^2\}<\infty$.
Define the functional
\[\mu(F)= \mathbb{E}_{m}\{g(Z, \theta_0, \varphi(F))\}.
\]
We assume that   $\mu: \mathscr {F}  \mapsto\mathsf {R}^r$ is differentiable at $F_0$ in the sense of Van der Vaart (1991).  Then
under certain regularity conditions the function $\Xi(Z, \theta_0, \varphi(F_0))$ to be used as the augmentation term is uniquely determined by
\begin{eqnarray}\label{con-influ}
\dfrac{\partial \mu(F_{\alpha})}{\partial \alpha}\bigg|_{\alpha = 0} = \mathbb{E}_m\{\Xi(Z, \theta_0, \varphi(F_0))\mathfrak{F}(Z)\}.
\end{eqnarray}
Equation (\ref{con-influ}) is useful for deriving the expression for the function $\Xi$ when $\varphi_{\N}=\varphi_{\N}(\cdot,\theta_{\N})$ is the finite population version of  a regression function  or a density.

The function $\Xi(Z, \theta_0, \varphi(F_0))$ is called influence function of $\mu(F_0)$. In the model-based context, the explicit or numerical computation of the influence function has been discussed  extensively in the literature, see, for example, Bickel et al. (1993), Newey (1994a, 1994b),
Chen et al. (2003), Chen (2007),
 Ichimura and Newey (2022),  Bravo et al. (2020), Chernozhukov et al. (2022) and references therein. It follows from above that these model-based approaches can be readily extended to the problem of complex survey data.

\subsection{Census estimating equation based approach}
\label{cee-approach}
\noindent
We next consider cases where the nuisance functional $\varphi_{\N}$ can be explicitly defined via the  following census   estimating  equations
\begin{equation}  \label{pee-nui}
\mathscr{T}_{\N}(\varphi_{\N})=\dfrac{1}{N}\sum_{i=1}^{N}\mathfrak{T}(Z_i,\varphi_{\N})=0.
\end{equation}
We assume that the equation system (\ref{pee-nui}) for defining  the function  $\varphi_{\N}$ is possibly over-identified, i.e., dim($\mathfrak{T}$) $\geq$ dim($\varphi$).
Given the  set of sampled units $\mS$ and survey weights $\{\pi_i^{-1},i\in\mS\}$,    the  design-based GEL estimator for $\varphi_{\N}$ can be  obtained as
\begin{equation*}  \label{gel-nui}
\hat{\varphi}_{\GEL}=\arg\inf_{\varphi\in\Psi}\sup_{\lambda\in\hat{\Lambda}_{\N,\mathfrak{T}}(\varphi)}
\hat{P}_{\N}(\varphi,\vartheta),
\end{equation*}
where $\hat{P}_{\N}(\varphi,\vartheta)=\sum_{i\in \mS}\{\rho(\vartheta^{\top}\pi_i^{-1}\mathfrak{T}(Z_i, \varphi))-\rho_0\},
$ and
$\hat{\Lambda}_{\N,\mathfrak{T}}(\varphi)=\{\vartheta:\vartheta^{\top}\pi_i^{-1}
\mathfrak{T}(Z_i,\varphi)\in\mathcal {V}, i\in\mS\}$.
By applying Theorem \ref{thm2}, we have that
\begin{equation*}  \label{nui-influ}
\hat{\varphi}_{\GEL}-\varphi_{\N}=-\mathbb{K}(\varphi_{\N})\mathbb{H}(\varphi_{\N})^{\top}
\mathbb{W}(\varphi_{\N})^{-1}\dfrac{1}{N}\sum_{i\in\mS}\pi_{i}^{-1}\mathfrak{T}(Z_i,\varphi_{\N})+o_p(n_{\B}^{-1/2}),
\end{equation*}
where $\mathbb{K}(\varphi)=[\mathbb{H}(\varphi)^{\top}\mathbb{W}(\varphi)^{-1}\mathbb{H}(\varphi)]^{-1}$, $\mathbb{H}(\varphi)=\partial \mathscr{T}(\varphi)/\partial \varphi^{\top}$ with  $\mathscr{T}(\varphi)$ satisfying 
$$\sup_{\varphi\in\Psi}\|\mathscr{T}_{\N}(\varphi)-\mathscr{T}(\varphi)\|=o(1),$$   and $\mathbb{W}(\varphi)=(n_{\B}/ N^2)\sum_{i=1}^{N}\pi_i^{-1}\mathfrak{T}(Z_i,\varphi)^{\otimes 2}$.
The augmentation term is therefore given by
$$
\Xi(Z,\theta,\varphi)=-D(\theta,\varphi) \mathbb{K}(\varphi)\mathbb{H}(\varphi)^{\top}\mathbb{W}(\varphi)^{-1}\mathfrak{T}(Z,\varphi),
$$
where the derivative $D(\theta,\varphi)$ is defined in Condition  A5  presented  in  Appendix A.
When $\varphi_{\N}$ is  just-identified by (\ref{pee-nui}), i.e., dim($\mathfrak{T}$) =dim($\varphi$),  the result is  simplified  to
$\Xi(Z,\theta,\varphi) = -D(\theta,\varphi) \mathbb{H}(\varphi)^{-1}\mathfrak{T}(Z,\varphi)$.
It is straightforward to show that the augmented estimating functions $\psi(Z,\theta,\varphi)=g(Z,\theta,\varphi)+\Xi(Z,\theta,\varphi)$ satisfy the following 
invariance property:
\begin{equation*}  \label{pels2.2}
\frac{1}{N}\sum\limits_{i\in \mS}\pi_i^{-1}\psi(Z_i,\theta_{\N},\hat{\varphi}_{\GEL})=\frac{1}{N}\sum\limits_{i\in \mS}\pi_i^{-1}\psi(Z_i,\theta_{\N},\varphi_{\N})
+o_p(n_{\B}^{-1/2}).
\end{equation*}
This observation intuitively justifies the standard chi-square limiting distributions of augmented two-step GEL ratio statistics presented in the paper under commonly used survey designs.

\subsection{Illustrative examples}

\noindent
We now  apply the general results to three examples: the Gini coefficient, Lorenz curves and  quantile shares, all involving a nuisance functional.  These three examples have important implications to the theory and practice in economics on inequality measures. The Gini coefficient, also called the Gini index, measures the degree of the inequality in income distributions,
the Lorenz curve depicts concentration and inequality in distribution of resources and in size distributions,  while the quantile share is used to detect  perturbations at different levels of a distribution (Beach and Davidson, 1983). If the variable under study represents income, then the quantile shares are also called income shares.

\vspace{3mm}

\noindent {\bf Example 1 (Gini Coefficient). } \
Let  $Z\in\mathsf {R}$ be a  nonnegative random variable on a probability space $(\Omega, \mathscr{A}, \mathbb{P}_m)$.  The cumulative distribution function  of   $Z$ is    $F_0(z)=\mathbb{P}_m(Z\leq z)$.
The  general family of Gini coefficients (Nyg\aa rd and  Sandstr\"{o}m, 1989)  is defined as
$$\theta_0=\dfrac{1}{\mu_0}\int_{0}^{\infty}\psi\{F_0(z)\}zdF_0(z),$$
where
$\mu_0=\mathbb{E}_{m}[Z]$, $\psi$ is a bounded and continuous function. Here, $\mathbb{E}_m[\cdot]$  denotes  the expectation taken with respect to  the probability measure $\mathbb{P}_m$. For the original Gini coefficient,   $\psi\{u\} = 2u-1$. The nuisance functional in this case is the cumulative distribution function $F_0(z)$. Let  $\mathcal {F}_{\N}=(Z_1,\cdots,Z_{\N})\in\mathsf {R}^{N}$ be a finite population from $\mathbb{P}_m$.
The finite population distribution function is given by $F_{\N}(z)=N^{-1}\sum_{i=1}^{N}I(Z_i\leq z)$, where $I(\cdot)$ is the indicator function, and the finite population mean is $\mu_{\N}=N^{-1}\sum_{i=1}^{N}Z_i$. Then, the finite population Gini coefficient is defined as
$\theta_{\N}= N^{-1}\sum_{i=1}^{N}\mu_{\N}^{-1}\psi\{F_{\N}(Z_i)\}Z_i$,
which  satisfies
\[
U_{\N}(\theta_{\N},F_{\N})= \dfrac{1}{N} \sum_{i=1}^Ng(Z_i,\theta_{\N},F_{\N}(Z_i))= 0 \,,
\]
where $g(Z,\theta,F)=\psi\{F\}Z-\theta Z$.

Denote
  $U(\theta,F) = \mathbb{E}_m[\psi\{F\}Z-\theta Z]$. Standard empirical process methods   can be used to show that
  $U_{\N}(\theta,F)$ converges uniformly in $(\theta,F)$   to $U(\theta,F)$. The  pathwise derivative of $U(\theta,F_{\N})$ in the direction $F-F_{\N}$
has the form  $D(\theta,F_{\N})[F-F_{\N}(Z)]=\mathbb{E}_m[\psi'\{F_{\N}(Z)\}Z\{F-F_{\N}(Z)\}]$, where $\psi'\{u\} = \partial\psi\{u\}/\partial u$.
Given the set of sampled units $\mS$ and first   order inclusion probabilities   $\pi_i$,  the survey weighted estimator of $F_{\N}(z)$ is obtained  by  $\hat{F}_{\N}(z) = \hat{N}^{-1}\sum_{i\in \mS}\pi_i^{-1}I(Z_i\leq z),$ where
$\hat{N}=\sum_{i\in \mS}\pi_i^{-1}$.  It can be shown that
\[
D(\theta_{\N},F_{\N})[\hat{F}_{\N}(Z)-F_{\N}(Z)]
= \frac{1}{N} \sum_{i\in \mS}\pi_i^{-1}\Xi(Z_i, F_{\N})+o_p(n_{\B}^{-1/2}) \,,
\]
where $\Xi(Z_i,F_{\N})=\mathbb{E}_m[Z\psi'\{F_{\N}(Z)\}\{I(Z\geq Z_i)-F_{\N}(Z)\}]$, which is used to construct the augmentation term.
For the original Gini coefficient,  we have $\Xi(Z_i,F_{\N})=2\mathbb{E}_m[Z\{I(Z\geq Z_i)-F_{\N}(Z)\}]$.

 \vspace{3mm}

\noindent {\bf Example 2 (Lorenz Curves). } \
Assume that $F(z)$ is differentiable and $f(z)$ is its density function.  For a given $\tau\in[0,1]$, the Lorenz curve   of $\mathbb{P}_m$   is defined as
\[
\theta_0(\tau) = \frac{1}{\mu_0}\int_{0}^{\xi_0(\tau)}zdF(z) \,,
\]
where 
$\xi_0(\tau) = F_0^{-1}(\tau)=\inf\{z:F_0(z)\ge\tau\}$, which is a nuisance functional.
The finite population Lorenz curve is defined as
$\theta_{\N}(\tau)= N^{-1}\sum_{i=1}^{N}\mu_{\N}^{-1}Z_iI\{Z_i\leq\xi_{\N}(\tau)\}$,  where $\xi_{\N}(\tau)=F_{\N}^{-1}(\tau)=\inf\{z:F_{\N}(z)\ge\tau\}$, the $\tau$th finite population quantile.
Note that  $\theta_{\N}(\tau)$ is the solution to
\begin{eqnarray*}
U_{\N}(\theta,\xi_{\N}(\tau)) =\frac{1}{N}\sum_{i=1}^Ng(Z_i,\theta,\xi_{\N}(\tau)) = 0,
\end{eqnarray*}
where
$
g(Z,\theta,\xi)=Z\{I(Z\leq \xi)-\theta\}.
$

Denote $U(\theta,\xi) = \mathbb{E}_m[Z\{I(Z\leq \xi)-\theta\}]$.   It can be shown that      $U_{\N}(\theta,\xi)$ converges uniformly in $(\theta,\xi)$   to $U(\theta,\xi)$, and that the  pathwise derivative of $U(\theta,\xi_{\N})$ in direction $\xi-\xi_{\N}$
is of the form
$
D(\theta,\xi_{\N}(\tau))[\xi-\xi_{\N}(\tau)]=\xi_{\N}(\tau)f(\xi_{\N}(\tau))[\xi-\xi_{\N}(\tau)] \,.
$
The survey weighted estimator of $\xi_{\N}(\tau)$ is given by
$\hat{\xi}(\tau)=\hat{F}^{-1}_{\N}(\tau)=\inf\{z:\hat{F}_{\N}(z)\ge\tau\}$.  Using the
Bahadur representation  established in Chen and Wu (2002), we obtain
 $$
 D(\theta_{\N},\xi_{\N}(\tau))[\hat{\xi}(\tau)-\xi_{\N}(\tau)] = \dfrac{1}{N}\sum_{i\in \mS}\pi_i^{-1}\Xi(Z_i,\xi_{\N}(\tau))+o_p(n_{\B}^{-1/2}),
 $$ where $\Xi(Z,\xi)=-\xi\{I(Z\leq \xi)-  \tau\}$.
The GEL-based estimation and  inference for  $\theta_{\N}(\tau)$ can consequently be conducted using     the   augmented   estimating function
$
\psi(Z,\theta,\xi)=g(Z,\theta,\xi)+\Xi(Z,\xi).
$

 \vspace{3mm}

\noindent {\bf Example 3 (Quantile Shares). } \
 For  two fixed quantile levels $\tau_1,\tau_2\in[0,1]$ with  $\tau_1\leq\tau_2$,
the quantile share of  $\mathbb{P}_m$  is defined as
$$
\theta_0(\tau_1,\tau_2)=\theta_0(\tau_2)-\theta_0(\tau_1) \,,
$$
where $\theta_0(\tau)$ is defined in Example 2. If $Z$ is an income variable, the income share $\theta_0(\tau_1,\tau_2)$  is
the percentage of total income shared by the population allocated to the income interval $[\xi_0(\tau_1), \; \xi_0(\tau_2)]$.
The finite population quantile share is defined as
$\theta_{\N}(\tau_1,\tau_2)=N^{-1}\sum_{i=1}^{N}\mu_{\N}^{-1}Z_iI\{\xi_{\N}(\tau_1)<Z_i\leq\xi_{\N}(\tau_2)\},$
 which is obtained by solving the   census estimating equation
\begin{eqnarray*}
U_{\N}(\theta,\xi_1,\xi_2) = \frac{1}{N}\sum_{i=1}^Ng(Z_i,\theta,\xi_1,\xi_2) = 0,
\end{eqnarray*}
with
$
g(Z,\theta,\xi_1,\xi_2)=Z\{I(\xi_1<Z\leq \xi_2)-\theta\}.
$
Using the same arguments given in Example 2 for the Lorenz curve, we obtain the following augmented estimating function
$
\psi(Z,\theta,\xi_1,\xi_2)=g(Z,\theta,\xi_1,\xi_2)+\Xi(Z,\xi_1,\xi_2),
$
where $\Xi(Z,\xi_1,\xi_2)=-\xi_2\{I(Z\leq \xi_2)-  \tau_2\}+\xi_1\{I(Z\leq \xi_1)-  \tau_1\}$.

\section{Survey Designs and Variance Estimation}
\label{sec.design}

\setcounter{equation}{0}

\noindent
We  give detailed  illustrations of how our results  can be readily applied to   a class of complex survey designs, along with  discussions on  design-based variance estimation of  the proposed efficient GEL estimators.
It follows from Theorem \ref{thm2}  that the point estimators $\hat{\theta}_{\GEL}$ and its estimated standard errors can be   used to construct Wald-type confidence regions. However, estimation of the asymptotic design-based variance-covariance matrix $V_2$ of $n_{\B}^{1/2}(\hat{\theta}_{\GEL}-\theta_{\N})$ is not straightforward for a general unequal probability survey design. 
We consider three commonly used survey
designs: single-stage unequal probability sampling single-stage survey designs, stratified sampling and cluster sampling. General discussions on variance estimation for complex survey designs can be found in Wu and Thompson (2020).

\subsection{Single-stage unequal probability sampling}
\label{sec.single}
If the survey design is single-stage PPS sampling with replacement or single-stage PPS sampling without replacement with negligible sampling fractions, we have the following approximation formula for the variance-covariance matrix $\Omega$:
\begin{eqnarray*}
\Omega= \dfrac{n_{\B}}{N^2}{\rm Var}\Big\{\sum_{i\in \mS}\pi_i^{-1}\psi(Z_i,\theta_{\N},\varphi_{\N})\mid \mF_{\N}\Big\}
=\dfrac{n_{\B}}{N^2}\sum_{i=1}^N\pi_i^{-1}\psi(Z_i,\theta_{\N},\varphi_{\N})^{\otimes 2}+o(1).
\end{eqnarray*}
Consequently, a design-based consistent estimator of $\Omega$ in the survey designs mentioned above can be obtained by
\begin{eqnarray*}
\hat{\Omega}=\dfrac{n}{N^2}\sum_{i\in\mS}\Big[\pi_i^{-1}\psi(Z_i,\hat{\theta}_{\GEL},\hat{\varphi})
-N n^{-1}\hat{\mathbb{U}}_{\N}(\hat{\theta}_{\GEL},\hat{\varphi})\Big]^{\otimes 2},
\end{eqnarray*}
where $\hat{\mathbb{U}}_{\N}(\theta,\varphi)= N^{-1}\sum_{i\in \mS}\pi_i^{-1}\psi(Z_i,\theta,\varphi)$.

However, for general survey designs, estimating $\Omega$ requires second order inclusion probabilities $\pi_{ij}={\rm Pr}(i,j\in\mS)$, which may not be available.
Approximate variance formulas not involving the $\pi_{ij}$ are often used in practice; see Haziza {\em et al.} (2008) for further discussion. 
For single-stage PPS sampling with non-negligible sampling fractions,  we can estimate $\Omega$ by the H\'{a}jek variance estimator
\[
\hat{\Omega} = \dfrac{n}{N^2}\sum\limits_{i\in \mS} c_i\big [\pi_i^{-1}\psi(Z_i,\hat{\theta}_{\GEL},\hat{\varphi})-\hat{\mathscr{B}}\big]^{\otimes 2}\,,
\]
where
\[
\hat{\mathscr{B}} = \Big \{\sum_{i\in\mS}c_i\pi_i^{-1}\psi(Z_i,\hat{\theta}_{\GEL},\hat{\varphi})\Big\}/\sum_{i\in\mS}c_i  \;\;\; {\rm and} \;\;\;
c_i = \{n(1-\pi_i)\}/(n-1) \,.
\]
Simulation results reported in
Haziza {\em et al.} (2008) showed that  the approximate variance estimator has good finite sample performances for commonly used single-stage survey designs.

We now discuss design consistent estimators of the weight matrix $W_2$ and the derivative $\Gamma_2$ under a general single-stage sampling design. The weight matrix $W_2$ can be consistently estimated by
\begin{eqnarray*}\label{hatw2}
\hat{W}_{\N}=\dfrac{n}{N^2}\sum_{i\in\mS}\pi_i^{-2}\psi(Z_i,\hat{\theta}_{\GEL},\hat{\varphi})^{\otimes 2}.
\end{eqnarray*}
If the function $\psi(Z,\theta,\varphi)$ is differentiable in $\theta$, it is easily to see that
$$\hat{\Gamma}_2=\dfrac{1}{N}\sum_{i\in\mS}\pi_i^{-1}
\dfrac{\partial\psi(Z,\hat{\theta}_{\GEL},\hat{\varphi})}{\partial\theta^{\top}}$$
is a design consistent estimator for $\Gamma_2$. For non-smooth functions, we employ the method of random perturbation proposed in Chen and Liao (2015). Denote by  $\mathscr{V}$
a large enough compact set in $\mathsf{R}^{p}$ and define
\begin{eqnarray*}\label{randper}
\begin{split}
\mathcal {D}_{\N,\theta}(\mathscr{V},\hat{\theta}_{\GEL},\hat{\varphi})
=\sqrt{N}\hat{\mathbb{U}}_{\N}(\hat{\theta}_{\GEL}+N^{-1/2}\mathscr{V},\hat{\varphi})
-\sqrt{N}\hat{\mathbb{U}}_{\N}(\hat{\theta}_{\GEL},\hat{\varphi}).
\end{split}
\end{eqnarray*}
Under the conditions presented  in Appendix, $\{\hat{\mathbb{U}}_{\N}(\theta,\varphi)-\mathbb{U}(\theta,\varphi), N=1,2, \cdots\}$ is   stochastically  equicontinuous. This, together with the differentiability of the limiting function $\mathbb{U}(\theta,\varphi)$ with respect to $\theta$, implies that
\begin{eqnarray}\label{approximate-ls}
\begin{split}
\mathcal {D}_{\N,\theta}(\mathscr{V},\hat{\theta}_{\GEL},\hat{\varphi})
=\Gamma_2(\tilde{\theta},\hat{\varphi})\mathscr{V}+o_p(1).
\end{split}
\end{eqnarray}
where $\tilde{\theta}$ is on the line segment between $\hat{\theta}_{\GEL}$ and $\hat{\theta}_{\GEL}+N^{-1/2}\mathscr{V}$.
Motivated by the expression (\ref{approximate-ls}),  we propose the following resampling procedure based on the least squares.
\begin{itemize}
\item[1.] Generate independent and identically distributed random samples, $\{\mathscr{V}_{b}: b = 1,\cdots, B\}$,
from some known multivariate distribution with mean zero and variance $I_{p}$.
\item[2.]  Compute $\mathcal {D}_{\N,\theta}(\mathscr{V}_{b},\hat{\theta}_{\GEL},\hat{\varphi})$ for $b = 1,\cdots, B$.
\item[3.]  Calculate
\begin{eqnarray}\label{resample-ls}
\hat{\Gamma}_{2,j}=\left(\dfrac{1}{B}\sum_{b=1}^{B}
\mathscr{V}_{b}\mathscr{V}_{b}^{\top}\right)^{-1}\dfrac{1}{B}\sum_{b=1}^{B}
\mathcal {D}_{j\N,\theta}(\mathscr{V}_{b},\hat{\theta}_{\GEL},\hat{\varphi})\mathscr{V}_{b},
\end{eqnarray}
for  $j=1,\cdots,r$, where $\mathcal {D}_{j\N,\theta}$ denotes the $j$th coordinate of $\mathcal {D}_{\N,\theta}$. The value of $\Gamma_2$ is then estimated by $\hat{\Gamma}_2$ with $\hat{\Gamma}_2=(\hat{\Gamma}_{2,1},\cdots,\hat{\Gamma}_{2,r})^{\top}$.
\end{itemize}

Note that $\hat{\Gamma}_{2,j}$ in (\ref{resample-ls}) is the least squares estimate from regressing $\mathcal {D}_{j\N,\theta}(\mathscr{V},\hat{\theta}_{\GEL},\hat{\varphi})$  over $\mathscr{V}$ based on (\ref{approximate-ls}). 
Denote by $E_{\mathscr{V}}[\cdot]$   the expectation with respect to $\mathscr{V}$.
The following theorem presents the consistency of the resampling estimator $\hat{\Gamma}_2$.

\begin{theorem}
\label{thm6}
Suppose that (i) $\mathscr{V}$ is a random vector with mean zero and variance $I_{d}$, independent of the survey sample $\mS$; (ii) for all sequences   
$\delta_{\N}=o(1)$, 
$$
\sup_{(\theta,\varphi)\in\Theta(\delta_{\N})\times\Psi(\delta_{\N})}
    \|\Gamma_2(\theta,\varphi)-\Gamma_2(\theta_{\N},\varphi_{\N})\|=o(1)
$$ 
and
$$
\sup_{(\theta,\varphi)\in\Theta(\delta_{\N})\times\Psi(\delta_{\N})}
\|E_{\mathscr{V}}[\mathfrak{D}_{\N,\mathscr{V}}(\mathscr{V},
\theta+N^{-1/2}\mathscr{V},\varphi)]\|
=o_p(N^{-1/2}),$$
where $\mathfrak{D}_{\N,\mathscr{V}}(\mathscr{V},\theta,\varphi)
=[\hat{\mathbb{U}}_{\N}(\theta,\varphi)-
\mathbb{U}(\theta,\varphi)]
\mathscr{V}^{\top}$.
Then $\hat{\Gamma}_2
\stackrel{p}{\rightarrow}\Gamma_2$.
\end{theorem}

Consequently, the variance-covariance matrix of $n_{\B}^{1/2}(\hat{\theta}_{\GEL}-\theta_{\N})$ can be consistently estimated by
$
\hat{V}_2=(\hat{\Gamma}_2^{\top}\hat{W}_{\N}^{-1}\hat{\Gamma}_2)^{-1}
\hat{\Gamma}_2^{\top}\hat{W}_{\N}^{-1}\hat{\Omega} \hat{W}_{\N}^{-1}\hat{\Gamma}_2 (\hat{\Gamma}_2^{\top}\hat{W}_{\N}^{-1}\hat{\Gamma}_2)^{-1}.
$

\subsection{Stratified sampling}
\label{sec.stra}

Suppose that the finite population $\mathcal {U}_{\N}$ is divided into  $H$ strata indexed by $h=1,\cdots,H$. Let $N=\sum_{h=1}^{H}N_h$ be the overall population size
where $N_h$ is the population size of the $h$th stratum.
Let $(hi)$ be the index for unit $i$ in stratum $h$. The parameter of interest  $\theta_{\N}$ is defined through the following stratified population estimating equations
 \begin{equation} \label{str-un}
U_{\N}(\theta_{\N},\varphi_{\N})=\dfrac{1}{N}
\sum_{h=1}^H\sum_{i=1}^{N_{\H}}g(Z_{hi},\theta_{\N},\varphi_{\N})=0.
\end{equation}
Assume that the stratum  sample $\mS_h$ of size $n_h$ is selected with first order inclusion probabilities $\{\pi_{hi},i\in \mS_h\}$, $h=1,\cdots,H$, independent across different strata. Let  $n=\sum_{h=1}^{H}n_h$ be the overall size of the stratified sample.
Assume that a design consistent estimator for $\varphi_{\N}$, denoted by $\hat{\varphi}$, can be obtained in advance by using the stratified samples.
The following two regularity conditions are imposed on the population estimating functions defined in (\ref{str-un}) and the first-step estimator $\hat{\varphi}$:
\begin{itemize}
\item [C1.]
There exists a function $U(\theta,\varphi)$ such that $\sup_{(\theta,\varphi)\in\Theta\times\Psi}\|U_{\N}(\theta,\varphi)-U(\theta,\varphi)\|=o(1)$,  and $U(\theta,\varphi)$  satisfies conditions A4 and A5 presented  in  Appendix A.

\item [C2.]
The  pathwise derivative $D(\theta_{\N},\varphi_{\N})[\hat{\varphi}-\varphi_{\N}]$   of $U(\theta_{\N},\varphi_{\N})$  is of the  following form:
$$D(\theta_{\N},\varphi_{\N})[\hat{\varphi}-\varphi_{\N}]=\dfrac{1}{N}\sum_{h=1}^H\sum_{i\in \mS_h}\pi_{hi}^{-1}\Xi(Z_{hi},\theta_{\N},\varphi_{\N})+o_p(n_{\B}^{-1/2}),$$
where  $\Xi(\cdot)$ satisfies the following
conditions: (i) $\Xi(Z_{hi},\theta_{\N},\varphi_{\N})$ has finite fourth population moments; and
(ii) $\sum_{h=1}^H\sum_{i\in \mS_h}\pi_{hi}^{-1}\Xi(Z_{hi},\theta_{\N},\varphi_{\N})$ is asymptotically normally distributed with mean zero and variance-covariance matrix at the order $O(n_{\B}^{-1}N^2)$.
\end{itemize}

Under stratified sampling designs, the efficient  two-step   GEL estimator for $\theta_{\N}$ satisfying
(\ref{str-un}) is defined as
\begin{equation}  \label{gele-str}
\hat{\theta}_{\GEL}=\arg\inf_{\theta\in\Theta}
\sup_{\eta\in\hat{\Lambda}_{\N,\psi}(\theta,\hat{\varphi})}\hat{P}_{\N}(\theta,
\eta,\hat{\varphi}),
\end{equation}
where $\hat{P}_{\N}(\theta,\eta,\varphi)
=\sum_{h=1}^H\sum_{i\in \mS_h}\{\rho(\eta^{\top}\pi_{hi}^{-1}\psi(Z_{hi},\theta,\hat\varphi))-\rho_0\}$ and
$\hat{\Lambda}_{\N,\psi}(\theta,\varphi)=\{\eta:\eta^{\top}\pi_{hi}^{-1}
\psi(Z_{hi},\theta,\varphi)\in\mathcal {V}, i\in\mS_h,h=1,\cdots,H\}$. We call $\hat{\theta}_{\GEL}$ under stratified sampling the pooled GEL estimator.
It follows from Theorem 3.2 that  the pooled GEL estimator $\hat{\theta}_{\GEL}$ defined in (\ref{gele-str})  is asymptotically normally distributed with mean $\theta_{\N}$ and variance-covariance matrix $n_{\B}^{-1}V_2$,
where $V_2$ has the same form given in Theorem 3.2 with the matrices $W_2$ and $\Omega$ replaced respectively by 
 $
 W_2= n_{\B} N^{-2}
 \sum_{h=1}^H\sum_{i=1}^{N_{\H}}g(Z_{hi},\theta_{\N},\varphi_{\N})^{\otimes 2}
$ and
 $$
\Omega= \dfrac{n_{\B}}{N^2}\sum_{h=1}^H{\rm Var}\Big\{\sum_{i\in \mS_h}\pi_{hi}^{-1}[\psi(Z_{hi},\theta_{\N},\varphi_{\N})] \mid \mF_{\N}\Big\}.
$$
If the stratum samples $\mS_h$ are selected by a PPS sampling design with small sampling fractions, we can estimate $\Omega$ by
\[
\hat{\Omega}=\frac{n}{N^2}\sum_{h=1}^{\H}\sum_{i\in \mS_h}\Big [\pi_{hi}^{-1}\psi(Z_{hi},\hat{\theta}_{\GEL},\hat{\varphi})-\hat{U}_{h}(\hat{\theta}_{\GEL},\hat\varphi)\Big]^{\otimes 2},
\]
where   $\hat{U}_{h}(\theta,\varphi)= n_h^{-1}\sum_{i\in \mS_h} \pi_{hi}^{-1} \psi(Z_{hi},\theta,\varphi)$. In cases where  the sampling fractions are not negligible, using arguments similar to those presented in Section 5 of the main paper, we can estimate $\Omega$ by
\[
\hat{\Omega} = \dfrac{n}{N^2}\sum_{h=1}^{\H}\sum_{i\in \mS_h} c_{ih}\big [\pi_{ih}^{-1}\psi(Z_{ih},\hat{\theta}_{\GEL},\hat{\varphi})-\hat{\mathscr{B}_h}\big]^{\otimes 2}\,,
\]
where
$
\hat{\mathscr{B}}_h = \sum_{i\in\mS_h}c_{ih}\pi_{ih}^{-1}\psi(Z_{ih},\hat{\theta}_{\GEL},\hat{\varphi})/\sum_{i\in\mS_h}c_{ih}$ and  $
c_{ih} = \{n_h(1-\pi_{ih})\}/\{n_h-1\}.
$

Under stratified sampling designs, the weight matrix $W_2$ and the derivative $\Gamma_2$ can be consistently estimated by using  the same estimators for  single-stage sampling designs,  with  $Z_i$, $\pi_{i}$ and $\sum_{i\in \mS}$ respectively replaced by $Z_{hi}$, $\pi_{ih}$ and $\sum_{h=1}^{\H}\sum_{i\in \mS_h}$.

\subsection{Cluster sampling}
\label{sec.cluster}
We now consider cluster sampling. Suppose that the population is divided into $K$
clusters and that the $i$th cluster has $M_i$ elements. The overall population size is  $N=\sum_{i=1}^{K}M_i$. Let $Z_{(ij)}$ be the value of $Z$ for the $j$th element in the $i$th cluster. Then, the true parameter    $\theta_{\N}$  satisfies
 \begin{equation} \label{clst-cl}
U_{\N}(\theta_{\N},\varphi_{\N})=\dfrac{1}{N}\sum_{i=1}^K\sum_{j=1}^{M_i}g(Z_{(ij)},\theta_{\N},\varphi_{\N})=0.
\end{equation}
We consider two-stage cluster sampling designs where the first stage sample $\mS_c$ is a set of $k$ clusters selected  from the population with inclusion probabilities $\pi_{1i} = {\rm Pr}(i\in \mS_c)$, and the second stage sample $\mS_i$ is 
a set of $m_i$ ($\le M_i$) units drawn from cluster  $i\in \mS_c$ with second-stage inclusion probabilities  $\pi_{j | i}={\rm Pr}(j\in \mS_i \mid i\in \mS_c)$.
The final first order inclusion probability for selecting unit $(ij)$ is given by $\pi_{(ij)} = {\rm Pr}(i \in \mS_c, j\in \mS_i) = \pi_{1i}\pi_{j | i}$. 
A popular  two-stage sampling design is the so-called self-weighting design for which $\pi_{1i} = kM_i/N$ and   $\pi_{j | i} = m/M_i$ such that
the final first order inclusion probabilities are the same for all units.

We  assume that  under two-stage cluster sampling the population estimating functions defined in (\ref{clst-cl}) and the first-step estimator $\hat{\varphi}$ satisfy the  following two regularity conditions:
\begin{itemize}
\item [D1.]
There exists a function $U(\theta,\varphi)$ such that $\sup_{(\theta,\varphi)\in\Theta\times\Psi}\|U_{\N}(\theta,\varphi)-U(\theta,\varphi)\|=o(1)$,  and
   $U(\theta,\varphi)$  satisfies conditions A4 and A5 presented  in Appendix A.

\item [D2.]
The  pathwise derivative $D(\theta_{\N},\varphi_{\N})[\hat{\varphi}-\varphi_{\N}]$  is of   the following form:
$$D(\theta_{\N},\varphi_{\N})[\hat{\varphi}-\varphi_{\N}]=\dfrac{1}{N}\sum_{i\in \mS_c}\sum_{j\in \mS_i} \pi_{(ij)}^{-1} \Xi(Z_{(ij)},\theta_{\N},\varphi_{\N})+o_p(n_{\B}^{-1/2}),$$
where  $\Xi(\cdot)$ satisfies the following
conditions:  (i) $\Xi(Z_{(ij)},\theta_{\N},\varphi_{\N})$ has finite fourth population moments; and (ii)
$\sum_{i\in \mS_c}\sum_{j\in \mS_i} \pi_{(ij)}^{-1} \Xi(Z_{(ij)},\theta_{\N},\varphi_{\N})$ is asymptotically normally distributed with mean zero and variance-covariance matrix at the order $O(n_{\B}^{-1}N^2)$.
\end{itemize}

Under two-stage cluster sampling, the parameter $\theta_{\N}$ is defined by
(\ref{clst-cl}). The efficient  two-step   GEL estimator $\hat{\theta}_{\GEL}$ is  defined as in (\ref{gele-str}) but replaces  $\hat{P}_{\N}(\theta,\eta,\hat{\varphi})$ by
\begin{equation*}
\hat{P}_{\N}(\theta,\eta,\hat{\varphi})
=\sum_{i\in \mS_c}\sum_{j\in \mS_i}\{\rho(\eta^{\top}\pi_{(ij)}^{-1}\psi(Z_{(ij)},\theta,\hat\varphi))-\rho_0\},
\end{equation*}
where $\psi(Z_{(ij)},\theta,\varphi)=g(Z_{(ij)},\theta,\varphi)+\Xi(Z_{(ij)},\theta,\varphi)$.
We can show that $
n_{\B}^{1/2}(\hat{\theta}_{\GEL}-\theta_{\N})\stackrel{{\cal L}}{\rightarrow}N(0,V_2),
$
where
 $
V_2=\Sigma_2\Gamma_2^{\top}W_2^{-1}\Omega W_2^{-1}\Gamma_2 \Sigma_2 \,,
\label{V1}
$
$\Sigma_2=(\Gamma_2^{\top}W_2^{-1}\Gamma_2)^{-1}$,    $\Gamma_2=\Gamma_2(\theta_{\N},\varphi_{\N})$,   $\Gamma_2(\theta,\varphi)$ is the ordinary derivate of $U(\theta,\varphi)$ with respect to $\theta$,
 $$
 W_2= \dfrac{n_{\B}}{N^2}\sum_{i=1}^K\sum_{j=1}^{M_i}\pi_{(ij)}^{-1}\psi(Z_{(ij)},\theta_{\N},\varphi_{\N})^{\otimes 2}\,,
 $$
 and
\begin{equation*}
\label{omega-ts}
\Omega =\dfrac{n_{\B}}{N^2} {\rm Var}\Big\{\sum_{i\in \mS_c}\sum_{j\in \mS_i} \pi_{(ij)}^{-1} \psi(Z_{(ij)},\theta_{\N},\varphi_{\N})\mid \mF_{\N}\Big\} \,.
\end{equation*}
For self-weighting two-stage sampling designs,  we can estimate  $\Omega$  directly by
\[
\hat{\Omega}=\frac{1}{k(k-1)}\sum_{i\in \mS_c}\Big (\bar{G}_i - \bar{G}\Big )^{\otimes 2}\,,
\]
where   $\bar{G}_i = \sum_{j\in \mS_i}\psi(Z_{(ij)},\hat{\theta}_{\GEL},\hat{\varphi})/m$ and
$\bar{G} = \sum_{i\in \mS_c}\bar{G}_i/k$.
Design consistent estimators of  weight matrix $W_2$ and derivative $\Gamma_2$ can be easily obtained under general two-stage cluster sampling with suitable changes in notation.

The results  described in Sections \ref{sec.single}  for single-stage unequal probability sampling  and  those for stratified sampling and  cluster sampling can be combined for
variance estimation under the more general stratified multi-stage sampling designs.

\section{Simulation Studies}
\label{sec.sim}

\setcounter{equation}{0}

\noindent
In this section, we report results from a simulation study on the finite  sample performances of our proposed augmented two-step GEL  estimators and the GEL ratio confidence intervals when the sample data are selected from a finite population by a probability sampling method.  The finite population $\mathcal{F}_{\N}=(Z_1,\cdots, Z_{\N})$ of size $N$ is generated from the model
$$
Z_i=X_i+\varepsilon_i,\,\,\,\,i=1,\cdots,N \, ,
$$
where  $X_i \sim 0.25+ {\rm Weibull}(2,2)$  and $\varepsilon_i$ follows the $\chi^2$ distribution with 3 degrees of freedom.
The parameter of interest is the finite population quantile share $\theta_{\N}(\tau_1,\tau_2)$  of  the  population  $\mathcal{F}_{\N}$ as discussed in
Section \ref{sec.exam}.
We consider four  scenarios for the quantile levels:
$(\tau_1,\tau_2)=(0,0.25)$,  $(0.25,0.5)$, $(0.5,0.75)$, $(0.75,1)$.

The finite population, once generated, is held fixed and  repeated simulation samples are selected   from  the finite population using  the following   four sampling methods:

\vspace{1.5mm}
\noindent (A)  Single-stage randomized systematic PPS sampling without replacement with small sampling fractions. The  finite population size  and the sample size are taken to be  $(N, n) = (20000, 300)$.  The sampling fraction is $n/N=1.5\%$, which can be viewed as negligible.

\vspace{1.5mm}
\noindent (B)  Single-stage Rao-Sampford PPS  sampling without replacement with large sampling fractions.  The  finite population size and the sample size  are chosen as  $(N, n) = (3000, 300)$.
The sampling fraction    is   $10\%$, which is non-negligible.

\vspace{1.5mm}
\noindent (C) Stratified Rao-Sampford PPS  sampling. The  finite population is divided into  $H = 3$ strata
with stratum sizes $(N_1,N_2,N_3)=(4000, 6000, 10000)$. Stratum  samples of size $n_h$  are selected  by the randomized systematic PPS sampling, for $h=1,2,3$.
The stratum  sample  sizes  are chosen as $(n_1,n_2,n_3)=(50,100,150)$.
The total sample size is   $n=n_1+n_2+n_3$.

\vspace{1.5mm}
\noindent (D) Two-stage cluster sampling. The finite population is split into  $1350$ clusters, with $200$ clusters having equal cluster size $M_j=30$, $250$ clusters having size $M_j=20$, and $900$ clusters having size $M_j=10$.
In the  first stage
sampling $k=n/5$  clusters are selected  by the randomized systematic
PPS sampling, and in   the second stage sampling $m = 5$ units are selected within each
selected cluster, independent among different clusters, by simple random sampling
without replacement.  The overall sample size is taken to be $n = 300$.

For each of  the four sampling methods, a total of 1000 simulation samples are selected from the finite population.
For each selected sample, we calculate the survey weighted point estimator, and use the following six different methods to construct  the  $95\%$ confidence intervals for the quantile share $\theta_{\N}(\tau_1,\tau_2)$ at  each quantile levels
$(\tau_1,\tau_2)$: the GEL ratio confidence intervals using the standard chi-square limiting distributions for each of
EL, ET, CU and GMM; the normal approximation confidence interval using the estimating equation based point estimator and a bootstrap estimate of the standard error (BC$_{n}$); and the bootstrap percentile interval with the estimating equation based point estimator (BC$_{p}$).
In all simulations, the proposed method (Augmented SWEE) is compared with the method of Zhao et al. (2020) (Conventional SWEE) under an assumed standard chi-square limiting distribution for the GEL ratio statistic. The latter is
based on
$
g(Z,\theta,\xi_1,\xi_2)=Z\{I(\xi_1<Z\leq \xi_2)-\theta\}
$ without the augmentation term, where $\xi_1$ and $\xi_1$ are the nuisance parameters with true values being the $\tau_1$th and  $\tau_2$th population quantile of $\mathcal{F}_{\N}$, respectively. Our  proposed augmented SWEE method is based on
$\psi(Z,\theta,\xi_1,\xi_2)=g(Z,\theta,\xi_1,\xi_2) -\xi_2\{I(Z\leq \xi_2)-  \tau_2\}+\xi_1\{I(Z\leq \xi_1)-  \tau_1\}$.

 \vspace{3mm}
 \noindent {\em  Point estimation.} \
Tables \ref{tab1}   presents the Monte Carlo   biases,  standard deviation (SD) and standard error (SE) of the survey weighted   estimator of $\theta_{\N}(\tau_1,\tau_2)$ at four  different values of $(\tau_1,\tau_2)$  for each of the four sampling methods. Here, the SD is the square root of the simulated true variance of the point estimator and the SE is the square root of the bootstrap variance estimator.
From Table \ref{tab1}, we have the following  observations. (i) Under all the settings,  the proposed augmented survey weighted estimators have negligible biases, and the values of SE are all close to the corresponding values of SD;
(ii) The proposed estimator has values of SD similar to the conventional estimator without the augmentation term.

\vspace{3mm}
 \noindent {\em  Confidence intervals.} \
 The finite sample performances of the six confidence intervals  are evaluated using the following criteria:
the  average lengths (AL),
 the coverage probabilities (CP), the lower tail error (LE) and the upper tail error (UE), which are respectively computed as
 \[\begin{array}{lllllllllll}
 \mbox{AL}& =& \dfrac{1}{M}\sum\limits_{m=1}^M \Bigl\{P_{{\rm \U}}^{(m)}(\tau_1,\tau_2) - P_{{\rm \L}}^{(m)}(\tau_1,\tau_2)\Bigr\},\\
\mbox{CP} & =&\dfrac{1}{M}\sum\limits_{m=1}^M I\Bigl\{P_{{\rm \L}}^{(m)}(\tau_1,\tau_2) < \theta_{\N}(\tau_1,\tau_2) < P_{{\rm \U}}^{(m)}(\tau_1,\tau_2)\Bigr\}, \\
 \mbox{LE} & =& \dfrac{1}{M}\sum\limits_{m=1}^M I\Bigl\{\theta_{\N}(\tau_1,\tau_2) \leq  P_{{\rm \L}}^{(m)}(\tau_1,\tau_2) \Bigr\},\\
\mbox{UE}& =&\dfrac{1}{M}\sum\limits_{m=1}^M I\Bigl\{ \theta_{\N}(\tau_1,\tau_2) \geq P_{{\rm \U}}^{(m)}(\tau_1,\tau_2)\Bigr\} ,
\end{array}
\]
where $P_{{\rm \L}}^{(m)}(\tau_1,\tau_2)$ and $P_{{\rm \U}}^{(m)}(\tau_1,\tau_2)$ are respectively  the
lower and upper boundaries of the   $95\%$ confidence interval computed from the $m$th simulation sample,  and $M$ is the  number of simulation runs.

Tables \ref{tab2}--\ref{tab5} report the simulation results computed  from  $M=1000$ simulation runs.
Using the proposed augmented estimating functions,
the GEL and GMM confidence intervals  have excellent performance  in terms of all the criteria listed above. Both
have better coverage accuracy than the normal approximation and bootstrap based confidence intervals.
The proposed GEL and GMM confidence intervals have coverage probabilities which are closer to the nominal level, and have shorter lengths than those of  normal approximation and bootstrap  methods.
Without using the augmentation term, the GEL and GMM  approaches under an assumed standard chi-square limiting distributions give  invalid results as coverage probabilities are completely off  the target nominal value.

\section{An Application}
\label{sec.data}

\setcounter{equation}{0}

\noindent
The proposed methods are further illustrated with an application to the   New York City Social Indicators Survey (NYSIS).
The  NYSIS
was a biennial survey of New York City
residents conducted   by the Columbia University
School of Social Work.
The core survey is designed  to demonstrate
the use of  several social indicators to
answer questions about inequality and wellbeing.  The survey also measures the
sources and extent of external supports from government, family and friends, community
and religious programs, and employers.
We use data  from the   2002 NYSIS survey (Teitler et al., 2004), which  examined the period between March and June, 2002.
The original data   were collected from 1501 adults  through telephone interviews and census individual weight  was assigned to each survey respondent.

In this example we are interested in making
statistical inference for quintile shares     on the  respondent's earnings in 2002. We consider a subset of the  2002 survey sample consisting of $n=956$ respondents who   have positive  earnings.
For analytical purposes, we re-scale the survey weights by $w_i=n\tilde{w}/\sum_{j=1}^{n}\tilde{w}_j$ such that $\sum_{i=1}^{n}w_i=n$, where $\tilde{w}_i$ is the original weight for $i$th survey respondent.
Four different quantile shares are considered: $\theta_{\N}(0,0.25)$,  $\theta_{\N}(0.25,0.5)$, $\theta_{\N}(0.5,0.75)$ and $\theta_{\N}(0.75,1)$.
The point estimator and their standard errors are computed in the same way as the simulation study described in Section \ref{sec.sim}.
In line with the simulation study, we also use  the same six methods  to construct    $95\%$ confidence intervals for quantile shares:
EL, ET, CU, GMM,  BC$_{n}$ and   BC$_{p}$.
We include both  the augmented and the conventional SWEE  methods for all the cases considered.

The analysis results are reported in Tables \ref{tab6}--\ref{tab7}.   In terms of point estimators, the two approaches produce very similar values. However, the $95\%$ confidence intervals from the
the conventional SWEE analysis are much wider than those from the augmented SWEE analysis. This is consistent with the theoretical results as well as results from the simulation studies.

\section{Discussion}
\label{sec.dis}

\setcounter{equation}{0}

\noindent
Semiparametric modeling techniques using estimating equations
combine the flexibility and robustness of nonparametric models and the interpretability of parametric models, and
provide a powerful general framework for analytical use of complex survey data. In the presence of nuisance functions, however, the conventional two-step semiparametric estimation approach with simple plug-in estimators for the nuisance function is not only inefficient but also sensitive to the plug-in estimator. Moreover, the conventional approach lacks the asymptotic pivotalness and is difficult to use in practice.
Our proposed augmented estimating functions tackle the weaknesses of the conventional approach in dealing with nuisance functions and complex survey data, and lead to more efficient estimation of the main parameters of interest and more desirable features on confidence intervals and hypothesis tests.
We show that the augmented two-step GEL ratio statistic is asymptotically pivotal under some commonly used survey designs, and the resulting maximum  GEL estimators achieve the semiparametric efficiency bound. The inferential framework developed in this paper for design-based inferences using survey data is generally applicable to parameters defined through estimating equations in the presence of nuisance functions. The proposed methods do not follow from any work in the existing literature and are especially attractive to problems in economic studies on inequality measures.

Applications of machine learning methods to the semiparametric estimation problems have received considerable attention in recent years.
Under the model-based framework with independent samples, Chernozhukov et al. (2018) proposed double/debiased machine learning estimators for treatment and structural parameters;
Chernozhukov et al. (2022)  considered  debiased and robust semiparametric GMM  estimation for plug-in semiparametric estimating equations; Chang (2020)  developed double/debiased machine learning estimators for difference-in-differences
models. In the  survey context,  Dagdoug et al. (2021)  proposed using random forests to construct
a new class of model-assisted estimators  for finite population parameters.  In this paper, we show that the limiting distribution of the point estimator of the main parameters of interest is invariant
to the first step plug-in estimator for the nuisance parameters under our proposed augmented approach. It is a challenge research question on how to deal with scenarios where machine learning methods are used in the first-step estimation under the current setting.
The development of appropriate machine learning methods  for general semiparametric models with survey data requires future investigation.

\section*{Acknowledgments}

\noindent
The authors thank the Editor and  anonymous referees
for their comments and suggestions that led to major improvement of the paper.
This research is supported by the scientific research fund for high-level talents of  Yunnan province, the National Natural Science Foundation of China (grant NO. 12071416), Yunnan Fundamental Research Projects (grant NO. 202201AV070006), and grants from the Natural Sciences and Engineering Research Council of Canada and The Canadian Statistical Sciences Institute.


\begin{table}[h]
\centering
\caption{Point Estimates of Quantile Shares}
\label{tab1}
\begin{tabular}{cccccccccccccccccc}\hline\hline
 &&\multicolumn{3}{c} {Augmented SWEE} &&\multicolumn{3}{c} {Conventional SWEE}\\
\cline{3-5} \cline{7-9}
Design &$\;\;\;$ $(\tau_1,\tau_2)$ $\;\;\;$ & Bias& SD&SE && Bias& SD&SE\\ \hline
A&(0.00,\,0.25)&	0.000 &0.005 &0.005  &&0.000 &0.006& 0.006 \\
&(0.25,\,0.50)&	-0.001 &0.005 &0.005  &&0.000 &0.005 &0.006 \\
&(0.50,\,0.75)&	-0.002 &0.005 &0.006  &&0.000 &0.005 &0.006 \\
&(0.75,\,1.00)&	-0.002 &0.010 &0.010  &&-0.002 &0.010 &0.010 \\

B&(0.00,\,0.25)&	0.000& 0.005 &0.005  &&0.000 &0.005 &0.005 \\
&(0.25,\,0.50)&	-0.005& 0.004 &0.006  &&-0.001& 0.004& 0.006 \\
&(0.50,\,0.75)&	-0.007 &0.005 &0.007  &&0.000 &0.005 &0.006 \\
&(0.75,\,1.00)&	-0.007 &0.009 &0.010  &&-0.001& 0.009& 0.010 \\

C&(0.00,\,0.25)&	0.000 &0.006 &0.006  &&0.000& 0.006& 0.006 \\
&(0.25,\,0.50)&	-0.001& 0.004 &0.005  &&0.000& 0.004 &0.005 \\
&(0.50,\,0.75)&	-0.002& 0.004 &0.005  &&0.000 &0.004 &0.005\\
&(0.75,\,1.00)&	-0.003& 0.009 &0.009  &&-0.001 &0.009 &0.009 \\

D&(0.00,\,0.25)&	0.000& 0.004 &0.004  &&0.000 &0.004& 0.004 \\
&(0.25,\,0.50)&	0.000 &0.004& 0.004  &&0.000 &0.004 &0.005\\
&(0.50,\,0.75)&0.000 &0.004 &0.005 &&0.000 &0.004& 0.006\\
&(0.75,\,1.00)&	-0.001 &0.009 &0.009  &&-0.001 &0.009& 0.009 \\

\hline\hline
\end{tabular}
\end{table}

\begin{table}[h]
\centering
\caption{$95\%$ Confidence Intervals of Quantile Shares Under the Survey Design A}
\label{tab2}
\begin{tabular}{cccccccccccccccccc}\hline\hline
 &&\multicolumn{4}{c} {Augmented SWEE} &&\multicolumn{4}{c} {Conventional SWEE}\\
\cline{3-6}  \cline{8-11}
Methods&$(\tau_1,\tau_2)$&LE&CP&UE&AL&&LE&CP&UE&AL\\ \hline


EL&(0.00,\,0.25)&	0.037& 0.940& 0.023 &0.024 && 0.000 &1.000& 0.000 &0.068 \\
&(0.25,\,0.50)&	0.031& 0.947 &0.022 &0.019 && 0.000 &1.000 &0.000 &0.091\\
&(0.50,\,0.75)&	0.032 &0.940 &0.028 &0.020 && 0.000& 1.000& 0.000& 0.111 \\
&(0.75,\,1.00)&	0.017 &0.944 &0.039 &0.040 && 0.000 &1.000& 0.000& 0.137 \\

ET&(0.00,\,0.25)&	0.051& 0.932 &0.017& 0.023 && 0.000 &1.000 &0.000 &0.068 \\
&(0.25,\,0.50)&	0.033 &0.944 &0.023 &0.019 && 0.000& 1.000& 0.000 &0.091 \\
&(0.50,\,0.75)&	0.031 &0.940 &0.029 &0.020 && 0.000& 1.000 &0.000& 0.111 \\
&(0.75,\,1.00)&	0.013 &0.943 &0.044 &0.039 && 0.000 &1.000 &0.000& 0.138 \\

CU&(0.00,\,0.25)&	0.057 &0.932& 0.011& 0.023 && 0.000& 1.000 &0.000& 0.068 \\
&(0.25,\,0.50)&	0.031& 0.948 &0.021 &0.019 && 0.000 &1.000 &0.000 &0.091\\
&(0.50,\,0.75)&	0.031 &0.942 &0.027 &0.020 && 0.000 &1.000 &0.000 &0.112\\
&(0.75,\,1.00)&	0.009 &0.944 &0.047 &0.040 && 0.000 &1.000 &0.000 &0.139 \\

GMM&(0.00,\,0.25)&	0.059 &0.929& 0.012 &0.023 && 0.000 &1.000 &0.000 &0.068 \\
&(0.25,\,0.50)&	0.033& 0.944 &0.023& 0.019 && 0.000& 1.000 &0.000 &0.091 \\
&(0.50,\,0.75)&	0.031& 0.940 &0.029& 0.020 && 0.000& 1.000& 0.000 &0.111 \\
&(0.75,\,1.00)&	0.010& 0.942 &0.048 &0.039 && 0.000& 1.000 &0.000 &0.138 \\

BC$_{n}$&(0.00,\,0.25)&	0.059 &0.928& 0.013& 0.023 && 0.057& 0.930 &0.013 &0.023 \\
&(0.25,\,0.50)&	0.014& 0.969 &0.017 &0.021 && 0.021 &0.966 &0.013 &0.021 \\
&(0.50,\,0.75)&	0.008 &0.966 &0.026 &0.023 && 0.022 &0.958 &0.020 &0.023 \\
&(0.75,\,1.00)&	0.008 &0.945 &0.047 &0.041 && 0.011 &0.951 &0.038& 0.041 \\

BC$_{p}$&(0.00,\,0.25)&	0.060& 0.926 &0.014 &0.022 && 0.045 &0.945 &0.010 &0.024 \\
&(0.25,\,0.50)&	0.010& 0.972& 0.018& 0.021 && 0.012 &0.981& 0.007 &0.024 \\
&(0.50,\,0.75)&	0.007 &0.965 &0.028 &0.023 && 0.004 &0.992 &0.004 &0.026 \\
&(0.75,\,1.00)&	0.007 &0.936 &0.057 &0.041 && 0.009 &0.949& 0.042 &0.041\\

\hline\hline
\end{tabular}
\end{table}

\begin{table}[h]
\centering
\caption{$95\%$ Confidence Intervals of Quantile Shares Under the Survey Design B}
\label{tab3}
\begin{tabular}{cccccccccccccccccc}\hline\hline
 &&\multicolumn{4}{c} {Augmented SWEE} &&\multicolumn{4}{c} {Conventional SWEE}\\
\cline{3-6}  \cline{8-11}
Methods&$(\tau_1,\tau_2)$&LE&CP&UE&AL&&LE&CP&UE&AL\\ \hline


EL&(0.00,\,0.25)&	0.039 &0.933 &0.028 &0.023 && 0.000 &1.000& 0.000 &0.067 \\
&(0.25,\,0.50)&	0.022& 0.963& 0.015 &0.020 && 0.000 &1.000 &0.000 &0.090 \\
&(0.50,\,0.75)&	0.026 &0.952& 0.022 &0.020 && 0.000 &1.000 &0.000 &0.112 \\
&(0.75,\,1.00)&	0.015 &0.956 &0.029 &0.038 && 0.000 &1.000 &0.000 &0.136 \\

ET&(0.00,\,0.25)&	0.044 &0.932& 0.024 &0.022 && 0.000& 1.000& 0.000 &0.066\\
&(0.25,\,0.50)&	0.022 &0.963 &0.015 &0.020 && 0.000 &1.000 &0.000 &0.090 \\
&(0.50,\,0.75)&	0.025 &0.953& 0.022 &0.019 && 0.000 &1.000 &0.000& 0.112 \\
&(0.75,\,1.00)&	0.012 &0.956 &0.032 &0.038 && 0.000& 1.000 &0.000& 0.136 \\

CU&(0.00,\,0.25)&	0.048 &0.940 &0.012 &0.022 && 0.000& 1.000& 0.000& 0.067 \\
&(0.25,\,0.50)&	0.022 &0.964& 0.014& 0.020 && 0.000& 1.000 &0.000& 0.091 \\
&(0.50,\,0.75)&	0.022& 0.959 &0.019 &0.020 && 0.000 &1.000 &0.000& 0.113 \\
&(0.75,\,1.00)&	 0.010& 0.957 &0.033& 0.038 && 0.000& 1.000& 0.000& 0.138 \\

GMM&(0.00,\,0.25)&	0.046& 0.942& 0.012 &0.022 && 0.000 &1.000& 0.000& 0.066 \\
&(0.25,\,0.50)&	0.023& 0.962 &0.015& 0.020 && 0.000 &1.000 &0.000 &0.090 \\
&(0.50,\,0.75)&	0.022& 0.958 &0.020 &0.020 && 0.000 &1.000 &0.000 &0.112 \\
&(0.75,\,1.00)&	0.011& 0.956 &0.033 &0.038 && 0.000 &1.000 &0.000 &0.137 \\

BC$_{n}$&(0.00,\,0.25)&	0.050& 0.936 &0.014 &0.021 && 0.054 &0.934& 0.012& 0.021 \\
&(0.25,\,0.50)&	0.000& 0.953 &0.047 &0.025 && 0.007 &0.987 &0.006 &0.025\\
&(0.50,\,0.75)&	0.000 &0.931 &0.069 &0.028 && 0.004& 0.991 &0.005 &0.028 \\
&(0.75,\,1.00)&	0.001 &0.932 &0.067 &0.042 && 0.009& 0.966 &0.025 &0.042 \\

BC$_{p}$&(0.00,\,0.25)&	0.057& 0.927& 0.016 &0.021 && 0.044 &0.946& 0.010 &0.023 \\
&(0.25,\,0.50)&	0.000 &0.958 &0.042 &0.025 && 0.007& 0.984 &0.009 &0.024 \\
&(0.50,\,0.75)&	0.000 &0.932& 0.068 &0.028 && 0.004& 0.994 &0.002 &0.026 \\
&(0.75,\,1.00)&	0.001 &0.921& 0.078 &0.042 && 0.010 &0.956 &0.034 &0.040 \\

\hline\hline
\end{tabular}
\end{table}

\begin{table}[h]
\centering
\caption{$95\%$ Confidence Intervals of Quantile Shares Under the Survey Design C}
\label{tab4}
\begin{tabular}{cccccccccccccccccc}\hline\hline
 &&\multicolumn{4}{c} {Augmented SWEE} &&\multicolumn{4}{c} {Conventional SWEE}\\
\cline{3-6}  \cline{8-11}
Methods&$(\tau_1,\tau_2)$&LE&CP&UE&AL&&LE&CP&UE&AL\\ \hline


EL&(0.00,\,0.25)&	0.049& 0.916& 0.035 &0.026 && 0.000 &1.000 &0.000& 0.070 \\
&(0.25,\,0.50)&	0.033 &0.946 &0.021 &0.018 && 0.000 &1.000 &0.000& 0.087 \\
&(0.50,\,0.75)&	0.026 &0.942 &0.032 &0.018 && 0.000 &1.000 &0.000 &0.101 \\
&(0.75,\,1.00)&	0.028 &0.939 &0.033 &0.036 && 0.000 &1.000 &0.000 &0.117 \\

ET&(0.00,\,0.25)&	0.068 &0.912 &0.020& 0.025 && 0.000& 1.000 &0.000 &0.070 \\
&(0.25,\,0.50)&	0.033 &0.947& 0.020 &0.018 && 0.000 &1.000 &0.000& 0.087 \\
&(0.50,\,0.75)&	0.025 &0.943& 0.032 &0.018 && 0.000 &1.000& 0.000 &0.101 \\
&(0.75,\,1.00)&	0.025 &0.941 &0.034 &0.035 && 0.000 &1.000& 0.000 &0.117 \\

CU&(0.00,\,0.25)&	0.078& 0.915& 0.007& 0.025 && 0.000 &1.000 &0.000 &0.070 \\
&(0.25,\,0.50)&	0.034 &0.946& 0.020& 0.018 && 0.000 &1.000 &0.000 &0.088 \\
&(0.50,\,0.75)&	0.025 &0.943 &0.032 &0.018 && 0.000 &1.000 &0.000 &0.102 \\
&(0.75,\,1.00)&	0.019 &0.944 &0.037 &0.036 && 0.000 &1.000 &0.000 &0.118 \\

GMM&(0.00,\,0.25)&	0.079 &0.913& 0.008 &0.025 && 0.000 &1.000 &0.000 &0.070 \\
&(0.25,\,0.50)&	0.034 &0.946 &0.020 &0.018 && 0.000 &1.000& 0.000 &0.087 \\
&(0.50,\,0.75)&	0.025 &0.941& 0.034 &0.018 && 0.000 &1.000 &0.000 &0.101 \\
&(0.75,\,1.00)&	0.020 &0.943& 0.037 &0.035 && 0.000 &1.000 &0.000& 0.118 \\

BC$_{n}$&(0.00,\,0.25)&	0.083 &0.901 &0.016 &0.024 && 0.081& 0.905& 0.014 &0.024 \\
&(0.25,\,0.50)&	0.008& 0.975 &0.017 &0.020 && 0.017& 0.971& 0.012 &0.020 \\
&(0.50,\,0.75)&	0.005 &0.952 &0.043& 0.020 && 0.017 &0.961 &0.022 &0.020 \\
&(0.75,\,1.00)&	0.011& 0.926 &0.063 &0.037 && 0.024 &0.943& 0.033 &0.037\\

BC$_{p}$&(0.00,\,0.25)&	0.084& 0.900 &0.016 &0.024 && 0.066 &0.923 &0.011 &0.026 \\
&(0.25,\,0.50)&	0.003& 0.978& 0.019 &0.020 && 0.006 &0.990& 0.004 &0.022 \\
&(0.50,\,0.75)&	0.001& 0.954& 0.045& 0.020 && 0.006 &0.989 &0.005 &0.022 \\
&(0.75,\,1.00)&	0.011 &0.933 &0.056 &0.036 && 0.022 &0.946 &0.032 &0.036 \\

\hline\hline
\end{tabular}
\end{table}

\begin{table}[h]
\centering
\caption{$95\%$ Confidence Intervals of Quantile Shares Under the Survey Design D}
\label{tab5}
\begin{tabular}{cccccccccccccccccc}\hline\hline
 &&\multicolumn{4}{c} {Augmented SWEE} &&\multicolumn{4}{c} {Conventional SWEE}\\
\cline{3-6}  \cline{8-11}
Methods&$(\tau_1,\tau_2)$&LE&CP&UE&AL&&LE&CP&UE&AL\\ \hline


EL&(0.00,\,0.25)&	0.022 &0.962 &0.016 &0.017 && 0.000 &1.000& 0.000 &0.056 \\
&(0.25,\,0.50)&	0.024 &0.954& 0.022 &0.018 && 0.000 &1.000 &0.000 &0.083 \\
&(0.50,\,0.75)&	0.030 &0.951 &0.019& 0.019 && 0.000 &1.000 &0.000 &0.106 \\
&(0.75,\,1.00)&	0.019 &0.945 &0.036 &0.037 && 0.000 &1.000 &0.000 &0.132 \\

ET&(0.00,\,0.25)&	0.026& 0.960& 0.014 &0.017 && 0.000 &1.000 &0.000 &0.056 \\
&(0.25,\,0.50)&	0.025 &0.954 &0.021 &0.018 && 0.000 &1.000 &0.000 &0.083 \\
&(0.50,\,0.75)&	0.031 &0.952 &0.017 &0.019 && 0.000 &1.000 &0.000 &0.106 \\
&(0.75,\,1.00)&	0.019 &0.940 &0.041& 0.036 && 0.000 &1.000 &0.000 &0.132 \\

CU&(0.00,\,0.25)&	0.032 &0.955& 0.013 &0.017 && 0.000 &1.000 &0.000 &0.057 \\
&(0.25,\,0.50)&	0.026 &0.956 &0.018& 0.018 && 0.000 &1.000& 0.000& 0.084 \\
&(0.50,\,0.75)&	0.031 &0.953 &0.016 &0.019 && 0.000 &1.000 &0.000& 0.107 \\
&(0.75,\,1.00)&	0.016& 0.941& 0.043 &0.036 && 0.000 &1.000 &0.000 &0.134 \\

GMM&(0.00,\,0.25)&	0.031& 0.955& 0.014& 0.017 && 0.000 &1.000 &0.000 &0.056 \\
&(0.25,\,0.50)&	0.025 &0.956 &0.019 &0.018 && 0.000& 1.000& 0.000 &0.083 \\
&(0.50,\,0.75)&	0.031 &0.952 &0.017& 0.019 && 0.000 &1.000& 0.000 &0.106 \\
&(0.75,\,1.00)&	0.016 &0.941 &0.043 &0.036 && 0.000 &1.000& 0.000 &0.132 \\

BC$_{n}$&(0.00,\,0.25)&	0.033 &0.952 &0.015& 0.017 && 0.033 &0.952 &0.015& 0.017 \\
&(0.25,\,0.50)&	0.015 &0.970 &0.015 &0.019 && 0.015 &0.970 &0.015 &0.019 \\
&(0.50,\,0.75)&	0.025 &0.963 &0.012 &0.021 && 0.025 &0.963 &0.012 &0.021 \\
&(0.75,\,1.00)&	0.014 &0.951 &0.035 &0.038 && 0.014 &0.951& 0.035 &0.038 \\

BC$_{p}$&(0.00,\,0.25)&	0.038& 0.950 &0.012 &0.017 && 0.029 &0.961& 0.010 &0.018\\
&(0.25,\,0.50)&	0.018& 0.966& 0.016& 0.019 && 0.013& 0.976 &0.011& 0.020 \\
&(0.50,\,0.75)&	0.025& 0.964 &0.011 &0.021 && 0.017 &0.979& 0.004 &0.023 \\
&(0.75,\,1.00)&	0.015 &0.947 &0.038 &0.038 && 0.013& 0.949 &0.038& 0.037 \\

\hline\hline
\end{tabular}
\end{table}

\begin{table}[h]
\centering
\caption{NYSIS Study:  Point Estimates of Quantile Shares on Earnings}
\label{tab6}
\begin{tabular}{cccccccccccccccccc}\hline\hline
 &\multicolumn{2}{c} {Augmented SWEE} &&\multicolumn{2}{c} {Conventional SWEE}\\
\cline{2-3} \cline{5-6}
$\;\;\;$ $(\tau_1,\tau_2)$ $\;\;\;$ & Estimate&Standard Error && Estimate&Standard Error\\ \hline
(0.00,\,0.25)&	0.037 &0.003& $\;\;\;\;\;\;$ & 0.036 &0.004\\
(0.25,\,0.50)&	0.107 &0.010&& 0.118 &0.009 \\
(0.50,\,0.75)&	0.198 &0.020 &&0.215 &0.017 \\
(0.75,\,1.00)&	0.615 &0.034 &&0.628 &0.026 \\
\hline\hline
\end{tabular}
\end{table}

\begin{table}[h]
\centering
\caption{NYSIS Study:  $95\%$ Confidence Intervals of Quantile Shares on Earnings }
\label{tab7}
\begin{tabular}{cccccccccccccccccc}\hline\hline
 &&\multicolumn{2}{c} {Augmented SWEE} &&\multicolumn{2}{c} {Conventional SWEE}\\
\cline{3-4}  \cline{6-7}
Methods&$(\tau_1,\tau_2)$&Confidence Interval&Length&&Confidence Interval&Length\\ \hline


EL&(0.00,\,0.25)&	(0.028, 0.043)& 0.015& $\;\;$ & (0.027, 0.046)& 0.019 \\
&(0.25,\,0.50)&	(0.093, 0.131)& 0.038&& (0.091, 0.145)& 0.054 \\
&(0.50,\,0.75)&	(0.178, 0.240)& 0.062 && (0.167, 0.258)& 0.091 \\
&(0.75,\,1.00)&	(0.588, 0.696)& 0.108 && (0.569, 0.705)& 0.136 \\

ET&(0.00,\,0.25)&	(0.029, 0.044)& 0.015 && (0.028, 0.046)& 0.018 \\
&(0.25,\,0.50)&	(0.097, 0.133)& 0.036 && (0.094, 0.146)& 0.052 \\
&(0.50,\,0.75)&	(0.186, 0.243)& 0.057 && (0.173, 0.260)& 0.087 \\
&(0.75,\,1.00)&	(0.584, 0.683)& 0.099 && (0.564, 0.693)& 0.129 \\

CU&(0.00,\,0.25)&	(0.030, 0.045)& 0.015 && (0.028, 0.047)& 0.019 \\
&(0.25,\,0.50)&	(0.099, 0.137)& 0.038 && (0.095, 0.149)& 0.054 \\
&(0.50,\,0.75)&	(0.191, 0.253)& 0.062 && (0.175, 0.266)& 0.091 \\
&(0.75,\,1.00)&	(0.567, 0.675)& 0.108 && (0.550, 0.687)& 0.137 \\

GMM&(0.00,\,0.25)&	(0.029, 0.044)& 0.015 && (0.027, 0.045)& 0.018 \\
&(0.25,\,0.50)&	(0.098, 0.133)& 0.035 && (0.092, 0.145)& 0.053 \\
&(0.50,\,0.75)&	(0.190, 0.246)& 0.056 && (0.171, 0.260)& 0.089 \\
&(0.75,\,1.00)&	(0.573, 0.682)& 0.109 && (0.562, 0.695)& 0.133 \\

BC$_{n}$&(0.00,\,0.25)&	(0.031, 0.042)& 0.011 && (0.028, 0.043)& 0.015 \\
&(0.25,\,0.50)&	(0.087, 0.126)& 0.039 && (0.100, 0.135)& 0.035 \\
&(0.50,\,0.75)&	(0.158, 0.237)& 0.079 && (0.181, 0.248)& 0.067 \\
&(0.75,\,1.00)&	(0.548, 0.681)& 0.133 && (0.577, 0.678)& 0.101 \\

BC$_{p}$&(0.00,\,0.25)&	(0.030, 0.045)& 0.015 && (0.030, 0.045)& 0.015 \\
&(0.25,\,0.50)&	(0.086, 0.128)& 0.042 && (0.094, 0.132)& 0.038\\
&(0.50,\,0.75)&	(0.159, 0.238)& 0.079 && (0.181, 0.247)& 0.066 \\
&(0.75,\,1.00)&	(0.555, 0.683)& 0.128 && (0.584, 0.688)& 0.104 \\

\hline\hline
\end{tabular}
\end{table}

\section*{Appendix}
\renewcommand{\theequation}{A.\arabic{equation}}
\renewcommand{\thesection}{A\arabic{section}}
\renewcommand{\thelemma}{A\arabic{lemma}}

\setcounter{equation}{0}

 \subsection*{{\bf Appendix A. Regularity Conditions for Proposition 2.1}}
\label{intro}

The asymptotic properties of the conventional survey weighted two-step empirical likelihood estimator presented in Zhao et al. (2020) are established under the following regularity conditions.

\begin{itemize}

\item [A1.]  (i) The finite population parameter vector  $\theta_{\N}  \in \Theta$  is the unique solution to  $U_{\N}(\theta_{\N},\varphi_{\N})$\\$=0$; (ii)  The parameter space $\Theta$ is a compact set in the $p$-dimensional Euclidean space;
(iii) The $\varphi_{\N}\in\Psi$ is a nuisance parameter   and
    $\Psi$  is a vector space of functions.

\item[A2.]
There exists a function $U(\theta,\varphi)$ such that as $N \rightarrow \infty$,
$\sup_{(\theta,\varphi) \in \Theta\times\Psi(\delta_{\N})}\|U_{\N}(\theta,\varphi) - U(\theta,\varphi)\|=o(1)$ for all sequences of positive numbers $\{\delta_{\N}\}$ with $\delta_{\N} = o(1)$. The limiting function $U(\theta,\varphi)$ also satisfies the following conditions:
\begin{itemize}
\item[(i)] There exists unique $\theta_0\in\Theta$ satisfying   \ $U(\theta_0,\varphi_0)=0$, where  $\varphi_0=\varphi_0(\cdot,\theta_0)\in\Psi$;

\item[(ii)] Uniformly for all  $\theta\in\Theta$, $U(\theta,\varphi)$ is continuous (with respect to the metric $\|\cdot\|_{\Psi}$) in $\varphi$ at $\varphi = \varphi_0$.
    \end{itemize}

\item[A3.]
(i)  $\max_{i\in \mS}\sup_{\theta\in\Theta,\varphi\in\Psi}\|g(Z_i,\theta,\varphi)\|=o_p(n_{\B}^{1/2})$;
(ii) For any sequence of positive numbers $\{\delta_{\N}\}$ with $\delta_{\N}=o(1)$,
$$\sup_{(\theta,\varphi),(\theta',\varphi') \in \Theta(\delta_{\N})\times\Psi(\delta_{\N})} \| U_{\N}(\theta,\varphi)  - U(\theta,\varphi)-[U_{\N}(\theta',\varphi')  - U(\theta',\varphi')] \| = o(N^{-1/2});$$
(iii) For all $\delta>0$ and some positive constant $c$,
$$
\sup_{(\theta,\varphi),(\theta',\varphi') \in \Theta(\delta)\times\Psi(\delta)}
{\rm Var}\Big\{[\hat{U}_{\N}(\theta,\varphi)-\hat{U}_{\N}(\theta',\varphi')]\mid \mathcal {F}_{\N}\Big\}\leq cn_{\B}^{-1}|\delta| \,,
$$
where $\hat{U}_{\N}(\theta,\varphi)=N^{-1}\sum_{i\in \mS}\pi_i^{-1}g(Z_i,\theta,\varphi)$, the commonly used survey weighted estimating equations.

\item [A4.]
For any  $(\theta,\varphi)\in\Theta(\delta)\times\Psi(\delta)$,
the  ordinary derivative $\Gamma_1(\theta,\varphi)$ in $\theta$ of  the limiting  functions $U(\theta,\varphi)$ exists and
  satisfies
$$\Gamma_1(\theta,\varphi)(\bar{\theta}-\theta)=\lim_{t\rightarrow 0}\frac{1}{t}[U(\theta+t(\bar{\theta}-\theta),\varphi(\cdot,\theta+t(\bar{\theta}-\theta)))
-U(\theta,\varphi(\cdot,\theta))]$$
for $\bar{\theta}\in\Theta$; the  derivative $\Gamma_1(\theta,\varphi)$  is continuous  at $\theta=\theta_{\N}$; and the matrix $\Gamma_1(\theta,\varphi)$ has
full column rank $p$.

\item [A5.]
For any $\theta\in \Theta(\delta)$,  the limiting function $U(\theta,\varphi)$ is pathwise differentiable at $\varphi\in \Psi(\delta)$ in the direction $[\bar{\varphi}-\varphi]$   in the sense that  the limit
$$D(\theta,\varphi)[\bar{\varphi}-\varphi]=\lim_{t\rightarrow 0}\frac{1}{t}[U(\theta,\varphi(\cdot,\theta)+t(\bar{\varphi}(\cdot,\theta)-\varphi(\cdot,\theta)))
-U(\theta,\varphi(\cdot,\theta))]$$ exists for $\{\varphi+t(\bar{\varphi}-\varphi):t\in[0,1]\}\subset \Psi$;
for all $\theta\in\Theta(\delta)$ and $(\theta', \varphi')\in\Theta(\delta)\times\Psi(\delta)$, the pathwise derivative  $D(\theta,\varphi')[\varphi-\varphi']$   exists in all directions $[\varphi-\varphi']\in\Psi$; and for all $(\theta, \varphi), (\theta', \varphi')\in\Theta(\delta_{\N})\times\Psi(\delta_{\N})$ with a positive sequence $\delta_{\N}=o(1)$: (i) $\|U(\theta,\varphi)-U(\theta,\varphi')-D(\theta,\varphi')[\varphi-\varphi']\|\leq c\|\varphi-\varphi'\|_{\Psi}^2$ for some constant $c\ge0$; (ii) $\|D(\theta,\varphi')[\varphi-\varphi']-D(\theta',\varphi')[\varphi-\varphi']\| = o(\delta_{\N})$.

\item [A6.] The  estimator $\hat{\varphi}$ satisfies the following conditions:
 (i) $\|\hat{\varphi}-\varphi_{\N}\|_{\Psi}=o_p(n_{\B}^{-1/4})$;  \ (ii)  $D(\theta_{\N},\varphi_{\N})[\hat{\varphi}-\varphi_{\N}]=N^{-1}\sum_{i\in \mS}\pi_i^{-1}\Xi(Z_i,\theta_{\N},\varphi_{\N})+o_p(n_{\B}^{-1/2})$, \ where  \ $\Xi(Z,\theta_{\N},\varphi_{\N})$ \ has finite fourth population moments (ie.,$N^{-1}\sum_{i=1}^N\|\Xi(Z_i,\theta_{\N},\varphi_{\N})\|^4<\infty$) and  $\sum_{i\in \mS}\pi_i^{-1}\Xi(Z_i,\theta_{\N},\varphi_{\N})$ is asymptotically normally distributed with mean zero and variance-covariance matrix at the order $O(n_{\B}^{-1}N^2)$.

\item [A7.] The sampling design along with the expected sample size $n_{\B}$ satisfies
(i) $n_{\B} = O(N^{\beta})$ for some $\beta$ such that $1/2 < \beta \leq 1$;
and (ii) $c_1 < \pi_i N n_{\B}^{-1} < c_2$, $i\in \mS$ for some positive constants $c_1$ and $c_2$.

 \item [A8.] Let  $\bar{\mathcal {Z}}_{\N} = N^{-1}\sum_{i=1}^N\mathcal {Z}_i$ and $\hat{\mathcal {Z}}_{\N} = N^{-1}\sum_{i\in \mS}\pi_i^{-1}\mathcal {Z}_i$. (i) For
    any vector $\mathcal {Z}$ satisfying
      $N^{-1}\sum_{i=1}^{N}\|\mathcal {Z}_i\|^{2+\sigma}<\infty$ (i.e., finite $2+\sigma$ population monents) with some small $\sigma>0$,
${\rm Var}(\hat{\mathcal {Z}}_{\N} \mid \mathcal {F}_{\N}) \leq c_0 n_{\B}^{-1}(N-1)^{-1}\sum_{i=1}^{N}(\mathcal {Z}_i - \bar{\mathcal {Z}}_{\N})(\mathcal {Z}_i - \bar{\mathcal {Z}}_{\N})^{\top}$ for some constant $c_0$;
(ii) For any $\mathcal {Z}$ with finite fourth population moment,
$\hat{\mathcal {Z}}_{\N}-\mathcal {Z}_{\N}$ is asymptotically normally distributed with mean zero and variance-covariance matrix at the order $O(n_{\B}^{-1})$.
\end{itemize}

Discussions and interpretations of these regularity conditions can be found in Zhao et al. (2020).

\subsection*{{\bf Appendix B. Technical Details and Proofs}}
\label{sec.proofs}
Let $f_{\N}=n_{\B}/N$ and
define the following alternative design-based GEL criterion function
\begin{eqnarray}\label{gel.fun.dual}
\hat{\mathcal{P}}_{\N}(\theta,\eta,\varphi)
=\dfrac{1}{n}\sum\limits_{i\in \mS}[\rho(\eta^{\top}\pi_i^{-1}f_{\N}\psi(Z_i,\theta,\varphi))-\rho_0].
\end{eqnarray}
As discussed in the Section 2.3 of the main paper,   the proposed  two-step GEL estimator $\hat{\theta}_{\GEL}$ can be equivalently defined as
\begin{equation}  \label{gele-new}
\hat{\theta}_{\GEL}=\arg\inf_{\theta\in\Theta}\sup_{\eta\in\hat{\Lambda}_{\N,\psi}(\theta,\hat{\varphi})}
\hat{\mathcal{P}}_{\N}(\theta,
\eta,\hat{\varphi}),
\end{equation}
where  $\hat{\Lambda}_{\N,\psi}(\theta,\varphi)=\{\eta:\eta^{\top}\pi_i^{-1}f_{\N}
\psi(Z_i,\theta,\varphi)\in\mathcal {V}, i\in\mS\}$. In addition, we define $\hat{\eta}_{\GEL}
=\arg\max_{\eta\in\hat{\Lambda}_{\N,\psi}(\hat\theta_{\GEL},\hat{\varphi})}
\hat{\mathcal{P}}_{\N}(\hat\theta_{\GEL},\eta,\hat{\varphi})$.

Recall that $\phi(Z,\theta,\varphi)=(\psi(Z,\theta,\varphi)^{\top},q(Z,\theta)^{\top})^{\top}$.
Define the following alternative design-based GEL criterion function
\begin{eqnarray}\label{rgel.fun.dual}
\hat{\mathcal{P}}_{\N}^{\R}(\theta,
\nu,\varphi)=\dfrac{1}{n}\sum_{i\in\mS}(\rho(\nu^{\top}\pi_i^{-1}f_{\N}\phi(Z_i,\theta,\varphi))-\rho_0).
\end{eqnarray}
The duality results in section 2.3 of the main paper shows that the  restricted two-step GEL estimator $\hat{\theta}_{\GEL}^{\R}$ can be also equivalently defined as
\begin{equation}  \label{rgele-new}
\hat{\theta}_{\GEL}^{\R}=\arg\inf_{\theta\in\Theta^{\R}}\sup_{\nu\in\hat{\Lambda}_{\phi,\N}(\theta,\hat{\varphi})}
\hat{\mathcal{P}}_{\N}^{\R}(\theta,
\nu,\hat{\varphi}),
\end{equation}
where
$\hat{\Lambda}_{\N,\phi}(\theta,\varphi)=\{\nu:\nu^{\top}\pi_i^{-1}f_{\N}
\phi(Z_i,\theta,\varphi)\in\mathcal {V}, i\in\mS\}$. Also define 
$$\hat{\nu}_{\GEL}^{\R}=\arg\max_{\nu\in\hat{\Lambda}_{\N,\phi}(\theta,\hat{\varphi})}
\hat{\mathcal{P}}_{\N}^{\R}(\hat{\theta}_{\GEL}^{\R},\nu,\hat{\varphi}).$$

To facilitates the usual asymptotic orders in the context of complex surveys,  we focus on GEL criteria and  GEL estimator   as in (\ref{gel.fun.dual})--(\ref{rgele-new})
to establish the large sample results of the paper. Let ``w.p.a.1'' denote ``with probability approaching 1''.
Let $C$ denote a generic positive constant which may vary depending on the context.


To facilitate the theoretical proofs, we provide four useful lemmas.

\begin{lemma}\label{lemma1}
Suppose  that conditions  A7 and B2(i) hold.  Then for any $\delta$ with $1/\alpha<\delta<1/2$ and $\hat{\Lambda}_{\N} = \{\eta: \|\eta\| \leq  n_{\B}^{-\delta}\}$,
$\sup_{\theta \in \Theta, \varphi\in\Psi, \eta \in \hat{\Lambda}_{\N}, i \in \mS}|\eta^{\top}
\pi_i^{-1}f_{\N}\psi(Z_i,\theta,\varphi)| =o_p(1)$ and w.p.a.1, $\hat{\Lambda}_{\N}\subseteq\hat{\Lambda}_{\N,\varphi}(\theta,\varphi)$ for all $(\theta,\varphi)\in\Theta\times \Psi$.
\end{lemma}
\proof
It follows from Condition A7 that $1/c_2<\pi_i^{-1}f_{\N}<1/c_1$ uniformly in $i$. Combining  this  with condition B2(i)   yields
$\max_{i\in \mS}\sup_{(\theta,\varphi)\in\Theta\times \Psi}\|\pi_i^{-1}f_{\N}\psi(Z_i,\theta,\varphi)\|=o_p(n_{\B}^{1/\alpha})$ for some $\alpha>2$.
Applying  the Cauchy-Schwarz inequality, we have that
\begin{eqnarray*}
\sup_{\theta \in \Theta, \varphi\in\Psi, \lambda \in \hat{\Lambda}_{\N}, i \in \mS}|\eta^{\top}
\pi_i^{-1}f_{\N}\psi(Z_i,\theta,\varphi)| &\leq& \|\eta\|\max_{i\in \mS}\sup_{\theta\in\Theta,\varphi\in\Psi}\|\pi_i^{-1}f_{\N}\psi(Z_i,\theta,\varphi)\|\\
&=&O_p(n_{\B}^{-\delta+1/\alpha})\stackrel{p}{\rightarrow}0,
\end{eqnarray*}
provided  $-\delta+1/\alpha<0$.
This further implies that  w.p.a.1, $\eta^{\top}\pi_i^{-1}f_{\N}\psi(Z_i,\theta,\varphi)\in \mathcal {V}$ for all $(\theta,\varphi)\in\Theta\times \Psi$ and $\|\eta\|\leq n_{\B}^{-\delta}$.
\hfill$\square$\\

\begin{lemma}\label{lemma2}
Suppose  that conditions  A7 and B2(i) hold, $\bar{\theta}\in\Theta$, $\bar{\theta}\stackrel{p}{\rightarrow} \theta_{\N}$ and $\|\hat{\mathbb{U}}_{\N}(\bar{\theta},\hat{\varphi})\| = O_p(n_{\B}^{-1/2})$. Then, w.p.a.1, $\bar\eta=\arg \sup_{\eta\in\hat{\Lambda}_{\N,\psi}(\bar\theta,\hat{\varphi})}\hat{\mathcal{P}}_{\N}(\bar\theta,\eta,\hat{\varphi})$ exists, $\bar\eta=O_p(n_{\B}^{-1/2})$, and $\sup_{\eta\in\hat{\Lambda}_{\N,\psi}(\bar\theta,\hat{\varphi})}\hat{\mathcal{P}}_{\N}(\bar\theta,\eta,\hat{\varphi})\leq O_p(n_{\B}^{-1})$.
\end{lemma}
\proof
Let $\hat{W}_{\N}(\theta,\varphi)=n_{\B}^{-1}\sum_{i\in \mS}\pi_i^{-2}f_{\N}^2\psi(Z_i,\theta,\varphi)^{\otimes2}$.
 It is easy to show that $\|\hat{W}_{\N}(\bar{\theta},\hat{\varphi}) - W_2\|=o_p(1)$,
 where $W_2=n_{\B}N^{-2}\sum_{i=1}^{N}\pi_i^{-1}\psi(Z_i,\theta_{\N},\varphi_{\N})^{\otimes2}$.
 Recall that $\hat{\Lambda}_{\N} = \{\eta: \|\eta\| \leq  n_{\B}^{-\delta}\}$, where $\delta$ is as defined in Lemma \ref{lemma1}.
It follows from the nonsingularity of  the matrix $W_2$ that  the smallest eigenvalue of $W_{\N}(\bar{\theta},\hat{\varphi})$ is bounded away from zero   w.p.a.1. This, coupled with the twice continuous differentiability of $\rho(v)$  in a neighborhood of zero and the results of Lemma \ref{lemma1}, shows that
$\hat{\mathcal{P}}_{\N}(\bar{\theta},\eta,\hat{\varphi})$ admits a second order Taylor expansion on $\hat{\Lambda}_{\N}$   with nonsingular second derivative matrix.
Then,  $\tilde\eta=\arg \sup_{\eta\in\hat{\Lambda}_{\N}}\hat{\mathcal{P}}_{\N}(\bar\theta,\eta,\hat{\varphi})$ exists w.p.a.1. Now using a second order Taylor's expansion for $\hat{\mathcal{P}}_{\N}(\bar\theta,\tilde{\eta},\hat{\varphi})$ around $\tilde{\eta}=0$, we obtain
\[
\begin{array}{llll}
\hat{\mathcal{P}}_{\N}(\bar\theta,\tilde{\eta},\hat{\varphi})= -\tilde{\eta}^{\top}\hat{\mathbb{U}}_{\N}(\bar{\theta},\hat{\varphi})
 + \dfrac{1}{2}\tilde{\eta}^{\top}\left[\dfrac{1}{n_{\B}}\sum\limits_{i\in \mS}\rho_2\Big(\dot{\eta}^{\top}f_{\N}\psi(Z_i,\bar{\theta},\hat{\varphi})\Big)
\pi_i^{-2}f_{\N}^2\psi(Z_i,\bar{\theta},\hat{\varphi})^{\otimes 2}\right]\tilde{\eta},
\end{array}
\]
where $\dot{\eta}$ is on the line segment between $\tilde{\eta}$ and $0$.
It follows from Lemma \ref{lemma1} and the fact $\rho_2(0)=-1$ that $\max_{i\in \mS}\rho_2(\dot{\eta}^{\top}f_{\N}\psi(Z_i,\bar{\theta},\hat{\varphi}))<-1/2$ w.p.a.1.
Combining this with the above expansion, we obtain that
\[\begin{array}{lllll}
\hat{\mathcal{P}}_{\N}(\bar\theta,\tilde{\eta},\hat{\varphi}) \leq -\tilde{\eta}^{\top}\hat{\mathbb{U}}_{\N}(\bar{\theta},\hat{\varphi}) - \dfrac{1}{4}\tilde{\eta}^{\top}\hat{W}_{\N}(\bar{\theta},\hat{\varphi})\tilde{\eta}\leq\|\tilde{\eta}\|
\|\hat{\mathbb{U}}_{\N}(\bar{\theta},\hat{\varphi})\|-C\|\tilde{\eta}\|^2.
\end{array}\]
By the definition of $\tilde{\eta}$,  we have that $0=\hat{\mathcal{P}}_{\N}(\bar\theta,0,\hat{\varphi})
\leq\hat{\mathcal{P}}_{\N}(\bar\theta,\tilde{\eta},\hat{\varphi})$. This leads to $0\leq \|\tilde{\eta}\|
\|\hat{\mathbb{U}}_{\N}(\bar{\theta},\hat{\varphi})\|-C\|\tilde{\eta}\|^2$, and thus $C\|\tilde{\eta}\|\leq
\|\hat{\mathbb{U}}_{\N}(\bar{\theta},\hat{\varphi})\|$,  w.p.a.1. It follows from the assumption $\hat{\mathbb{U}}_{\N}(\bar{\theta},\hat{\varphi})=O_p(n_{\B}^{-1/2})$ that $\|\tilde{\eta}\|=O_p(n_{\B}^{-1/2})=o_p(n_{\B}^{-\delta})$. Therefore,  w.p.a.1. $\tilde{\eta}\in {\rm int}(\hat{\Lambda}_{\N})$ and hence the maximum $\tilde{\eta}$ also satisfies the first order conditions $\partial \hat{\mathcal{P}}_{\N}(\bar\theta,\tilde{\eta},\hat{\varphi})/\partial\eta=0$. Using the results of Lemma \ref{lemma1}, it can be further conclude that $\tilde{\eta}\in \hat{\Lambda}_{\N,\psi}(\bar{\theta},\hat{\varphi})$. This, together with the concavity of $\hat{\mathcal{P}}_{\N}(\bar\theta,\eta,\hat{\varphi})$ and convexity of $\hat{\Lambda}_{\N,\psi}(\bar{\theta},\hat{\varphi})$, implies that $\hat{\mathcal{P}}_{\N}(\bar\theta,\tilde{\eta},\hat{\varphi})
=\sup_{\eta\in\hat{\Lambda}_{\N,\psi}(\bar{\theta},\hat{\varphi})}\hat{\mathcal{P}}_{\N}(\bar\theta,\eta,\hat{\varphi})$ and hence $\bar{\eta}=\tilde{\eta}$.
This yields the conclusions that  w.p.a.1, $\bar\eta=\arg \sup_{\eta\in\hat{\Lambda}_{\N,\psi}(\bar\theta,\hat{\varphi})}
\hat{\mathcal{P}}_{\N}(\bar\theta,\eta,\hat{\varphi})$ exists and $\bar\eta=O_p(n_{\B}^{-1/2})$. Since $\|\hat{\mathbb{U}}_{\N}(\bar{\theta},\hat{\varphi})\|=O_p(n_{\B}^{-1/2})$ and $\|\tilde\eta\|=O_p(n_{\B}^{-1/2})$, together with the above inequality,  it is easy to show that
$\hat{\mathcal{P}}_{\N}(\bar\theta,\bar{\eta},\hat{\varphi}) \leq\|\bar{\eta}\|
\|\hat{\mathbb{U}}_{\N}(\bar{\theta},\hat{\varphi})\|-C\|\bar{\eta}\|^2=O_p(n_{\B}^{-1})$.
\hfill$\square$\\

\begin{lemma}\label{lemma3}
Suppose  that conditions  A7 and B2(i) hold.  Then   $\|\hat{\mathbb{U}}_{\N}(\hat{\theta}_{\GEL},\hat{\varphi})\| = O_p(n_{\B}^{-1/2})$.
\end{lemma}
\proof
 Let ${\tilde \eta} = -n_{\B}^{-\delta}\hat{\mathbb{U}}_{\N}(\hat{\theta}_{\GEL},\hat{\varphi})
 /\|\hat{\mathbb{U}}_{\N}(\hat{\theta}_{\GEL},\hat{\varphi})\|$, where $\delta$ is as defined in Lemma \ref{lemma1}. It then follows from  Lemma \ref{lemma1} that   $\max_{i \in \mS}|{\tilde \eta}^{\top}\pi_i^{-1}f_{\N}\psi(Z_i,\hat{\theta}_{\GEL},\hat{\varphi})| \xrightarrow{p} 0$, and thus ${\tilde \eta} \in {\hat \Lambda}_{\N,\psi}(\hat{\theta}_{\GEL},\hat{\varphi})$. Then, we have  that for any $\dot{\eta}$ on the line segment between $\tilde{\eta}$ and $0$, w.p.a.1 $\rho_2(\dot{\eta}^{\top}\psi(Z_i,\hat{\theta}_{\GEL},\hat{\varphi}))\geq -C$ uniformly in $i$.
By condition B6, 
$$\dfrac{n_{\B}}{N^2}\sum_{i=1}^N
\pi_i^{-1}\sup_{\theta\in\Theta,\varphi\in\Psi}\|\psi(Z_i,\theta,\varphi)\|^2=O(1).$$ Moreover, we can show the following uniform convergence property
$$\dfrac{1}{n_{\B}}\sum\limits_{i\in \mS}\pi_i^{-2}f_{\N}^2\sup_{\theta\in\Theta,\varphi\in\Psi}\|\psi(Z_i,\theta,\varphi)\|^2=
\dfrac{n_{\B}}{N^2}\sum\limits_{i=1}^N
\pi_i^{-1}\sup_{\theta\in\Theta,\varphi\in\Psi}\|\psi(Z_i,\theta,\varphi)\|^2+o_p(1).$$
By the Cauchy-Schwarz inequality, we have
\[\begin{array}{lllll}
\dfrac{1}{n_{\B}}\sum\limits_{i\in \mS}\pi_i^{-2}f_{\N}^2\psi(Z_i,\hat{\theta}_{\GEL},\hat{\varphi})^{\otimes 2}
\leq \dfrac{1}{n_{\B}}\sum\limits_{i\in \mS}\pi_i^{-2}f_{\N}^2\sup_{\theta\in\Theta,\varphi\in\Psi}\|\psi(Z_i,\theta,\varphi)\|^2I_q
\xrightarrow{p} CI_q,
 \end{array}\] where $I_q$ is the $q\times q$ identity matrix.
Using a  second order Taylor expansion,
\[
\begin{array}{llll}
\hat{\mathcal{P}}_{\N}(\hat{\theta}_{\GEL},\tilde{\eta},\hat{\varphi})&=& -\tilde{\eta}^{\top}\hat{\mathbb{U}}_{\N}(\hat{\theta}_{\GEL},\hat{\varphi})\\
 &+& \dfrac{1}{2}\tilde{\eta}^{\top}\left[\dfrac{1}{n_{\B}}\sum\limits_{i\in \mS}\rho_2\Big(\dot{\eta}^{\top}f_{\N}\psi(Z_i,\hat{\theta}_{\GEL},\hat{\varphi})\Big)
\pi_i^{-2}f_{\N}^2\psi(Z_i,\hat{\theta}_{\GEL},\hat{\varphi})^{\otimes 2}\right]\tilde{\eta}\\
&\geq&n_{\B}^{-\delta}\|\psi(Z_i,\hat{\theta}_{\GEL},\hat{\varphi})\|-\dfrac{C}{2}
\tilde{\eta}^{\top}\left[\dfrac{1}{n_{\B}}\sum\limits_{i\in \mS}
\pi_i^{-2}f_{\N}^2\psi(Z_i,\hat{\theta}_{\GEL},\hat{\varphi})^{\otimes 2}\right]\tilde{\eta}\\
&\geq& n_{\B}^{-\delta}\|\psi(Z_i,\hat{\theta}_{\GEL},\hat{\varphi})\|-Cn_{\B}^{-2\delta}.
\end{array}
\]
This, combined with Lemma \ref{lemma2} and the fact the  $(\hat{\theta}_{\GEL},\hat{\eta}_{\GEL})$ is a saddle point,   shows that
\begin{eqnarray*}\label{ineq1}
\begin{split}
n_{\B}^{-\delta} \|\hat{\mathbb{U}}_{\N}(\hat{\theta}_{\GEL},\hat{\varphi})\|-Cn_{\B}^{-2\delta}\leq \hat{\mathcal{P}}_{\N}(\hat{\theta}_{\GEL},\tilde{\eta},\hat{\varphi}) \leq \hat{\mathcal{P}}_{\N}(\hat{\theta}_{\GEL},\hat{\eta}_{\GEL},\hat{\varphi})\\
\leq\sup_{\eta\in\hat{\Lambda}_{\N,\psi}(\theta_{\N},\hat{\varphi})}
\hat{\mathcal{P}}_{\N}(\theta_{\N},\eta,\hat{\varphi})\leq O_p(n_{\B}^{-1}).
\end{split}
\end{eqnarray*}
Solving above equation for $\|\hat{\mathbb{U}}_{\N}(\hat{\theta}_{\GEL},\hat{\varphi})\|$ then gives
$\|\hat{\mathbb{U}}_{\N}(\hat{\theta}_{\GEL},\hat{\varphi})\|\leq O_p(n_{\B}^{\delta-1})+ C n_{\B}^{-\delta}\leq  O_p(n_{\B}^{-\delta}).$  Now we  consider ${\bar \eta} = \varepsilon_n\hat{\mathbb{U}}_{\N}(\hat{\theta}_{\GEL},\hat{\varphi})$ with any $\varepsilon_n \rightarrow 0$. Obviously, ${\bar \eta} = o_p(n_{\B}^{-\delta})$, and thus $\bar{\eta}\in\hat{\Lambda}_{\N}$ w.p.a.1.  Similarly, we have that
 \[\begin{array}{lllll}
 -\bar{\eta}^{\top}\hat{\mathbb{U}}_{\N}(\hat{\theta}_{\GEL},\hat{\varphi})
 -C\|\bar{\eta}\|^2=
 \varepsilon_n \|\hat{\mathbb{U}}_{\N}(\hat{\theta}_{\GEL},\hat{\varphi})\|^2-C\varepsilon_{n}^2
 \|\hat{\mathbb{U}}_{\N}(\hat{\theta}_{\GEL},\hat{\varphi})\|^2\leq O_p(n_{\B}^{-1}).
\end{array}\]
It can be shown that  $\varepsilon_n\|\hat{\mathbb{U}}_{\N}(\hat{\theta}_{\GEL},\hat{\varphi})\|^2 = O_p(n_{\B}^{-1})$ by noting that
$1 - C\varepsilon_n>0$  for  $n$ large enough. Then we can show  $\|\hat{\mathbb{U}}_{\N}(\hat{\theta}_{\GEL},\hat{\varphi})\| = O_p(n_{\B}^{-1/2})$.
\hfill$\square$\\

\begin{lemma}\label{lemma4}
Suppose
that  conditions  A1, A6--A8, B1 and B3--B4  hold.  Then, uniformly in  $\theta$ and $\eta$, for $\theta-\theta_{\N}=O_p(n_{\B}^{-1/2})$ and $\eta=O_p(n_{\B}^{-1/2})$, we have
\begin{eqnarray*}
|\hat{\mathcal{P}}_{\N}(\theta,\eta,\hat{\varphi})-\mathcal {L}_{\N}(\theta,\eta)|=o_p(n_{\B}^{-1}) \,,
\end{eqnarray*}
where $\mathcal {L}_{\N}(\theta,\eta)=[-\hat{\mathbb{U}}_{\N}(\theta_{\N},\varphi_{\N})-\Gamma_2 (\theta-\theta_{\N})]^{\top}\eta - \frac{1}{2}\eta^{\top}W_2\eta$.
\end{lemma}
\proof
Using a second-order Taylor series expansion for  $\hat{\mathcal{P}}_{\N}(\theta,\eta,\hat{\varphi})$ around $\eta = 0$, we obtain that, for $\theta-\theta_{\N}=O_p(n_{\B}^{-1/2})$ and $\eta=O_p(n_{\B}^{-1/2})$,
\[
\begin{array}{llll}
\hat{\mathcal{P}}_{\N}(\theta,\eta,\hat{\varphi})= -\eta^{\top}\hat{\mathbb{U}}_{\N}(\theta,\hat{\varphi})
+ \dfrac{1}{2}\eta^{\top}\left[\dfrac{1}{n_{\B}}\sum\limits_{i\in \mS}\rho_2\Big(\dot{\eta}^{\top}f_{\N}\psi(Z_i,\theta,\hat{\varphi})\Big)
\pi_i^{-2}f_{\N}^2\psi(Z_i,\theta,\hat{\varphi})^{\otimes 2}\right]\eta,
\end{array}
\]
where $\dot{\eta}$ is on the line segment between $\eta$ and $0$. By the triangle inequality,
\[
\begin{array}{llllll}
 |\hat{\mathcal{P}}_{\N}(\theta,\eta,\hat{\varphi})-\mathcal {L}_{\N}(\theta,\eta)|\\
 ~~~~\leq  |-[\hat{\mathbb{U}}_{\N}(\theta,\hat{\varphi})-\hat{\mathbb{U}}_{\N}(\theta_{\N},\varphi_{\N})
-\Gamma_2(\theta_{\N},\varphi_{\N}) (\theta-\theta_{\N})]^{\top}\eta| \\
~~~~~~~~~~ \; +\dfrac{1}{2}\bigg|\eta^{\top}\bigg(\dfrac{1}{n_{\B}}\sum\limits_{i\in \mS}\rho_2(\dot{\eta}^{\top}f_{\N}\psi(Z_i,\theta,\hat{\varphi}))
\pi_i^{-2}f_{\N}^2\psi(Z_i,\theta,\hat{\varphi})^{\otimes 2} + W_2\bigg)\eta\bigg| \\
 ~~~~=:R_{\N1}+R_{\N2}\,.
\end{array}
\]
For $R_{\N1}$, we have that $R_{\N1}\leq \|\hat{\mathbb{U}}_{\N}(\theta,\hat{\varphi})-\hat{\mathbb{U}}_{\N}(\theta_{\N},\varphi_{\N})
-\Gamma_2 (\theta-\theta_{\N})\|\|\eta\|$. Note that
\[\begin{array}{lllll}
\|\hat{\mathbb{U}}_{\N}(\theta,\hat{\varphi})-\hat{\mathbb{U}}_{\N}(\theta_{\N},\varphi_{\N})
-\Gamma_2 (\theta-\theta_{\N})\|\\
~~~\leq\|[\hat{\mathbb{U}}_{\N}(\theta,\hat{\varphi})-\mathbb{U}(\theta,\hat{\varphi})]  - [\hat{\mathbb{U}}_{\N}(\theta_{\N},\varphi_{\N}) -\mathbb{U}(\theta_{\N},\varphi_{\N})]\|
\\
~~~~~~~~~+\|\mathbb{U}(\theta,\varphi_{\N})
-\mathbb{U}(\theta_{\N},\varphi_{\N})-\Gamma_2 (\theta-\theta_{\N})\|+
\|\mathbb{U}(\theta,\hat{\varphi})
-\mathbb{U}(\theta,\varphi_{\N})\|.\\
\end{array}\]
The first term on the right of  above inequality can be rewritten as
$\|[\hat{\mathbb{U}}_{\N}(\theta,\hat{\varphi})-\mathbb{U}(\theta,\hat{\varphi})]  - [\hat{\mathbb{U}}_{\N}(\theta_{\N},\varphi_{\N}) -\mathbb{U}(\theta_{\N},\varphi_{\N})]\|=\|\mathcal {J}_{1}(\theta,\hat{\varphi},\theta_{\N},\varphi_{\N})+\mathcal {J}_{2}(\theta,\hat{\varphi},\theta_{\N},\varphi_{\N})\|$,
where
\begin{eqnarray*}
\mathcal {J}_{1}(\theta,\hat{\varphi},\theta_{\N},\varphi_{\N})&=&\hat{\mathbb{U}}_{\N}(\theta,\hat{\varphi})  - \hat{\mathbb{U}}_{\N}(\theta_{\N},\varphi_{\N}) -\mathbb{U}_{\N}(\theta,\hat{\varphi}) +\mathbb{U}_{\N}(\theta_{\N},\varphi_{\N}),\\
\mathcal {J}_{2}(\theta,\hat{\varphi},\theta_{\N},\varphi_{\N})&=& \mathbb{U}_{\N}(\theta,\hat{\varphi}) -\mathbb{U}_{\N}(\theta_{\N},\varphi_{\N})-\mathbb{U}(\theta,\hat{\varphi})  +\mathbb{U}(\theta_{\N},\varphi_{\N}) \,.
\end{eqnarray*}
Using  Condition B3, we can show that both $\{\hat{\mathbb{U}}_{\N}(\theta,\varphi)-\mathbb{U}_{\N}(\theta,\varphi), N=1,2, \cdots\}$ and $\{\mathbb{U}_{\N}(\theta,\varphi)-\mathbb{U}(\theta,\varphi), N=1,2, \cdots\}$ are    stochastically  equicontinuous.  Therefore, we
obtain that $\|\mathcal {J}_{j}(\theta,\hat{\varphi},\theta_{\N},\varphi_{\N})\|=o(n_{\B}^{-1/2})$, $j=1,2$,
uniformly in $\theta$
for $\theta-\theta_{\N}=O_p(n_{\B}^{-1/2})$. Then, by triangle inequality,
$\|[\hat{\mathbb{U}}_{\N}(\theta,\hat{\varphi})-\mathbb{U}(\theta,\hat{\varphi})]  - [\hat{\mathbb{U}}_{\N}(\theta_{\N},\varphi_{\N}) -\mathbb{U}(\theta_{\N},\varphi_{\N})]\|=o_p(n_{\B}^{-1/2})$.
With assumption $\theta-\theta_{\N}=O_p(n_{\B}^{-1/2})$ and the differentiability of $\mathbb{U}(\theta,\varphi)$ with respect to $\theta$, we have
$\|\mathbb{U}(\theta,\varphi_{\N})
-\mathbb{U}(\theta_{\N},\varphi_{\N})-\Gamma_2 (\theta-\theta_{\N})\|=o_p(n_{\B}^{-1/2})$.
It follows from conditions A6 and B4 that $\|\mathbb{U}(\theta,\hat{\varphi})
-\mathbb{U}(\theta,\varphi_{\N})\|\leq C\|\hat{\varphi}-\varphi_{\N}\|_{\Psi}^2 =o_p(n_{\B}^{-1/2})$.
Combining above arguments, we obtain that $\|\hat{\mathbb{U}}_{\N}(\theta,\hat{\varphi})-\hat{\mathbb{U}}_{\N}(\theta_{\N},\varphi_{\N})
-\Gamma_2 (\theta-\theta_{\N})\|=o_p(n_{\B}^{-1/2})$.
This, together with assumption $\eta=O_p(n_{\B}^{-1/2})$,  immediately implies that $R_{\N1}=o_p(n_{\B}^{-1}).$

Next we consider $R_{\N2}$.
It is easy to show that uniformly in $\theta$
for $\theta-\theta_{\N}=O_p(n_{\B}^{-1/2})$,
$$\bigg\|\dfrac{1}{n_{\B}}\sum_{i\in \mS}\pi_i^{-2}f_{\N}^2\psi(Z_i,\theta,\hat{\varphi})^{\otimes2} - W_2\bigg\|=o_p(1).$$
This combined with Lemma \ref{lemma1} and the assumption   $\eta=O_p(n_{\B}^{-1/2})$, shows that
\[\begin{array}{lllll}
\bigg|\eta^{\top}\bigg(\dfrac{1}{n_{\B}}\sum\limits_{i\in \mS}\rho_2(\dot{\eta}^{\top}f_{\N}\psi(Z_i,\theta,\hat{\varphi}))
\pi_i^{-2}f_{\N}^2\psi(Z_i,\theta,\hat{\varphi})^{\otimes 2} + W_2\bigg)\eta\bigg| \\
~~~~~\leq\|\eta\|^2\bigg\|\dfrac{1}{n_{\B}}\sum\limits_{i\in \mS}\rho_2(\dot{\eta}^{\top}f_{\N}\psi(Z_i,\theta,\hat{\varphi}))
\pi_i^{-2}f_{\N}^2\psi(Z_i,\theta,\hat{\varphi})^{\otimes 2}+ W_2\bigg\|\\
~~~~~=O_p(n_{\B}^{-1})o_p(1)=o_p(n_{\B}^{-1}).
\end{array}\]
Then we have $|\hat{\mathcal{P}}_{\N}(\theta,\eta,\hat{\varphi})-\mathcal {L}_{\N}(\theta,\eta)|=o_p(n_{\B}^{-1})$, uniformly  for $\theta-\theta_{\N}=O_p(n_{\B}^{-1/2})$ and $\eta=O_p(n_{\B}^{-1/2})$.
\hfill$\square$\\


\noindent
\textbf{Proof of  Theorem 3.1:}
Note that $\Theta=\{\theta:\|\theta-\theta_{\N}\|\geq \epsilon\}\cup\{\theta:\|\theta-\theta_{\N}\|< \epsilon\}$ for any $\epsilon>0$. Obviously,  $\{\theta:\|\theta-\theta_{\N}\|< \epsilon\}$ is also a compact subset of $\Theta$.  Thus, there exists $\theta_1\in\{\theta:\|\theta-\theta_{\N}\|\geq \epsilon\}$ such that
\[
\inf_{\theta:\|\theta-\theta_{\N}\|\geq \epsilon}\|\mathbb{U}_{\N}(\theta,\varphi_{\N})\|=\|\mathbb{U}_{\N}(\theta_1,\varphi_{\N})\|.
\]
By the identification of $\theta_{\N}$, we conclude that  $\|\mathbb{U}_{\N}(\theta_1,\varphi_{\N})\|>0$ for any $\theta_1\neq \theta_{\N}$. This combined with the above equality, implies that $\inf_{\theta:\|\theta-\theta_{\N}\|\geq \epsilon}\|\mathbb{U}_{\N}(\theta,\varphi_{\N})\|>0$ for all $\epsilon>0$. It follows that for every $\epsilon>0$, there exists a number $c(\epsilon)>0$ such that $\|\mathbb{U}_{\N}(\theta,\varphi_{\N})\|\geq c(\epsilon)>0$ for every $\theta$ with $\|\theta-\theta_{\N}\|>\epsilon$.
Thus, the event $\{\|\theta-\theta_{\N}\|>\epsilon\}$ is contained in the event $\{\|\mathbb{U}_{\N}(\theta,\varphi_{\N})\|\geq c(\epsilon)>0\}$, and ${\rm Pr}(\|\hat{\theta}_{\GEL}-\theta_{\N}\|>\epsilon\mid \mathcal {F}_{\N})\leq {\rm Pr}(\|\mathbb{U}_{\N}(\hat{\theta}_{\GEL},\varphi_{\N})\|\geq c(\epsilon)\mid \mathcal {F}_{\N})$ for all $\epsilon>0$. Therefore, to establish the consistency for the proposed GEL estimators $\hat{\theta}_{\GEL}$, it suffices to show that $\|\mathbb{U}_{\N}(\hat{\theta}_{\GEL},\varphi_{\N})\|=o_p(1)$.
Using the triangle inequality, we obtain that
\[\begin{array}{lllll}
\|\mathbb{U}_{\N}(\hat{\theta}_{\GEL},\varphi_{\N})\|&\leq&
\|\mathbb{U}_{\N}(\hat{\theta}_{\GEL},\varphi_{\N})-\mathbb{U}_{\N}(\hat{\theta}_{\GEL},\hat{\varphi})\|\\
&&+\|\mathbb{U}_{\N}(\hat{\theta}_{\GEL},\hat{\varphi})
-\hat{\mathbb{U}}_{\N}(\hat{\theta}_{\GEL},\hat{\varphi})\|
+\|\hat{\mathbb{U}}_{\N}(\hat{\theta}_{\GEL},\hat{\varphi})\|.
\end{array}\]
For the first term on the right of  above inequality, we have the following inequality
\begin{eqnarray*}
\|\mathbb{U}_{\N}(\hat{\theta}_{\GEL},\varphi_{\N})-\mathbb{U}_{\N}(\hat{\theta}_{\GEL},\hat{\varphi})\|
\leq
\|\mathbb{U}_{\N}(\hat{\theta}_{\GEL},\varphi_{\N})-\mathbb{U}(\hat{\theta}_{\GEL},\varphi_{\N})\|\\
+\|\mathbb{U}(\hat{\theta}_{\GEL},\hat{\varphi})-\mathbb{U}_{\N}(\hat{\theta}_{\GEL},\hat{\varphi})\|
+\|\mathbb{U}(\hat{\theta}_{\GEL},\varphi_{\N})-\mathbb{U}(\hat{\theta}_{\GEL},\hat{\varphi})\|.
\end{eqnarray*}
By assumption B3(i), it can be shown that $\|\mathbb{U}_{\N}(\hat{\theta}_{\GEL},\varphi_{\N})-\mathbb{U}(\hat{\theta}_{\GEL},\varphi_{\N})\|=o_p(1)$ and $\|\mathbb{U}(\hat{\theta}_{\GEL},\hat{\varphi})-\mathbb{U}_{\N}(\hat{\theta}_{\GEL},\hat{\varphi})\|=o_p(1)$. Using assumption  B2(ii), we can establish the following uniform convergence results
\begin{eqnarray}
\label{appenda1}
\begin{split}
\sup_{\theta\in\Theta,\varphi\in\Psi(\delta_{\N})}\|\hat{\mathbb{U}}_{\N}(\theta,\varphi)
-\mathbb{U}_{\N}(\theta,\varphi)\|=o_p(1)\,\,\,\mbox{and}\\
\sup_{\theta\in\Theta,\varphi\in\Psi(\delta_{\N})}\|\mathbb{U}_{\N}(\theta,\varphi)
-\mathbb{U}(\theta,\varphi)\|=o_p(1),\\
\end{split}
\end{eqnarray}
for all positive sequences $\delta_{\N}=o(1)$. Combined this with assumption A6, implies that $\|\mathbb{U}_{\N}(\hat{\theta}_{\GEL},\hat{\varphi})
-\hat{\mathbb{U}}_{\N}(\hat{\theta}_{\GEL},\hat{\varphi})\|=o_p(1)$.
It follows that by assumption B1 and $\hat{\varphi}=\varphi_{\N}+o_p(1)$, $\|\mathbb{U}(\hat{\theta}_{\GEL},\varphi_{\N})-\mathbb{U}(\hat{\theta}_{\GEL},\hat{\varphi})\|=o_p(1)$. Then $\|\mathbb{U}_{\N}(\hat{\theta}_{\GEL},\varphi_{\N})-\mathbb{U}_{\N}(\hat{\theta}_{\GEL},\hat{\varphi})\|=o_p(1)$.
It follows from Lemma \ref{lemma3} that  $\|\hat{\mathbb{U}}_{\N}(\hat{\theta}_{\GEL},\hat{\varphi})\|=o_p(1)$.  Combining above arguments, we have that $\|\mathbb{U}_{\N}(\hat{\theta}_{\GEL},\varphi_{\N})\|=o_p(1)$, which is equivalent to ${\rm Pr}(\|\mathbb{U}_{\N}(\hat{\theta}_{\GEL},\varphi_{\N})\|\geq c(\epsilon)\mid \mathcal {F}_{\N})\rightarrow 0$ for all $c(\epsilon)>0$. Now the design-based consistency of $\hat{\theta}_{\GEL}$ follows.
\hfill$\square$\\

\noindent
\textbf{Proof of  Theorem 3.2:}
To establish the asymptotic normality of $\hat{\theta}_{\GEL}$, we first  show that $\|\hat{\theta}_{\GEL}-\theta_{\N}\|=O(n_{\B}^{-1/2})$.
By assumption B1, there exists a constant $C$ such that $\|\hat{\theta}_{\GEL}-\theta_{\N}\|\leq C \|\mathbb{U}(\hat{\theta}_{\GEL},\varphi_{\N})-\mathbb{U}(\theta_{\N},\varphi_{\N})\|$ w.p.a.1.
As discussed in the proof of Lemma \ref{lemma4},  both $\{\hat{\mathbb{U}}_{\N}(\theta,\varphi)-\mathbb{U}_{\N}(\theta,\varphi), N=1,2, \cdots\}$ and $\{\mathbb{U}_{\N}(\theta,\varphi)-\mathbb{U}(\theta,\varphi), N=1,2, \cdots\}$ are    stochastically  equicontinuous. Then we have
$\|[\mathbb{U}(\hat{\theta}_{\GEL},\hat{\varphi})-\hat{\mathbb{U}}_{\N}(\hat{\theta}_{\GEL},\hat{\varphi})]  - [\mathbb{U}(\theta_{\N},\varphi_{\N})-\hat{\mathbb{U}}_{\N}(\theta_{\N},\varphi_{\N})] \|=o_p(n_{\B}^{-1/2})$. By assumption B4, $\|\mathbb{U}(\hat{\theta}_{\GEL},\varphi_{\N})
-\mathbb{U}(\hat{\theta}_{\GEL},\hat{\varphi})\|=o_p(n_{\B}^{-1/2})$. By Lemma \ref{lemma3}, $\|\hat{\mathbb{U}}_{\N}(\hat{\theta}_{\GEL},\hat{\varphi})\|=O_p(n_{\B}^{-1/2})$.
By condition A8, $\|\hat{\mathbb{U}}_{\N}(\theta_{\N},\varphi_{\N})\|=O_p(n_{\B}^{-1/2})$.
Combining these facts with  the triangle inequality, implies that
\begin{eqnarray*}
\|\mathbb{U}(\hat{\theta}_{\GEL},\varphi_{\N})-\mathbb{U}(\theta_{\N},\varphi_{\N})\|
\leq
\|[\mathbb{U}(\hat{\theta}_{\GEL},\hat{\varphi})-\hat{\mathbb{U}}_{\N}(\hat{\theta}_{\GEL},\hat{\varphi})]  - [\mathbb{U}(\theta_{\N},\varphi_{\N})-\hat{\mathbb{U}}_{\N}(\theta_{\N},\varphi_{\N})] \|\\
+\|\mathbb{U}(\hat{\theta}_{\GEL},\varphi_{\N})
-\mathbb{U}(\hat{\theta}_{\GEL},\hat{\varphi})\|+\|\hat{\mathbb{U}}_{\N}(\hat{\theta}_{\GEL},\hat{\varphi})\|
+\|\hat{\mathbb{U}}_{\N}(\theta_{\N},\varphi_{\N})\|
=O_p(n_{\B}^{-1/2}).
\end{eqnarray*}
Consequently, $\|\hat{\theta}_{\GEL}-\theta_{\N}\|\leq C \|\mathbb{U}(\hat{\theta}_{\GEL},\varphi_{\N})-\mathbb{U}(\theta_{\N},\varphi_{\N})\|\leq O_p(n_{\B}^{-1/2})$.

Recall that $\mathcal {L}_{\N}(\theta,\eta)=[-\hat{\mathbb{U}}_{\N}(\theta_{\N},\varphi_{\N})-\Gamma_2 (\theta-\theta_{\N})]^{\top}\eta - \frac{1}{2}\eta^{\top}W_2\eta$. By Lemma \ref{lemma4},
$
|\hat{\mathcal{P}}_{\N}(\theta,\eta,\hat{\varphi})-\mathcal {L}_{\N}(\theta,\eta)|=o_p(n_{\B}^{-1})
$
uniformly in  $\theta$ and $\eta$, for $\theta-\theta_{\N}=O_p(n_{\B}^{-1/2})$ and $\eta=O_p(n_{\B}^{-1/2})$. We now consider the optimization problem $\inf_{\theta\in\Theta}\sup_{\eta\in\mathsf{R}^r}\mathcal {L}_{\N}(\theta,\eta)$.
It is clear that $\mathcal {L}_{\N}(\theta,\eta)$ is concave in $\eta$. This combined with the fact that  $\Theta$ is compact, implies that the first-order conditions  for an interior global
maximum are satisfied at $(\tilde{\theta}^{\top}, \tilde{\eta}^{\top})^{\top}$ and are given by
\begin{eqnarray}
-\Gamma_2^{\top}\tilde{\eta} = 0 \,,~~ -\hat{\mathbb{U}}_{\N}(\theta_{\N},\varphi_{\N})-\Gamma_2[\tilde{\theta} - \theta_{\N}]  - W_2\tilde{\eta} = 0.
\label{foc}
\end{eqnarray}
The two systems of equations can be combined and rewritten as
$$
-\left(\begin{array}{cc}
0&\Gamma_2^{\top}\\
\Gamma_2&W_2\end{array}\right)\left(\begin{array}{ccccc}
\tilde{\theta}-\theta_{\N}\\\tilde{\eta}-0\end{array}\right) = \left(\begin{array}{ccccc}
0\\
\hat{\mathbb{U}}_{\N}(\theta_{\N},\varphi_{\N})\end{array}\right)\,.
$$
Using the result on the inverse of a block matrix, we obtain that
\begin{equation*}
\left(\begin{array}{ccccc}
\tilde{\theta}-\theta_{\N}\\\tilde{\eta}-0\end{array}\right) = - \left(\begin{array}{cc}
\Sigma&H\\
H^{\top}&\mathscr{P}\end{array}\right)\left(\begin{array}{ccccc}
0\\
\hat{\mathbb{U}}_{\N}(\theta_{\N},\varphi_{\N})\end{array}\right) \,.
\label{order}
\end{equation*}
where  $\Sigma_2=(\Gamma_2^{\top}W_2^{-1}\Gamma_2)^{-1}$, $H=\Sigma_2\Gamma_2^{\top}W_2^{-1}$ and $\mathscr{P}=W_2^{-1}-W_2^{-1}\Gamma_2\Sigma_2\Gamma_2^{\top}W_2^{-1}$.
This immediately shows that  $\tilde{\theta}-\theta_{\N}=-\Sigma_2\Gamma_2^{\top}W_2^{-1}\hat{\mathbb{U}}_{\N}(\theta_{\N},\varphi_{\N})$ and $\tilde{\eta}=-\mathscr{P} \hat{\mathbb{U}}_{\N}(\theta_{\N},\varphi_{\N})$.
A little more work gives that $\hat{\theta}_{\GEL}-\tilde{\theta}=o_p(n_{\B}^{-1/2})$ and $\eta_{\GEL}-\tilde{\eta}=o_p(n_{\B}^{-1/2})$.
Combining this with condition B6 implies that
$
n_{\B}^{1/2}(\hat{\theta}_{\GEL}-\theta_{\N})\stackrel{{\cal L}}{\rightarrow}N(0,V_2),
$
where
 $
V_2=\Sigma_2\Gamma_2^{\top}W_2^{-1}\Omega W_2^{-1}\Gamma_2 \Sigma_2 \,,
\label{V1}
$
with $
\Omega= n_{\B}N^{-2}{\rm Var}\{\sum_{i\in \mS}\pi_i^{-1}\psi(Z_i,\theta_{\N},\varphi_{\N})\mid \mF_{\N}\}.
$
\hfill$\square$\\

\noindent
\textbf{Proof of  Corollary 3.1:}
Denote by
$X\in \mathsf {R}_{+}^{d_x}$ the size variable that contains information for the sampling design.
Under single stage
PPS sampling with replacement,  each unit $i$ in $\mS$ is selected from $\{1,2,\cdots,N\}$ with the given probabilities $p_i=X_i$ and the  first-order inclusion probabilities of unit $i$ can be written as $\pi_i=nX_i$.
This combined with the arguments of Wu and  Thompson (2020), implies that the
augmented survey weighted estimating equations  $\hat{\mathbb{U}}_{\N}(\theta_{\N},\varphi_{\N})$ could be rewritten as
\[\begin{array}{llll}
\hat{\mathbb{U}}_{\N}(\theta_{\N},\varphi_{\N})=
\dfrac{1}{Nn}\sum\limits_{i\in \mathcal {S}}X_i^{-1}\psi(Z_i,\theta_{\N},\varphi_{\N})=\dfrac{1}{Nn}\sum\limits_{i=1}^{n}R_i.
\end{array}
\]
where $R_1,\cdots, R_n$ are independent and identically distributed random variables, with the  common distribution of the random variable $R$  given by
$${\rm Pr}\Big(R=X_i^{-1}\psi(Z_i,\theta_{\N},\varphi_{\N})\Big)=X_i,~~i=1,\cdots,N.$$
Now, simple algebraic manipulations show that
$
E(R)=\sum_{i=1}^{N}X_i^{-1}\psi(Z_i,\theta_{\N},\varphi_{\N})X_i=\sum_{i=1}^{N}\psi(Z_i,\theta_{\N},\varphi_{\N})=0,$ and
$$
{\rm Var}(R)=E(R^2)=\sum_{i=1}^{N}X_i^{-1}\psi(Z_i,\theta_{\N},\varphi_{\N})^{\otimes 2}.
$$
Consequently, we have that
\begin{eqnarray*}
{\rm Var}\Big\{\hat{\mathbb{U}}_{\N}(\theta_{\N},\varphi_{\N})\mid \mF_{\N}\Big\}
=\dfrac{1}{n}{\rm Var}(R)
=\dfrac{1}{N^2n}\sum\limits_{i=1}^{N}X_i^{-1}\psi(Z_i,\theta_{\N},\varphi_{\N})^{\otimes 2}\\
=\dfrac{1}{N^2}\sum\limits_{i=1}^{N}\pi_i^{-1}\psi(Z_i,\theta_{\N},\varphi_{\N})^{\otimes 2}.
\end{eqnarray*}
Therefore, under single stage PPS sampling with replacement,
$$
\Omega= n_{\B}{\rm Var}\Big\{\hat{\mathbb{U}}_{\N}(\theta_{\N},\varphi_{\N})\mid \mF_{\N}\Big\}=\dfrac{n_{\B}}{N^2}\sum\limits_{i=1}^{N}\pi_i^{-1}\psi(Z_i,\theta_{\N},\varphi_{\N})^{\otimes 2}=W_2,
$$
and  the covariance matrix
$V_2$ reduce to $(\Gamma_2^{\top}W_2^{-1}
\Gamma_2)^{-1}=\Sigma_2$. Using the arguments of Zhao et al. (2022), we can show that the result is also
valid for single-stage PPS sampling without replacement with negligible sampling.

Denote by $I_i$ the indicator variable for unit $i$, and denote by $A_i$ the event ``if unit $i$ is in the sample". Then $I_i={\bf I}(A_i)$, where ${\bf I}(A)$ is
the indicator function of event A. One special PPS sampling without replacement is the Poisson sampling,  under which $I_1,\cdots,I_{\N}$ are independent Bernoulli random variables with  success probabilities $\pi_1,\cdots,\pi_{\N}$, and thus
\begin{eqnarray*}
{\rm Var}\Big\{\hat{\mathbb{U}}_{\N}(\theta_{\N},\varphi_{\N})\mid \mF_{\N}\Big\}
=\dfrac{1}{N^2}\sum\limits_{i=1}^{N}\pi_i^{-1}\psi(Z_i,\theta_{\N},\varphi_{\N})^{\otimes 2}-\dfrac{1}{N^2}\sum\limits_{i=1}^{N}\psi(Z_i,\theta_{\N},\varphi_{\N})^{\otimes 2}.
\end{eqnarray*}
If the sampling rate  is negligible, i.e.,  $n_{\B}/N=o(1)$,  then we have that
$$\dfrac{n_{\B}}{N^2}\sum\limits_{i=1}^{N}\psi(Z_i,\theta_{\N},\varphi_{\N})^{\otimes 2}=o(1),$$
which implies that
$
\Omega= n_{\B}{\rm Var}\{\hat{\mathbb{U}}_{\N}(\theta_{\N},\varphi_{\N})\mid \mF_{\N}\}=W_2+o(1).
$
Therefore, under Poisson sampling with negligible sampling fractions,  $
n_{\B}^{1/2}(\hat{\theta}_{\GEL}-\theta_{\N})\stackrel{{\cal L}}{\rightarrow}N(0,V_2),
$
where
 $
V_2=(\Gamma_2^{\top}W_2^{-1}
\Gamma_2)^{-1}
$.
\hfill$\square$\\

\noindent
\textbf{Proof of  Theorem 3.3:}
It follows from the duality results presented in Section 2.3 of the main paper that the GEL  ratio statistic ${\rm T}_{\N}(\theta)$ can be
equivalently defined as
\[\begin{array}{lllll}
{\rm T}_{\N}(\theta) &=& -2n_{\B}\{[\hat{\mathcal{P}}_{\N}(\hat{\theta}_{\GEL},\hat{\eta}_{\GEL},\hat{\varphi})- \hat{\mathcal{P}}_{\N}(\theta,\eta_{\theta},\hat{\varphi})\}.
\end{array}
\]
Combining the $n_{\B}^{1/2}$-consistence of $\hat{\theta}_{\GEL}$ with the conclusion of Lemma \ref{lemma4}, we have that
$$2n_{\B}|\hat{\mathcal{P}}_{\N}(\hat{\theta}_{\GEL},\hat{\eta}_{\GEL},\hat{\varphi})-\mathcal {L}_{\N}(\hat{\theta}_{\GEL},\hat{\eta}_{\GEL})|=o_p(1),$$
where $\mathcal {L}_{\N}(\theta,\eta)=[-\hat{\mathbb{U}}_{\N}(\theta_{\N},\varphi_{\N})-\Gamma_2 (\theta-\theta_{\N})]^{\top}\eta - \frac{1}{2}\eta^{\top}W_2\eta$.
This, together with the facts that $\hat{\theta}_{\GEL}-\tilde{\theta}=o_p(n_{\B}^{-1/2})$ and $\eta_{\GEL}-\tilde{\eta}=o_p(n_{\B}^{-1/2})$, as discussed in  the proof of Theorem 3.2, implies that
$
2n_{\B}\hat{\mathcal{P}}_{\N}(\hat{\theta}_{\GEL},\hat{\eta}_{\GEL},\hat{\varphi}) = 2n_{\B}\mathcal{L}_{\N}(\tilde{\theta}, \tilde{\eta}) + o_p(1).
$
By the first-order condition  (\ref{foc}), we obtain that
$
2n_{\B}\hat{\mathcal{P}}_{\N}(\hat{\theta}_{\GEL},\hat{\eta}_{\GEL},\hat{\varphi}) = n_{\B}\tilde{\eta}^{\top}W_2\tilde{\eta} + o_p(1).
$
Since $\tilde{\eta}=-\mathscr{P} \hat{\mathbb{U}}_{\N}(\theta_{\N},\varphi_{\N})$ from the proof of Theorem 3.3 and $\mathscr{P}W_2\mathscr{P}=\mathscr{P}$,
\begin{eqnarray}\label{expan1}
2n_{\B}\hat{\mathcal{P}}_{\N}(\hat{\theta}_{\GEL},\hat{\eta}_{\GEL},\hat{\varphi}) =n_{\B}\hat{\mathbb{U}}_{\N}(\theta_{\N},\varphi_{\N})^{\top}
\mathscr{P}\hat{\mathbb{U}}_{\N}(\theta_{\N},\varphi_{\N})+o_p(1) \,.
\end{eqnarray}
Furthermore, it follows from Lemma \ref{lemma3} that $2n_{\B}\hat{\mathcal{P}}_{\N}(\theta_{\N},\eta,\hat{\varphi}) =2n_{\B}\mathcal {L}_{\N}(\theta_{\N},\eta)+o_p(1)$,
where $\mathcal {L}_{\N}(\theta_{\N},\eta)=-\hat{\mathbb{U}}_{\N}(\theta_{\N},\varphi_{\N})^{\top}\eta - \frac{1}{2}\eta^{\top}W_2\eta$. The first order conditions with respect to $\eta=\eta(\theta_{\N},\hat{\varphi})$ is given by $-\hat{\mathbb{U}}_{\N}(\theta_{\N},\varphi_{\N})^{\top} - W_2\eta=0$. Then, we have
\begin{eqnarray}\label{expan2}
2n_{\B}\hat{\mathcal{P}}_{\N}(\theta_{\N},\eta, \hat{\varphi}) =n_{\B}\hat{\mathbb{U}}_{\N}(\theta_{\N},\varphi_{\N})^{\top}
W_2^{-1}\hat{\mathbb{U}}_{\N}(\theta_{\N},\varphi_{\N})+o_p(1).
\end{eqnarray}
Combining (\ref{expan1}) with (\ref{expan2}), we obtain
\[\begin{array}{lllll}
{\rm T}_{\N}(\theta) &=& -2n_{\B}\{[\hat{\mathcal{P}}_{\N}(\hat{\theta}_{\GEL},\hat{\eta}_{\GEL},\hat{\varphi})- \hat{\mathcal{P}}_{\N}(\theta,\eta_{\theta},\hat{\varphi})\}\\
&=&n_{\B}^{1/2}\hat{\mathbb{U}}_{\N}(\theta_{\N},\varphi_{\N})^{\top}W_2^{-1}
\Gamma_2\Sigma_2\Gamma_2^{\top}W_2^{-1}n_{\B}^{1/2}\hat{\mathbb{U}}_{\N}(\theta_{\N},\varphi_{\N})+o_p(1).\\
\end{array}
\]
The limiting distribution of statistic ${\rm T}_{\N}(\theta)$ can be consequently obtained by using the Slutsky's theorem  and  the asymptotic normality of $n_{\B}^{1/2}\hat{\mathbb{U}}_{\N}(\theta_{\N},\varphi_{\N})$.
\hfill$\square$\\

\noindent
\textbf{Proof of  Corollary 3.2:}
It follows from the proof of Corollary  3.1 that  $\Omega=W_2$ under the given sampling designs,  which leads to
$$
\Omega^{1/2}W_2^{-1}
\Gamma_2\Sigma_2\Gamma_2^{\top}W_2^{-1}\Omega_2^{1/2}=W_2^{-1/2}
\Gamma_2\Sigma_2\Gamma_2^{\top}W_2^{-1/2} \,.
$$
Simple algebraic manipulations show that
\[
{\rm trace}\{W_2^{-1/2}
\Gamma_2\Sigma_2\Gamma_2^{\top}W_2^{-1/2}\}={\rm trace}\{
\Gamma_2\Sigma_2\Gamma_2^{\top}W_2^{-1} \}={\rm trace}\{
\Sigma_2\Gamma_2^{\top}W_2^{-1}\Gamma_2\}
=p \,.
\]
Therefore, using Theorem
9.2.1 of Rao and Mitra (1971),  $
{\rm T}_{\N}(\theta_{\N})\stackrel{{\cal L}}{\rightarrow}\chi_p^2
$, under the given sampling designs.
\hfill$\square$\\

\noindent
\textbf{Proof of  Theorem 3.4:}
Recall that
$\mathscr{U}_{\N}(\theta,\varphi)=\sum_{i=1}^N\phi(Z_i,\theta,\varphi)/N$. Let $\hat{\mathscr{U}}_{\N}(\theta,\varphi)=\sum_{i\in\mS}\pi_{i}^{-1}\phi(Z_i,\theta,\varphi)/N$, and let $\mathscr{U}(\theta,\varphi)=(\mathbb{U}(\theta,\varphi)^{\top},\mathfrak{U}(\theta,\varphi)^{\top})^{\top}$.
We first show that the    restricted GEL estimator $\hat\theta_{\GEL}^{\R}$ defined in \eqref{rgele-new}  is design consistent for $\theta_{\N}$.
By the same kinds of arguments we used in  the proof of Theorem 3.1, all we need to verify now are the following results: (i)
$\max_{i\in \mS}\sup_{\theta\in\Theta,\varphi\in\Psi}\|\phi(Z_i,\theta,\varphi)\|=o_p(n_{\B}^{1/2})$; (ii)  $\hat{\mathscr{U}}_{\N}(\hat{\theta}_{\GEL}^{\R},\hat{\varphi})=O_p(n_{\B}^{-1/2})$; (iii) for any $\delta_{\N}=o(1)$, $\sup_{\theta\in\Theta,\varphi\in\Psi(\delta_{\N})}\|\hat{\mathscr{U}}_{\N}(\theta,\varphi)
-\mathscr{U}_{\N}(\theta,\varphi)\|=o_p(1)$ and $\sup_{\theta\in\Theta,\varphi\in\Psi(\delta_{\N})}\|\mathscr{U}_{\N}(\theta,\varphi)
-\mathscr{U}(\theta,\varphi)\|=o(1)$. Following the same arguments as used in the proof of
Lemma \ref{lemma3}, we can conclude (i). Note that $\|\phi(Z_i,\theta,\varphi)\|=\{\|\psi(Z_i,\theta,\varphi)\|^2+\|q(Z_i,\theta)\|^2\}^{1/2}$. This, together with conditions B2(i) and B7(i), implies that
\begin{eqnarray*}
\max_{i\in \mS}\sup_{\theta\in\Theta,\varphi\in\Psi}\|\phi(Z_i,\theta,\varphi)\|\leq
\Big\{(\max_{i\in \mS}\sup_{\theta\in\Theta,\varphi\in\Psi}\|\psi(Z_i,\theta,\varphi)\|)^2\\
+(\max_{i\in \mS}\sup_{\theta\in\Theta}\|q(Z_i,\theta)\|)^2\Big\}^{1/2}
=o_p(n_{\B}^{1/2}).
\end{eqnarray*}
By recycling arguments we have used in proof of Lemma \ref{lemma3}, we can readily establish (ii).
Using condition B7(ii),  we can show that
\begin{eqnarray*}
\sup_{\theta\in\Theta}\|\hat{\mathfrak{U}}_{\N}(\theta)
-\mathfrak{U}_{\N}(\theta)\|=o_p(1)\,\,\,\mbox{and}\,\,\,
\sup_{\theta\in\Theta}\|\mathfrak{U}_{\N}(\theta)
-\mathfrak{U}(\theta)\|=o(1).
\end{eqnarray*}
Combined this with results (\ref{appenda1}), implies that
\begin{eqnarray*}
\sup_{\theta\in\Theta,\varphi\in\Psi(\delta_{\N})}\|\hat{\mathscr{U}}_{\N}(\theta,\varphi)
-\mathscr{U}_{\N}(\theta,\varphi)\|=
\sup_{\theta\in\Theta,\varphi\in\Psi(\delta_{\N})}\Big\{\|\hat{\mathbb{U}}_{\N}(\theta,\varphi)
-\mathbb{U}_{\N}(\theta,\varphi)\|^2\\
+\|\hat{\mathfrak{U}}_{\N}(\theta,\varphi)
-\mathfrak{U}_{\N}(\theta,\varphi)\|^2\Big\}^{1/2}
\\
\leq
\bigg\{\Big(\sup_{\theta\in\Theta,\varphi\in\Psi(\delta_{\N})}\|\hat{\mathbb{U}}_{\N}(\theta,\varphi)
-\mathbb{U}_{\N}(\theta,\varphi)\|\Big)^2\\
+\Big(\sup_{\theta\in\Theta}\|\hat{\mathfrak{U}}_{\N}(\theta)
-\mathfrak{U}_{\N}(\theta)\|\Big)^2\bigg\}^{1/2}
=o_p(1).
\end{eqnarray*}
Similarly, we can verify that $\sup_{\theta\in\Theta,\varphi\in\Psi(\delta_{\N})}\|\mathscr{U}_{\N}(\theta,\varphi)
-\mathscr{U}(\theta,\varphi)\|=o(1)$.   The design consistency of restricted GEL estimator $\hat\theta_{\GEL}^{\R}$  is then established.

We now turn to investigate asymptotic normality of
the restricted estimators $\hat\theta_{\GEL}^{\R}$.
By condition B7(iv), $\|\hat{\mathfrak{U}}_{\N}(\theta)  - \hat{\mathfrak{U}}_{\N}(\theta') -\mathfrak{U}_{\N}(\theta) +\mathfrak{U}_{\N}(\theta')\|=o_p(n_{\B}^{-1/2})$
uniformly in  $\theta,\theta' \in \Theta(\delta_{\N})$  with $\delta_{\N}=o(1)$.
 Previous arguments can be recycled to established that
$\|\hat{U}_{\N}(\theta,\varphi)  - \hat{U}_{\N}(\theta',\varphi') -U_{\N}(\theta,\varphi) +U_{\N}(\theta',\varphi')\|=o_p(n_{\B}^{-1/2})$, uniformly in  $(\theta,\varphi),(\theta',\varphi') \in \Theta(\delta_{\N})\times\Psi(\delta_{\N})$ with $\delta_{\N}=o(1)$.
Furthermore, by condition B4, we can readily show that,  for all $(\theta, \varphi), (\theta', \varphi')\in\Theta(\delta_{\N})\times\Psi(\delta_{\N})$ with   $\delta_{\N}=o(1)$,
\begin{eqnarray*}
\|\mathscr{U}(\theta,\varphi)-\mathscr{U}(\theta,\varphi')\|
=\|\mathbb{U}(\theta,\varphi)-\mathbb{U}(\theta,\varphi')\|\leq c\|\varphi-\varphi'\|_{\Psi}^2,\,\,\,
\mbox{for some constant $c\ge0$}.
\end{eqnarray*}

Define $\hat{\nu}_{\GEL}^{\R}=\arg\max_{\nu\in\hat{\Lambda}_{\N,\phi}(\theta,\hat{\varphi})}
\hat{\mathcal{P}}_{\N}^{\R}(\hat{\theta}_{\GEL}^{\R},\nu,\hat{\varphi})$.
Note that $\hat\theta_{\GEL}^R$ defined in (\ref{gele-new}) and $\hat\nu_{\GEL}^R$ can be also viewed as  the optimizers of
$$\min_{\theta\in\Theta^{\R}}\sup_{\nu\in\mathsf{R}^{r+s}}\{\hat{\mathcal{P}}_{\N}^{\R}(\theta,\nu,\hat{\varphi})
+\Upsilon^{\top}R(\theta)\},$$ where  $\Upsilon$ is a $k\times 1$ vector of Lagrange multipliers.
Define   $\mathcal {L}_{\N}^{\R}(\theta,\nu)=[-\hat{\mathscr{U}}_{\N}(\theta_{\N},\varphi_{\N})-\Pi (\theta-\theta_{\N})]^{\top}\nu - \frac{1}{2}\nu^{\top}\mathscr{W}\nu$. Using similar arguments to the proof of Lemma \ref{lemma4}, we can show that
\begin{eqnarray*}
|\hat{\mathcal{P}}_{\N}^{\R}(\theta,\nu,\hat{\varphi})-\mathcal {L}_{\N}^{\R}(\theta,\nu)|=o_p(n_{\B}^{-1}) \,,
\end{eqnarray*}
uniformly in  $\theta$ and $\nu$, for $\theta-\theta_{\N}=O_p(n_{\B}^{-1/2})$ and $\nu=O_p(n_{\B}^{-1/2})$.
Let $\tilde{\theta}_{\GEL}^{\R}$ and $\tilde{\nu}_{\GEL}^{\R}$ be the optimizers of
$$\min_{\theta\in\Theta^{\R}}\sup_{\nu\in\mathsf {R}^{r+s}} \{\mathcal {L}_{\N}^{\R}(\theta,\nu)+\Upsilon^{\top}R(\theta)\}.$$
The first-order conditions for an interior global
maximum are given by
$$
\begin{array}{rrrrrr}
-\Pi^{\top}\tilde{\nu}_{\GEL}^{\R}+\Phi(\tilde{\theta}_{\GEL}^{\R})^{\top}\tilde{\Upsilon}^{\R} &=& 0 \,,\\
-\Pi(\tilde{\theta}_{\GEL}^{\R} - \theta_{\N}) - \hat{\mathscr{U}}_{\N}(\theta_{\N},\varphi_{\N}) - \mathscr{W}\tilde{\lambda}_{\GEL}^{\R} &=& 0 \,,\\
R(\tilde{\theta}_{\GEL}^{\R}) &=& 0 \,,
\end{array}
$$
where $\Phi(\theta) = \partial R(\theta)/\partial \theta$.
Following the same  arguments as used in  the proof of Theorem  3.2, we can  show that $\|\tilde{\theta}_{\GEL}^{\R}-\theta_{\N}\|=O_p(n_{\B}^{-1/2})$. This, together with the condition $R(\theta_{\N})=0$, implies that  $R(\tilde{\theta}_{\GEL}^{\R})=\Phi(\theta_{\N}) (\tilde{\theta}_{\GEL}^{\R} - \theta_{\N})+o_p(n_{\B}^{-1/2})$. Furthermore, we have $\Phi(\tilde{\theta}_{\GEL}^{\R})=\Phi(\theta_{\N})+O_p(n_{\B}^{-1/2})$.
It follows from $\tilde{\nu}_{\GEL}^{\R}=O_p(n_{\B}^{-1/2})$ that $\tilde{\Upsilon}^{\R}=O_p(n_{\B}^{-1/2})$.
The above first-order conditions can be further written as
$$
\begin{array}{cccccc}
-\Pi^{\top}\tilde{\nu}_{\GEL}^{\R}+\Phi^{\top}\tilde{\Upsilon}^{\R} &=& o_p(n_{\B}^{-1/2}) \,,\\
-\Pi(\tilde{\theta}_{\GEL}^{\R} - \theta_{\N}) - \hat{\mathscr{U}}_{\N}(\theta_{\N},\varphi_{\N}) - \mathscr{W}\tilde{\lambda}_{\GEL}^{\R} &=& 0 \,,\\
\Phi(\tilde{\theta}_{\GEL}^{\R}-\theta_{\N}) &=& o_p(n_{\B}^{-1/2}) \,.
\end{array}
$$
In matrix form, we have
$$
\left(\begin{array}{ccccc}
-\mathscr{W}&-\Pi&0\\-\Pi^{\top}&0&\Phi^{\top}\\
0&\Phi&0\end{array}\right)
\left(\begin{array}{ccccc}\tilde{\nu}_{\GEL}^{\R}\\
\tilde{\theta}_{\GEL}^{\R} - \theta_{\N}\\\tilde{\Upsilon}^{\R}\end{array}\right) =\left(\begin{array}{ccccc}
\hat{\mathscr{U}}_{\N}(\theta_{\N})\\
0\\0\end{array}\right)+ o_p(n_{\B}^{-1/2}) \,.
$$
Let
$$
\mathfrak{M}=\left(\begin{array}{ccccc}
-\mathscr{W}&-\Pi&0\\-\Pi^{\top}&0&\Phi^{\top}\\
0&\Phi&0\end{array}\right)=:\left(\begin{array}{ccccc}
\mathfrak{M}_{11}&\mathfrak{M}_{12}\\\mathfrak{M}_{21}&\mathfrak{M}_{22}\end{array}\right) \,,
$$
with $\mathfrak{M}_{11}=-\mathscr{W}$, $\mathfrak{M}_{12}=(-\Pi,0)$, $\mathfrak{M}_{21}=\mathfrak{M}_{12}^{\top}$ and
$$
\mathfrak{M}_{22}=\left(\begin{array}{ccccc}
0&\Phi^{\top}\\
\Phi&0\end{array}\right) \,.
$$
By the matrix algebra we  that
$$
\mathfrak{M}^{-1}=\left(\begin{array}{ccccc}
\mathfrak{M}_{11}^{-1}&0\\
0&0\end{array}\right)+\left(\begin{array}{ccccc}
-\mathfrak{M}_{11}^{-1}\mathfrak{M}_{12}\\
I\end{array}\right)\mathscr{D}^{-1}(-\mathfrak{M}_{21}\mathfrak{M}_{11}^{-1}~~ I) \,,
$$
where
$$
\mathscr{D}=\mathfrak{M}_{22}-\mathfrak{M}_{21}\mathfrak{M}_{11}^{-1}\mathfrak{M}_{12}=\left(\begin{array}{ccccc}
(\Sigma^{\R})^{-1} &\Phi^{\top}\\
\Phi&0\end{array}\right) \,.
$$
Furthermore, we  have
$$
\mathscr{D}^{-1}=\left(\begin{array}{ccccc}
\Sigma^{\R}-\Sigma^{\R} \Phi^{\top}(\Phi\Sigma^{\R} \Phi^{\top})^{-1}\Phi\Sigma^{\R}&-\Sigma^{\R} \Phi^{\top}(\Phi\Sigma^{\R} \Phi^{\top})^{-1}\\
-(\Phi\Sigma^{\R} \Phi^{\top})^{-1}\Phi\Sigma^{\R} &(\Phi\Sigma^{\R} \Phi^{\top})^{-1}\end{array}\right)\,.
$$
Consequently, we obtain that $\tilde{\nu}_{\GEL}^{\R}=[\mathfrak{M}_{11}^{-1}
+\mathfrak{M}_{11}^{-1}\mathfrak{M}_{12}\mathscr{D}^{-1}\mathfrak{M}_{21}\mathfrak{M}_{11}^{-1}]
\hat{\mathscr{U}}_{\N}(\theta_{\N},\varphi_{\N})+ o_p(n_{\B}^{-1/2})$, and
$$
\left(\begin{array}{ccccc}
\tilde{\theta}_{\GEL}^{\R} - \theta_{\N}\\\tilde{\Upsilon}^{\R}\end{array}\right) =-\mathscr{D}^{-1}\mathfrak{M}_{21}\mathfrak{M}_{11}^{-1}\hat{\mathscr{U}}_{\N}(\theta_{\N},\varphi_{\N})+ o_p(n_{\B}^{-1/2}) \,.
$$
Let $\mathscr{C}^{\R}=\Sigma^{\R}-\Sigma^{\R} \Phi^{\top}(\Phi\Sigma^{\R} \Phi^{\top})^{-1}\Phi\Sigma^{\R}$
and $\mathscr{P}^{\R}=\mathscr{W}^{-1}-\mathscr{W}^{-1}\Pi \mathscr{C}^{\R}\Pi^{\top}\mathscr{W}^{-1}$.
Simple algebraic manipulations show that
$\tilde{\nu}_{\GEL}^{\R}=-\mathscr{P}^{\R}\hat{\mathscr{U}}_{\N}(\theta_{\N},\varphi_{\N})+ o_p(n_{\B}^{-1/2})$,
$\tilde{\theta}_{\GEL}^{\R} - \theta_{\N}=-\mathscr{C}^{\R}\Pi^{\top}\mathscr{W}^{-1}\hat{\mathbb{U}}_{\N}(\theta_{\N},\varphi_{\N})
+o_p(n_{\B}^{-1/2})$, and $\tilde{\Upsilon}^{\R}=(\Phi\Sigma^{\R} \Phi^{\top})^{-1}\Phi\Sigma^{\R}\Pi^{\top}\mathscr{W}^{-1}
\hat{\mathbb{U}}_{\N}(\theta_{\N},\varphi_{\N})+o_p(n_{\B}^{-1/2})$. Therefore,
$$
n_{\B}^{1/2}(\tilde{\theta}_{\GEL}^{\R}-\theta_{\N})\stackrel{{\cal L}}{\rightarrow}N(0,V^{\R}),
$$
where
 $
V^{\R}=\mathscr{C}^{\R}\Pi^{\top} \mathscr{W}^{-1} \Omega^{\R} \mathscr{W}^{-1} \Pi \mathscr{C}^{\R}.
$
A little more work gives that $\hat{\theta}_{\GEL}^{\R}-\tilde{\theta}_{\GEL}^{\R}=o_p(n_{\B}^{-1/2})$ and $\nu_{\GEL}^{\R}-\tilde{\nu}_{\GEL}^{\R}=o_p(n_{\B}^{-1/2})$. The desired conclusion of Theorem 3.4 then follows directly.
\hfill$\square$\\

\noindent
\textbf{Proof of  Theorem 3.5:}
The duality results presented in section 2.3 of the main paper motivates us to  consider the following alternative  GEL  ratio statistic
\begin{equation*}
{\rm T}_{\N}^{\R}(\theta_{\N}) = -2n_{\B}\{\hat{\mathcal{P}}_{\N}(\hat{\theta}_{\GEL},\hat{\nu}_{\GEL},\hat{\varphi})- \hat{\mathcal{P}}_{\N}^{\R}(\hat{\theta}_{\GEL}^{\R},\hat{\nu}_{\GEL}^{\R},\hat{\varphi})\},
\end{equation*}
where $\hat{\mathcal{P}}_{\N}^{\R}(\theta,
\nu,\varphi)$ is defined in (\ref{rgel.fun.dual}),  $\hat\theta_{\GEL}^R$ is defined in (\ref{gele-new}), and $$\hat{\nu}_{\GEL}^{\R}=\arg\max_{\nu\in\hat{\Lambda}_{\N,\phi}(\theta,\hat{\varphi})}
\hat{\mathcal{P}}_{\N}^{\R}(\hat{\theta}_{\GEL}^{\R},\nu,\hat{\varphi}).$$
By recycling previous arguments,
we have that $2n_{\B}|\hat{\mathcal{P}}_{\N}(\hat{\theta}_{\GEL},\hat{\eta}_{\GEL},\hat{\varphi})-\mathcal {L}_{\N}(\hat{\theta}_{\GEL},\hat{\eta}_{\GEL})|=o_p(1)$
and $2n_{\B}|\hat{\mathcal{P}}_{\N}^{\R}(\hat{\theta}_{\GEL}^{\R},\hat{\nu}_{\GEL}^{\R},\hat{\varphi})-\mathcal {L}_{\N}^{\R}(\hat{\theta}_{\GEL}^{\R},\hat{\nu}_{\GEL}^{\R})|=o_p(1)$, where $\mathcal {L}_{\N}(\theta,\eta)$ is defined in Lemma \ref{lemma4} and $\mathcal {L}_{\N}^{\R}(\theta,\nu)$ is defined in the proof of Theorem 3.4. Then, for ${\rm T}_{\N}^{\R}(\theta)$, we have that
\begin{eqnarray*}
{\rm T}_{\N}^{\R}(\theta_{\N}) &=& -2n_{\B}\{\hat{\mathcal{P}}_{\N}(\hat{\theta}_{\GEL},\hat{\nu}_{\GEL},\hat{\varphi})-\mathcal {L}_{\N}(\hat{\theta}_{\GEL},\hat{\eta}_{\GEL})\}\\ &&-2n_{\B}\{\hat{\mathcal{P}}_{\N}^{\R}(\hat{\theta}_{\GEL}^{\R},\hat{\nu}_{\GEL}^{\R},\hat{\varphi})-\mathcal {L}_{\N}^{\R}(\hat{\theta}_{\GEL}^{\R},\hat{\nu}_{\GEL}^{\R})\}\\
&&-2n_{\B}\{\mathcal {L}_{\N}(\hat{\theta}_{\GEL},\hat{\eta}_{\GEL})-\mathcal {L}_{\N}^{\R}(\hat{\theta}_{\GEL}^{\R},\hat{\nu}_{\GEL}^{\R})\}\\
&=&-2n_{\B}\{\mathcal {L}_{\N}(\hat{\theta}_{\GEL},\hat{\eta}_{\GEL})-\mathcal {L}_{\N}^{\R}(\hat{\theta}_{\GEL}^{\R},\hat{\nu}_{\GEL}^{\R})\}+o_p(1).
\end{eqnarray*}
Recall that $\tilde{\theta}_{\GEL}^{\R}$ and $\tilde{\nu}_{\GEL}^{\R}$ are the optimizers of
$\min_{\theta\in\Theta^{\R}}\sup_{\nu\in\mathsf {R}^{r+s}} \mathcal {L}_{\N}^{\R}(\theta,\nu).$
Since $\hat{\theta}_{\GEL}^{\R}-\tilde{\theta}_{\GEL}^{\R}=o_p(n_{\B}^{-1/2})$ and $\nu_{\GEL}^{\R}-\tilde{\nu}_{\GEL}^{\R}=o_p(n_{\B}^{-1/2})$ as discussed in the proof of Theorem 3.4, $2n_{\B}\mathcal {L}_{\N}^{\R}(\hat{\theta}_{\GEL}^{\R},\hat{\nu}_{\GEL}^{\R})=
2n_{\B}\mathcal {L}_{\N}^{\R}(\tilde{\theta}_{\GEL}^{\R},\tilde{\nu}_{\GEL}^{\R})+o_p(1)$.
In the proof of Theorem 3.4, we have already showed  that  $-\Pi(\tilde{\theta}_{\GEL}^{\R} - \theta_{\N}) - \hat{\mathscr{U}}_{\N}(\theta_{\N},\varphi_{\N}) - \mathscr{W}\tilde{\lambda}_{\GEL}^{\R} = 0$ and $\tilde{\nu}_{\GEL}^{\R}=-\mathscr{P}^{\R}\hat{\mathscr{U}}_{\N}(\theta_{\N},\varphi_{\N})+ o_p(n_{\B}^{-1/2})$, where $\hat{\mathscr{U}}_{\N}(\theta,\varphi)=\sum_{i\in\mS}\pi_{i}^{-1}\phi(Z_i,\theta,\varphi)/N$. This, together with the fact $\mathscr{P}^{\R}\mathscr{W}\mathscr{P}^{\R}=\mathscr{P}^{\R}$, implies that
\begin{eqnarray*}
2n_{\B}\mathcal {L}_{\N}^{\R}(\tilde{\theta}_{\GEL}^{\R},\tilde{\nu}_{\GEL}^{\R}) &=& n_{\B}(\tilde{\nu}_{\GEL}^{\R})^{\top}\mathscr{W}\tilde{\nu}_{\GEL}^{\R}\\
&=&n_{\B}\hat{\mathscr{U}}_{\N}(\theta_{\N},\varphi_{\N})^{\top}
\mathscr{P}^{\R}\hat{\mathscr{U}}_{\N}(\theta_{\N},\varphi_{\N})+o_p(1) \,.
\end{eqnarray*}
Using the same arguments used in  the proof of Theorem 3.3, we can show
that
\[
2n_{\B}\mathcal {L}_{\N}(\hat{\theta}_{\GEL},\hat{\eta}_{\GEL}) =n_{\B}\hat{\mathbb{U}}_{\N}(\theta_{\N},\varphi_{\N})^{\top}
\mathscr{P}\hat{\mathbb{U}}_{\N}(\theta_{\N},\varphi_{\N})+o_p(1) \,,
\]
where $\mathscr{P}=W_2^{-1}-W_2^{-1}\Gamma_2\Sigma_2\Gamma_2^{\top}W_2^{-1}$. Now, define the $(r+s)\times r$ selection matrix
$\mathscr{S}_{\psi}=(I_r,0)^{\top}$. Then, we have that $\hat{\mathbb{U}}_{\N}(\theta_{\N},\varphi_{\N})=
\mathscr{S}_{\psi}^{\top}\hat{\mathscr{U}}_{\N}(\theta_{\N},\varphi_{\N})$, $\Omega=\mathscr{S}_{\psi}^{\top}\Omega^{\R}\mathscr{S}_{\psi}$ and $\mathscr{S}_{\psi}^{\top}\Pi=\Gamma_2$.
Thus,
\[
2n_{\B}\mathcal {L}_{\N}(\hat{\theta}_{\GEL},\hat{\eta}_{\GEL}) =n_{\B}\hat{\mathscr{U}}_{\N}(\theta_{\N},\varphi_{\N})^{\top}\mathscr{S}_{\psi}
\mathscr{P}\mathscr{S}_{\psi}^{\top}\hat{\mathscr{U}}_{\N}(\theta_{\N},\varphi_{\N})+o_p(1) \,.
\]
Consequently, we obtain that
\[
\begin{array}{llllll}
{\rm T}_{\N}^{\R}(\theta_{\N})
=n_{\B}\hat{\mathscr{U}}_{\N}(\theta_{\N},\varphi_{\N})^{\top}[\mathscr{P}^{\R}-\mathscr{S}_{\psi}
\mathscr{P}\mathscr{S}_{\psi}^{\top}]\hat{\mathscr{U}}_{\N}(\theta_{\N},\varphi_{\N})+o_p(1).
\end{array}
\]
Therefore, $
{\rm T}_{\N}^{\R}(\theta_{\N})\stackrel{{\cal L}}{\rightarrow}\mathcal{Q}^{\top} \Delta^{\R} \mathcal{Q},
$
where
$\mathcal{Q}\sim N(0, I_{r+s})$ and
$
\Delta^{\R} =(\Omega^{\R})^{1/2}[\mathscr{P}^{\R}-\mathscr{S}_{\psi}
\mathscr{P}\mathscr{S}_{\psi}^{\top}](\Omega^{\R})^{1/2}$.
\hfill$\square$\\

\noindent
\textbf{Proof of  Corollary 3.3:}
Using the same arguments as used in the proof of Corollary 3.1, we
can  show that $\Omega^{\R}=\mathscr{W}$ under the given sampling designs, which leads to $
V^{\R}=\mathscr{C}^{\R} (\Sigma^{\R})^{-1} \mathscr{C}^{\R}=\mathscr{C}^{\R}
$
and
\begin{eqnarray*}
\Omega^{\R}[\mathscr{P}^{\R}-\mathscr{S}_{\psi}
\mathscr{P}\mathscr{S}_{\psi}^{\top}]=I_{r+s}-\Pi \mathscr{C}^{\R}\Pi^{\top}\mathscr{W}^{-1}-\mathscr{W}\mathscr{S}_{\psi}
\mathscr{P}\mathscr{S}_{\psi}^{\top}.
\end{eqnarray*}
Simple algebraic manipulations show that
$$
{\rm trace}(\Pi \mathscr{C}^{\R}\Pi^{\top}\mathscr{W}^{-1})={\rm trace}\{\mathscr{C}^{\R}(\Sigma^{\R})^{-1}\}=p-k.
$$
It follows by the fact $\mathscr{S}_{\psi}^{\top}\mathscr{W}\mathscr{S}_{\psi}=W_2$ that
${\rm trace}(\mathscr{W}\mathscr{S}_{\psi}
\mathscr{P}\mathscr{S}_{\psi}^{\top})={\rm trace}(W_2
\mathscr{P})=r-p$. Therefore, we have that
\begin{eqnarray*}
{\rm trace}(\Delta^{\R})=r+s-(p-k)-(r-p)=s+k.
\end{eqnarray*}
Consequently, applying Theorem 9.2.1   of Rao and Mitra (1971),  we can conclude that  $
{\rm T}_{\N}^{\R}(\theta_{\N})\stackrel{{\cal L}}{\rightarrow}\chi_{s+k}^2
$ under the given sampling designs.
\hfill$\square$\\

\noindent
\textbf{Proof of  Theorem  5.1:}
To establish the consistency of the least-squares-based resampling estimate, it suffices to show that $E_{\mathscr{V}}[\mathscr{V}\mathscr{V}^{\top}]=I_{p}$ and $E_{\mathscr{V}}[\mathcal {D}_{\N,\theta}(\mathscr{V},\hat{\theta}_{\GEL},\hat{\varphi})\mathscr{V}^{\top}]
\stackrel{p}{\rightarrow}\Gamma_2$. The former is obvious. In what follows, we verify   the latter statement.

Recall that $\mathfrak{D}_{\N,\mathscr{V}}(\mathscr{V},\theta,\varphi)=[\hat{\mathbb{U}}_{\N}(\theta,\varphi)-
\mathbb{U}(\theta,\varphi)]
\mathscr{V}^{\top}$. By definition, we have that
\[
\begin{array}{llllll}
&&E_{\mathscr{V}}[\mathcal {D}_{\N,\theta}(\mathscr{V},\hat{\theta}_{\GEL},\hat{\varphi})\mathscr{V}^{\top}]-
E_{\mathscr{V}}[\Gamma_2\mathscr{V}\mathscr{V}^{\top}]\\
&=&
E_{\mathscr{V}}\Big[N^{\frac{1}{2}}\mathfrak{D}_{\N,\mathscr{V}}(\mathscr{V},
\hat{\theta}_{\GEL}+N^{-1/2}\mathscr{V},\hat{\varphi})-N^{\frac{1}{2}}\mathfrak{D}_{\N,\mathscr{V}}(\mathscr{V},
\hat{\theta}_{\GEL},\hat{\varphi})\Big]\\
&&+E_{\mathscr{V}}\Big[N^{\frac{1}{2}}
\Big(\mathbb{U}(\hat{\theta}_{\GEL}+N^{-1/2}\mathscr{V},\hat{\varphi})
-\mathbb{U}(\hat{\theta}_{\GEL},\hat{\varphi})\Big)\mathscr{V}^{\top}-
\Gamma_2\mathscr{V}\mathscr{V}^{\top}\Big].
\end{array}
\]
By the definition of $\mathscr{V}$ and  the second part of assumption (ii) of Theorem  5.1, we have
\begin{eqnarray}\label{lemma1-eq1}
\begin{split}
\sup_{(\theta,\varphi)\in\Theta(\delta_{\N})\times\Psi(\delta_{\N})}
\left\|E_{\mathscr{V}}\Big[N^{\frac{1}{2}}\mathfrak{D}_{\N,\mathscr{V}}(\mathscr{V},
\theta+N^{-1/2}\mathscr{V},\varphi)-N^{\frac{1}{2}}\mathfrak{D}_{\N,\mathscr{V}}(\mathscr{V},
\theta,\varphi)\Big]\right\|\\
=\sup_{(\theta,\varphi)\in\Theta(\delta_{\N})\times\Psi(\delta_{\N})}
\|N^{\frac{1}{2}}E_{\mathscr{V}}[\mathfrak{D}_{\N,\mathscr{V}}(\mathscr{V},
\theta+N^{-1/2}\mathscr{V},\varphi)]\|=o_p(1).
\end{split}
\end{eqnarray}
By the differentiability of $\mathbb{U}(\theta,\varphi)$ in the local neighborhood of $(\theta_{\N},\varphi_{\N})$ and  the first part of assumption (ii) of Theorem  5.1, we have that
\begin{eqnarray}\label{lemma1-eq2}
\begin{split}
\Big\|E_{\mathscr{V}}\Big[N^{\frac{1}{2}}
\Big(\mathbb{U}(\hat{\theta}_{\GEL}+N^{-1/2}\mathscr{V},\hat{\varphi})
-\mathbb{U}(\hat{\theta}_{\GEL},\hat{\varphi})\Big)\mathscr{V}^{\top}-
\Gamma_2\mathscr{V}\mathscr{V}^{\top}\Big]\Big\|\\
\leq E_{\mathscr{V}}\bigg[\sup\limits_{(\theta,\varphi)\in\Theta(\delta_{\N})\times\Psi(\delta_{\N})}
\Big\|[\Gamma_2(\theta,\varphi)-\Gamma_2]\mathscr{V}\mathscr{V}^{\top}\Big\|\bigg]\\
\leq E_{\mathscr{V}}\|\mathscr{V}\mathscr{V}^{\top}\|
\sup\limits_{(\theta,\varphi)\in\Theta(\delta_{\N})\times\Psi(\delta_{\N})}
\|\Gamma_2(\theta,\varphi)-\Gamma_2\|=o(1).
\end{split}
\end{eqnarray}
Combining (\ref{lemma1-eq1}) and (\ref{lemma1-eq2}), we can conclude
$E_{\mathscr{V}}[\mathcal {D}_{\N,\theta}(\mathscr{V},\hat{\theta}_{\GEL},\hat{\varphi})\mathscr{V}^{\top}]-
E_{\mathscr{V}}[\Gamma_2\mathscr{V}\mathscr{V}^{\top}]=o(1)$. This, together with the fact that
$E_{\mathscr{V}}[\Gamma_2\mathscr{V}\mathscr{V}^{\top}]=\Gamma_2$, implies that
$$E_{\mathscr{V}}[\mathcal {D}_{\N,\theta}(\mathscr{V},\hat{\theta}_{\GEL},\hat{\varphi})\mathscr{V}^{\top}]
\stackrel{p}{\rightarrow}\Gamma_2.$$
This completes the proof Theorem 5.1.
\hfill$\square$\\

\bigskip

\bigskip

\hrule

\hrule

\bigskip

\hfill February 15, 2023


\begin{thebibliography}{}

\bibitem[\protect\citeauthoryear{Ackerberg et al.}{Ackerberg et al.}{2014}]{Ackerberg}
Ackerberg, D., Chen, X., Hahn, J. and Liao, Z. (2014).  
\newblock  Asymptotic Efficiency of Semiparametric Two-step GMM. 
\newblock {\em Review of Economic Studies\/} {\em 81},  919--943.

\bibitem[\protect\citeauthoryear{Atkinson}{Atkinson}{1970}]{Atkinson}
Atkinson, A. B. (1970). 
\newblock  On the Measurement of Inequality. 
\newblock {\em Journal of Economic Theory\/} {\em 2}, 244--263.

\bibitem[\protect\citeauthoryear{Beach and Davidson}{Beach and Davidson}{1983}]{BeachDavidson}
Beach, C. M. and Davidson, R.  (1983). 
\newblock Distribution-Free Statistical Inference with Lorenz
Curves and Income Shares. 
\newblock {\em Review of Economic Studies\/} {\em 50}, 723--734.

\bibitem[\protect\citeauthoryear{Berger and De La Riva Torres}{Berger and De La Riva Torres}{2016}]{BergerTorres}
Berger, Y. G. and De La Riva Torres, O. (2016). 
\newblock Empirical Likelihood Confidence Intervals for Complex Sampling Designs. 
\newblock {\em Journal of the Royal Statistical Society,\/} Series B {\em  78}, 319--341.

\bibitem[\protect\citeauthoryear{Bhattacharya}{Bhattacharya}{2007}]{Bhattacharya}
Bhattacharya, D. (2007). 
\newblock Inference on Inequality from Household Survey Data. 
\newblock {\em Journal of Econometrics\/} {\em 137}, 674--707.

\bibitem[\protect\citeauthoryear{Bickel et al.}{Bickel et al.}{1993}]{Bickeletal1993}
Bickel, P. J., Klaassen, C. A., Bickel, P. J., Ritov, Y. A., Klaassen, J., Wellner, J. A., and Ritov, Y. A. (1993). 
\newblock {\em Efficient and adaptive estimation for semiparametric models}. Baltimore: Johns Hopkins University Press.

\bibitem[\protect\citeauthoryear{Binder}{Binder}{1983}]{Binder}
Binder, D. A. (1983). 
\newblock On the Variances of Asymptotically Normal Estimators from Complex Surveys. 
\newblock {\em International Statistical Review\/} {\em 51}, 279--292.

\bibitem[\protect\citeauthoryear{Bravo et al.}{Bravo et al.}{2020}]{BravoEscancianoKeilegom}
Bravo, F., Escanciano, J. C. and Van Keilegom, I. (2020). 
\newblock Two-Step Semiparametric Empirical Likelihood Inference. 
\newblock {\em The Annals of Statistics\/} {\em 48}, 1--26.

\bibitem[\protect\citeauthoryear{Cattaneo}{Cattaneo}{2010}]{Cattaneo}
Cattaneo, M. D. (2010). 
\newblock Efficient Semiparametric Estimation of Multi-Valued Treatment Effects Under Ignorability.   
\newblock {\em Journal of Econometrics\/} {\em 155}, 138--154.

\bibitem[\protect\citeauthoryear{Chang}{Chang}{2020}]{Chang}
Chang, N.  C. (2020). 
\newblock Double/Debiased Machine Learning for Difference-in-Differences Models. 
\newblock {\em The Econometrics Journal\/} {\em 23(2)}, 177--191.

\bibitem[\protect\citeauthoryear{Chen and  Sitter}{Chen and  Sitter}{1999}]{ChenSitter}
Chen, J. and  Sitter, R. R. (1999).
\newblock A Pseudo Empirical Likelihood Approach to the Effective Use of Auxiliary Information in Complex Surveys.
\newblock {\em Statistica Sinica\/} {\em 9}, 385--406.

\bibitem[\protect\citeauthoryear{Chen and Wu}{Chen and Wu}{2002}]{ChenWu}
Chen, J. and Wu, C. (2002). 
\newblock Estimation of Distribution Function and Quantiles Using the Model-Calibrated Pseudo Empirical Likelihood Method. 
\newblock {\em Statistica Sinica\/} {\em 12}, 1223--1239.

\bibitem[\protect\citeauthoryear{Chen and Tabri}{Chen and Tabri}{2021}]{ChenTabri}
Chen, R. and Tabri, R. V. (2021). 
\newblock Jackknife Empirical Likelihood for Inequality Constraints on Regular Functionals. 
\newblock {\em Journal of Econometrics\/} {\em 221}, 68--77.

\bibitem[\protect\citeauthoryear{Chen and  Kim}{Chen and  Kim}{2014}]{ChenKim}
Chen, S. and  Kim, J. K. (2014).
\newblock Population Empirical Likelihood for Nonparametric Inference in Survey Sampling.
\newblock {\em Statistica Sinica\/} {\em 24}, 335--355.

\bibitem[\protect\citeauthoryear{Chen}{Chen}{2007}]{Chen}
Chen, X. (2007). 
\newblock {\em Large Sample Sieve Estimation of Semi-Nonparametric Models\/}.
\newblock In: Heckman, J., Leamer, E. (Eds.), 
 {\em  Handbook of Econometrics, vol. VI}. Elsevier
Science B.V, pp. 5549--5632.

\bibitem[\protect\citeauthoryear{Chen  and Liao}{Chen  and Liao}{2015}]{ChenLiao}
Chen X. and Liao, Z. (2015). 
\newblock Sieve Semiparametric Two-Step GMM Under Weak
Dependence. 
\newblock {\em Journal of Econometrics\/} {\em 189}, 163--186.

\bibitem[\protect\citeauthoryear{Chen et al.}{Chen et al.}{2008}]{ChenHongTarozzi}
Chen, X., Hong, H. and Tarozzi, A. (2008). 
\newblock Semiparametric Efficiency in GMM Models With Auxiliary Data. 
\newblock {\em The Annals of Statistics\/} {\em 36}, 808--843.

\bibitem[\protect\citeauthoryear{Chen et al.}{Chen et al.}{2003}]{ChenLintonKeilegom}
Chen, X., Linton, O. B. and  Van Keilegom, I. (2003). 
\newblock Estimation of Semiparametric Models
When the Criterion Function is not Smooth. 
\newblock {\em Econometrica\/} {\em 71}, 1591--1608.


\bibitem[\protect\citeauthoryear{Chernozhukov et al.}{Chernozhukov et al.}{2018}]{ChernozhukovChetverikovDemirerRobins}
Chernozhukov, V., Chetverikov, D.,  Demirer, M.,  Duflo, E.,  Hansen, C.,  Newey, W.  and Robins, J.  (2018).
\newblock Double/Debiased Machine Learning for Treatment and Structural Parameters. 
\newblock {\em The Econometrics Journal\/} {\em 21}, C1--C68.

\bibitem[\protect\citeauthoryear{Chernozhukov et al.}{Chernozhukov et al.}{2021}]{ChernozhukovEscancianoIchimuraNewey}
Chernozhukov, V., Escanciano, J. C., Ichimura, H., Newey, W. K., and Robins, J. M. (2022). 
\newblock  Locally robust semiparametric estimation. 
\newblock {\em Econometrica\/} {\em 90(4)}, 1501--1535.


\bibitem[\protect\citeauthoryear{Corcoran}{Corcoran}{1998}]{Corcoran}
Corcoran, S. (1998). 
\newblock Bartlett Adjustment of Empirical Discrepancy Statistics. 
\newblock {\em Biometrika\/} {\em 85},
965--972.

\bibitem[\protect\citeauthoryear{Cressie and Read}{Cressie and Read}{1984}]{CressieRead}
Cressie, N. and T. Read. (1984). 
\newblock Multinomial Goodness-of-Fit Tests. 
\newblock {\em Journal of the Royal Statistical Society,\/} Series B {\em 46}, 440--464.

\bibitem[\protect\citeauthoryear{Davidson and Duclos}{Davidson and Duclos}{2000}]{DavidsonDuclos}
Davidson, R. and Duclos, J.  (2000). 
\newblock Statistical Inference for Stochastic Dominance and for the
Measurement of Poverty and Inequality. 
\newblock {\em Econometrica\/} {\em 68}, 1435--1464.

\bibitem[\protect\citeauthoryear{Dagdoug et al.}{Dagdoug et al.}{2021}]{DagdougGogaHaziza}
Dagdoug, M., Goga, C. and Haziza, D. (2021).  
\newblock Model-Assisted Estimation Through Random Forests in Finite Population Sampling. 
\newblock {\em Journal of the American Statistical Association}. DOI: 10.1080/01621459.2021.1987250


\bibitem[\protect\citeauthoryear{Frazier  and Renault}{Frazier  and Renault}{2017}]{FrazierRenault}
Frazier, D. T. and Renault, E.  (2017). 
\newblock Efficient Two-Step Estimation via Targeting. 
\newblock {\em Journal of Econometrics\/} {\em 201}, 212--227.

\bibitem[\protect\citeauthoryear{Fuller}{Fuller}{2009}]{Fuller}
Fuller, W. A. (2009). 
\newblock {\em Sampling Statistics}. Wiley, Hoboken, New Jersey.

\bibitem[\protect\citeauthoryear{Godambe and Thompson}{Godambe and Thompson}{1986}]{GodambeThompson}
Godambe, V. P.  and Thompson, M. E.  (1986).
\newblock Parameters of Superpopulation and Survey Population: Their Relationships and Estimation.
\newblock {\em International Statistical Review\/} {\em 54}, 127--138.


\bibitem[\protect\citeauthoryear{Goga and  Ruiz-Gazen}{Goga and  Ruiz-Gazen}{2014}]{GogaRuiz-Gazen}
Goga, C. and  Ruiz-Gazen, A. (2014). 
\newblock Efficient Estimation of Nonlinear Finite Population Parameters by Using Non-parametrics.  
\newblock {\em Journal of the Royal Statistical Society,\/} Series B {\em 76}, 113--140.



\bibitem[\protect\citeauthoryear{Hansen et al.}{Hansen et al.}{1996}]{HansenHeatonYaron}
Hansen, L., Heaton, J.  and  Yaron, A.  (1996).  
\newblock Finite-Sample Properties of Some Alternative GMM Estimators.
\newblock {\em Journal of Business and Economic Statistics\/} {\em 14}, 262--280.

\bibitem[\protect\citeauthoryear{Haziza et al.}{Haziza et al.}{2008}]{HazizaMecattiRao}
Haziza, D., Mecatti, F. and Rao, J. N. K. (2008). 
\newblock Evaluation of Some Approximate
Variance Estimators under the Rao-Sampford Unequal Probability
Sampling Design. 
\newblock {\em Metron\/} {\em 66}, 91--108.

\bibitem[\protect\citeauthoryear{Ichimura and Newey}{Ichimura and Newey}{2022}]{IchimuraNewey}
Ichimura, H., and Newey, W. K. (2022). 
\newblock The influence function of semiparametric estimators. 
\newblock {\em Quantitative Economics\/} {\em 13(1)},  29--61.

\bibitem[\protect\citeauthoryear{Imbens et al.}{Imbens et al.}{1998}]{ImbensSpadyJohnson}
Imbens, G., Spady, R.  and  Johnson, P.  (1998).  
\newblock Information Theoretic Approaches to Inference in Moment
Condition Models. 
\newblock {\em Econometrica\/} {\em 66}, 333--357.

\bibitem[\protect\citeauthoryear{Kitamura and Stutzer}{Kitamura and Stutzer}{1997}]{KitamuraStutzer}
Kitamura, Y. and Stutzer, M.  (1997).  
\newblock An Information-Theoretic Alternative to Generalized Method of Moments
Estimation. 
\newblock {\em Econometrica\/} {\em 65}, 861--874.

\bibitem[\protect\citeauthoryear{Matsushita and Otsu}{Matsushita and Otsu}{2020}]{MatsushitaOtsu}
Matsushita, Y. and Otsu, T. (2020). 
\newblock Likelihood Inference on Semiparametric Models
with Generated Regressors. 
\newblock {\em Econometric Theory\/} {\em 36}, 626--657.

\bibitem[\protect\citeauthoryear{Newey}{Newey}{1990}]{Newey}
Newey, W. (1990). 
\newblock Semiparametric Efficiency Bounds. 
\newblock {\em Journal of Applied Econometrics\/} {\em 5}, 99--135.

\bibitem[\protect\citeauthoryear{Newey}{Newey}{1994}]{Newey1994a}
Newey, W.  (1994a).
 \newblock  Kernel estimation of partial means and a general variance estimator. 
 \newblock {\em Econometric Theory\/} {\em  10},  233--253.

\bibitem[\protect\citeauthoryear{Newey}{Newey}{1994}]{Newey1994b}
Newey, W.  (1994b). 
\newblock The Asymptotic Variance of Semiparametric Estimators. 
\newblock {\em Econometrica\/} {\em 62}, 1349--1382.

\bibitem[\protect\citeauthoryear{Newey and Smith}{Newey and Smith}{2004}]{NeweySmith}
Newey, W. and Smith, R. J. (2004). 
\newblock Higher-Order Properties of GMM and Generalized Empirical Likelihood Estimators. 
\newblock {\em  Econometrica\/} {\em 72}, 219--255.

\bibitem[\protect\citeauthoryear{Nygard  and Sandstrom}{Nygard  and Sandstrom}{1989}]{NygardSandstrom}
Nyg\aa rd, F. and  Sandstr\"{o}m, A. (1989). 
\newblock Income Inequality Measures Based on Sample Surveys. 
\newblock {\em Journal of
Econometrics\/} {\em 42}, 81--95.

\bibitem[\protect\citeauthoryear{Oguz-Alper and Berger}{Oguz-Alper and Berger}{2016}]{Oguz-AlperBerger}
Oguz-Alper, M. and Berger, Y. G. (2016). 
\newblock Modelling Complex Survey Data with Population Level
information: An Empirical Likelihood Approach. 
\newblock {\em Biometrika\/} {\em 103}, 447--459.

\bibitem[\protect\citeauthoryear{Owen}{Owen}{1988}]{Owen}
Owen, A. B. (1988). 
\newblock Empirical Likelihood Ratio Confidence Intervals for a Single Functional.
\newblock {\em Biometrika\/} {\em 75}, 237--249.

\bibitem[\protect\citeauthoryear{Parente  and Smith}{Parente  and Smith}{2011}]{ParenteSmith}
Parente, P. M. and Smith, R. J. (2011). 
\newblock GEL Methods for Nonsmooth Moment Indicators. 
\newblock {\em Econometric Theory\/} {\em 27}, 74--113.

\bibitem[\protect\citeauthoryear{Qin and Lawless}{Qin and Lawless}{1994}]{QinLawless}
Qin, J. and Lawless, J. (1994).
\newblock  Empirical Likelihood and General Estimating Equations. 
\newblock {\em The Annals of Statistics\/} {\em 22}, 300--325.

\bibitem[\protect\citeauthoryear{Rao and Scott}{Rao and Scott}{1981}]{RaoScott}
Rao, J. N. K., and Scott, A. J. (1981).  
\newblock The Analysis of Categorical Data From Complex Sample Surveys: Chi-Squared Tests for Goodness of Fit and Independence in Two-Way Tables.
\newblock {\em Journal of the American Statistical Association \/} {\em 76}, 221--230.

\bibitem[\protect\citeauthoryear{Rao and Wu}{Rao and Wu}{1988}]{RaoWu}
Rao, J. N., and Wu, C. F. J. (1988). 
\newblock Resampling inference with complex survey data. 
\newblock {\em Journal of the American Statistical Association\/} {\em 83(401)}, 231--241.

\bibitem[\protect\citeauthoryear{Rao et al.}{Rao et al.}{1992}]{RaoWuYue}
Rao, J. N. K., Wu, C. F. J., and Yue, K. (1992). 
\newblock Some recent work on resampling methods for complex surveys. 
\newblock {\em Survey Methodology\/} {\em18(2)}, 209--217.

\bibitem[\protect\citeauthoryear{Rao and Mitra}{Rao and Mitra}{1971}]{RaoMitra}
Rao, C. R. and Mitra, S. K. (1971). 
\newblock  {\em Generalized Inverse of Matrices and Its Applications}. Wiley.

\bibitem[\protect\citeauthoryear{Rubin-Bleuer and Schiopu}{Rubin-Bleuer and Schiopu}{2005}]{Rubin-BleuerSchiopu}
Rubin-Bleuer, S. and Schiopu Kratina, I. (2005). 
\newblock On the Two-Phase Framework
for Joint Model and Design-Based Inference. 
\newblock {\em The Annals of Statistics\/} {\em 33}, 2789--2810.


\bibitem[\protect\citeauthoryear{Teitler et al. }{Teitler et al. }{2004}]{TeitlerGarfinkelGarciaKenney}
Teitler, J., Garfinkel, I., Garcia, S. and Kenney, S. (2004). 
\newblock {\em New York Social Indicators 2001: Growing Prosperity, Lingering Inequality. Social Indicators Survey Center}. Columbia University.


\bibitem[\protect\citeauthoryear{Van der Vaart}{Van der Vaart}{1991}]{Van der Vaart}
Van der Vaart, A. W. (1991). 
\newblock On Differentiable Functionals. 
\newblock {\em The Annals of Statistics\/} {\em 19}, 178--204.

\bibitem[\protect\citeauthoryear{Wu and Rao}{Wu and Rao}{2006}]{WuRao}
Wu, C. and Rao, J. N. K. (2006). 
\newblock Pseudo Empirical Likelihood Ratio Confidence Intervals
for Complex Surveys.
\newblock {\em The Canadian Journal of Statistics\/} {\em 34}, 359--375.

\bibitem[\protect\citeauthoryear{Wu and Thompson}{Wu and Thompson}{2020}]{WuThompson}
Wu, C. and Thompson, M. E. (2020).
\newblock  {\em Sampling Theory and Practice}. Springer, Cham.

\bibitem[\protect\citeauthoryear{Zhao and Wu}{Zhao and Wu}{2019}]{ZhaoWu}
Zhao, P. and Wu, C. (2019). 
\newblock Some Theoretical and Practical Issues with Empirical Likelihood Methods for Complex Surveys. 
\newblock {\em International Statistical Review\/} {\em  87}, S239--256.

\bibitem[\protect\citeauthoryear{Zhao et al. }{Zhao et al. }{2020}]{ZhaoHazizaWu}
Zhao, P.,   Haziza, D. and  Wu, C. (2020). 
\newblock  Survey Weighted Estimating Equation Inference With Nuisance Functionals. 
\newblock {\em Journal of Econometrics\/} {\em 216}, 516--536.

\bibitem[\protect\citeauthoryear{Zhao et al. }{Zhao et al. }{2022}]{ZhaoHazizaWu}
Zhao, P.,   Haziza, D. and  Wu, C. (2022). 
\newblock Sample Empirical Likelihood and the Design-Based
Oracle Variable Selection Theory. 
\newblock {\em Statistica Sinica\/} {\em  32}, 435--457. 

\bibitem[\protect\citeauthoryear{Zheng}{Zheng}{2002}]{Zheng}
Zheng, B. (2002). 
\newblock Testing Lorenz Curves with Non-Simple Random Samples. 
\newblock {\em Econometrica\/} {\em 70}, 1235--1243.

\bibitem[\protect\citeauthoryear{Zhong and  Rao}{Zhong and  Rao}{2002}]{ZhongRao2002}
Zhong, C.X.B. and Rao, J. N. K. (2000). 
\newblock  Empirical likelihood inference under stratified
sampling using auxiliary population information. 
\newblock {\em Biometrika\/} {\em 87}, 929--938.



 \end{thebibliography}
\end{document}